\documentclass[%
 reprint,
superscriptaddress,
 amsmath,amssymb,
 aps,
prb,
]{revtex4-2}

\usepackage{graphicx}% Include figure files
\usepackage{dcolumn}% Align table columns on decimal point
\usepackage{bm}% bold math
\usepackage{array}
\usepackage{amsmath,amsfonts,amssymb}
\usepackage[ansinew]{inputenc}
\usepackage{color}
\usepackage{calc}
\usepackage[normalem]{ulem}

\begin{document}

%\preprint{APS/123-QED}

\title{
Ultrafast optically induced tunneling in narrow metallic gaps from the time dependent density functional perspective\\
%Ultrafast optically induced tunneling in narrow metallic gaps addressed within the time dependent density functional theory\\
}

% Force line breaks with \\

\author{Boyang Ma}
\affiliation{Department of Physics, Technion---Israel Institute of Technology, Haifa 32000, Israel}
\affiliation{Solid State Institute, Technion---Israel Institute of Technology, Haifa 32000, Israel}
\affiliation{The Helen Diller Quantum Center, Technion---Israel Institute of Technology, Haifa 32000, Israel} 
\author{Antton Babaze}
\affiliation{Materials Physics Center, CSIC-UPV/EHU, Manuel de Lardizabal 5, 20018 Donostia, Spain}
\affiliation{Donostia International Physics Center, Manuel de Lardizabal 4, 20018 Donostia, Spain}
\affiliation{Department of Applied Physics, School of Architecture, UPV/EHU, 20018 Donostia, Spain}
\author{Michael Kr\"uger}
\affiliation{Department of Physics, Technion---Israel Institute of Technology, Haifa 32000, Israel}
\affiliation{Solid State Institute, Technion---Israel Institute of Technology, Haifa 32000, Israel}
\affiliation{The Helen Diller Quantum Center, Technion---Israel Institute of Technology, Haifa 32000, Israel}
\author{Javier Aizpurua}
\affiliation{Donostia International Physics Center, Manuel de Lardizabal 4, 20018 Donostia, Spain}
\affiliation{Department of Electricity and Electronics, FCT-ZTF, UPV/EHU, B$^o$ Sarriena s/n, 48940 Leioa, Spain}
\affiliation{IKERBASQUE, Basque Foundation for Science, 48009 Bilbao, Spain}
\author{Andrei G. Borisov}
 \email{andrei.borissov@universite-paris-saclay.fr}
 \affiliation{Institut des Sciences Mol\'eculaires d'Orsay (ISMO), UMR 8214, CNRS, Universit\'e Paris-Saclay, 91405 Orsay Cedex, France.}
      \affiliation{Donostia International Physics Center, Manuel de Lardizabal 4, 20018 Donostia, Spain}

\date{\today}% It is always \today, today,

\begin{abstract}

Electron tunneling through a potential barrier is a salient quantum effect behind 
multiple practical applications such as, for example, in electronics and scanning 
tunneling microscopy. Often considered within the quasi-static picture where the 
tunneling current flows through the system in response to an applied dc field, 
electron tunneling can be brought into the realm of ultrafast phenomenona  
when triggered by the electric field of a short optical pulse. 
Ultrafast scanning tunneling microscopy thus emerges as a result of the 
combination of the ultimate spatial and temporal resolution, offering 
unprecedented perspectives for studying electron and phonon dynamics at surfaces. 
%Thus ultrafast scanning tunneling microscopy can be introduced, which combines 
%the ultimate spatial and temporal resolution and offers unprecedented perspectives 
%for studying electron and phonon dynamics at surfaces. 
In this work, using the 
time-dependent density functional theory, we address the electron tunneling 
triggered by short (single-cycle and several-cycle) optical pulses in narrow 
metallic gaps under conditions relevant for actual experiments. We identify 
photon-assisted tunneling with one-photon, two-photon, and higher-order photon 
absorption, and discuss the effect of the tunneling barrier, applied bias, and 
strength of the optical field on the transition from photon-assisted tunneling 
(weak optical fields) to the optical field emission at strong optical fields. 
Numerical single-electron calculations and an analytical strong-field theory 
model are implemented to gain deeper insights into the results of the 
time-dependent density functional theory calculations. 
Additionally, our parameter-free calculations allow us to retrieve and explain
recent experimental results on optically induced transport in narrow metallic 
gaps under an applied dc bias.

%the one-step Tucker’s approach to photon assisted tunneling.   

%The electron tunneling through the potential barrier is a silent 
%quantum effect behind multiple practical applications such as, for 
%example, in electronics and scanning tunneling microscopy. Often looked at 
%within the quasi-static picture where the tunneling current flows 
%through the system in response to an applied dc field, the electron 
%tunneling can be brought into the realm of ultrafast phenomenona  
%when triggered by the electric field of the short optical pulse. Thus 
%introduced ultrafast scanning tunneling microscopy combines the 
%ultimate spatial and temporal resolution and offers unprecedented 
%perspectives for studying electron and phonon dynamics at surfaces. 
%In this work, using time-dependent density functional theory we 
%address the electron tunneling triggered by short (single-cycle and 
%several-cycle) optical pulses in narrow metallic gaps under conditions 
%relevant for actual experiments. 
%We identify the photon assisted tunneling with one-, two-, ... , and  
%n-photon absorption, and we discuss the effect of the tunneling barrier, 
%applied bias, and strength of the optical field on transition 
%from photon assisted tunneling (weak optical fields) to the optical 
%field emission at strong optical fields. Additionally, our parameter-free 
%calculations allow us to retrieve recent experimental results on the 
%optically induced transport in narrow metallic gaps. 

\end{abstract}

\maketitle

%\tableofcontents

\section{Introduction \label{sec:Intro}}

Combining the atomic-scale resolution of the scanning tunneling microscope (STM) 
with the time resolution offered by short and intense optical pulses has given 
rise to the field of ultrafast scanning tunneling microscopy (USTM), which offers 
numerous opportunities to study intrinsic and photon-induced dynamics at surfaces 
\cite{Yoshida2019,Arashida2022,Wolf_SciAdv2022,Garg2022,Luo2024,Luo2025,
maier2025STM2color,RevModPhys.92.025003,MULLER_REVIEW,Vogelsang_2025,Zhao2025}. 
This is without mentioning the burgeoning field of THz-assisted STM (THz STM)
\cite{cocker2013ultrafast,Yoshioka2016,Cocker2016,Roelcke2024,Jelic2024,Sheng2024}.
The development of experimental techniques calls for theoretical approaches 
capable to account for the main physical processes at play, to analyze and to explain 
the experimental data on a parameter-free basis, and to propose new experiments.  
Thus, the description of the USTM has to address the lightwave-driven 
electron currents in metallic junctions for a wide range of experimental 
conditions. The role of the size of the junction, of the applied bias, of the 
optical field strength, and of the waveform of the optical transient are of interest. 
% In this respect, the growing interest in phenomena evolving on attosecond time-scales 
% as well as in coherent aspects relevant for coherent control such as in two colour 
% experiments \cite{GargScience2020,maier2025STM2color,davidovich2025STM2color} sets the 
% explicitly time-domain quantum approaches as the methods of choice. 

\begin{figure*}[t!]
\centering
\includegraphics[width=1.\linewidth]{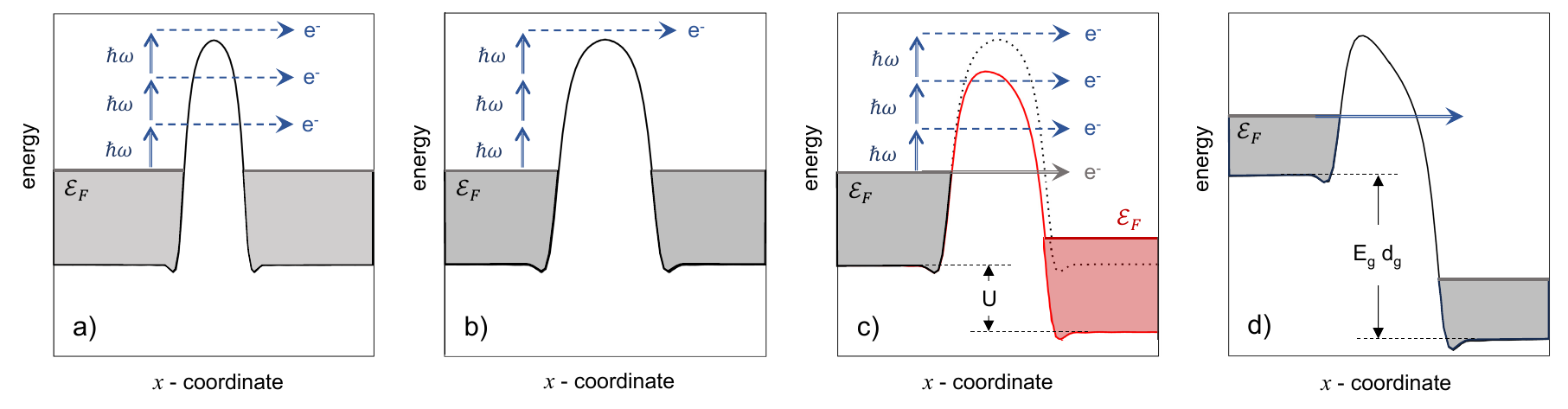}
\caption{Sketch of the processes behind the electron transport in the gap 
under different conditions. The potential of the junction between two metallic 
leads is shown as function of the $x$-coordinate perpendicular to the surface. 
$\mathcal{E}_F$ stands for the Fermi energy, $\hbar \omega$ is the energy of an 
absorbed photon.
\textbf{Panel a}: The multiphoton regime of electron transport for narrow 
junction (weak fields). No bias is applied. The $\ell$-photon absorption 
(vertical solid arrows, $\hbar \omega$) can lead to tunneling through 
the potential barrier reduced by $\ell \hbar \omega$ (here $\ell=1$ or $\ell=2$), 
or to a classically allowed over-the-barrier transition when $\ell \hbar \omega$ 
is larger than the height of the tunneling barrier (here $\ell=3)$. The different 
$\ell$ channels of electron transfer are shown with dashed blue arrows. 
\textbf{Panel b}: The same as panel a, but the width of the junction is large.  
The electron tunneling contribution is negligible. Instead, electron transport 
is dominated by classically allowed over-the-barrier transitions. 
The process can be seen as electron emission followed by electron 
propagation in the vacuum gap.
\textbf{Panel c}: The same as panel b, but a dc bias $U$ is applied. 
This reduces the potential barrier and permits tunneling. Along with 
over-the-barrier transitions assisted by multiphoton absorption, the dc tunneling 
(solid grey arrow) or tunneling induced by $\ell$-photon absorption is possible.
\textbf{Panel d}: The optical field emission regime (strong fields) where an 
electron tunnels through the potential barrier reduced by the optical field 
(solid blue arrow).  
}
\label{fig:Processes}
\end{figure*}

For a wide gap between metals, the lightwave-induced electron transport 
involves electron emission into the vacuum. This process can thus be understood
using theories of conventional and time-resolved photoemission from surfaces 
\cite{schattke2003solid,PetekPEEM,BAUER2015} (weak optical fields), or using 
theoretical approaches developed in the context of the interaction of 
short and intense laser pulses with metallic tips, plasmonic nanoparticles 
or plasmonic structures with gaps of $\gtrsim10$~nm widths \cite{Lemell_CEP_2003,
Ropers2007,Bormann2010,Kruger2012,park2012,Dombi2013,Passig2017,Kruger2018,
RevModPhys.92.025003,Keathley_VanishingCEP_2019,Ciappina2017AttoPhys,
Dienstbier2023,Heide2024}. While such a wide junction might be produced in USTM, 
the most interesting regime of operation of the device associated with atomic-scale 
resolution is the tunneling regime with a nanometric size of the junction. In 
this situation, the lightwave-induced electron transport in the system 
results not only from the electron emission above the tunneling barrier separating the 
metal surfaces, but also from electron tunneling assisted by 
absorption of one or several photons, as illustrated in Figure~\ref{fig:Processes}.

%
% FIGURE 1
%

Lightwave-assisted electron tunneling is often discussed within 
the framework of the perturbative continuous-wave (cw) approximation 
such as the Landauer-B\"{u}ttiker theory \cite{LandauerButtiker,PedersenButtiker} 
as well as the Tien-Gordon theory of photon-assisted tunneling 
(PAT) \cite{TienGordon,Tucker,Platero,KOHLER2005379}. 
Using the scattering theory formulation of Pedersen and B\"{u}ttiker \cite{PedersenButtiker},  
%Using the formulation of Tucker \cite{Tucker}, 
the zero-frequency electron current  $\mathcal{I}$ induced by cw light of frequency 
$\omega$ between two metallic leads separated by a distance $d_{\rm{gap}}$ results from 
transitions involving electronic states ``dressed" by the optical field. 
%
%\begin{align}\label{eq:Tucker}
%  \mathcal{I} = \sum_{\ell=-\infty}^{\infty} J_\ell^2\left(\frac{V_{\rm{g}}}{\hbar\omega} \right)~ 
%  I_{\rm{dc}}\left(U+\ell \hbar \omega \right) - I_{\rm{dc}}\left(U\right), 
%\end{align}
%
%
\begin{align}\label{eq:Pedersen}
  \mathcal{I}  
  = - I_{\rm{dc}}\left(U\right) +
  &\sum_{\ell=-\infty}^{\infty} J_\ell^2\left(\frac{V_{\rm{g}}}{\hbar\omega} \right)~ 
   \int d\mathcal{E} ~ T(\mathcal{E}) \nonumber \\
  &\times \left[f_L(\mathcal{E}-\ell \hbar \omega) - f_R(\mathcal{E}+U)\right], 
\end{align}
where the summation runs over the electron transport channels associated with 
absorption ($\ell>0$) or emission ($\ell<0$) of $\ell$ photons. 
In Eq.~\eqref{eq:Pedersen}, $\mathcal{E}$ is the electron energy, 
$\ell \hbar \omega$ is the electron energy change associated 
with the $\ell$ photons exchange with the field,  $J_\ell$ is the $\ell$th order 
Bessel function, $U$ is the applied dc bias, $V_{\rm{g}} = E_{\rm{g}} d_{\rm{gap}} $ 
is the optical bias (with $E_{\rm{g}}$  the optical field amplitude in the junction), 
$f_L(\mathcal{E})$ ($f_R(\mathcal{E})$) is the 
Fermi distribution of the left (right) lead, $T(\mathcal{E})$ is the elastic transmission
coefficient, and $I_{\rm{dc}}(U)$ is the dc current-voltage characteristic 
of the tunneling junction.  In the perturbative regime $E_{\rm{g}}$ is 
small, and the weight of the electron transport channels quickly decreases with 
increasing $\ell$, indeed 
$J^2_{\ell}(\frac{E_{\rm{g}}d_{\rm{gap}}}{\hbar \omega}) \propto  E_{\rm{g}}^{2\ell}$.
Obviously, the cw approximation is not suited 
for short optical pulses. Furthermore, a perturbative theory does not allow one 
to address strong optical fields.

Eq.~\eqref{eq:Pedersen} admits an interpretation where
$J_\ell^2\left(\frac{V_{\rm{g}}}{\hbar\omega} \right)$ can be seen as
the probability of electron excitation by $\ell$ photon absorption. 
As a result of the $\ell$ photon absorption, the energies of the 
occupied electronic states extend up to $\mathcal{E}_F+\ell \hbar \omega$ 
($\mathcal{E}_F$ stands for the Fermi energy). 
%, and the second term $I_{\rm{dc}}\left(U+\ell \hbar \omega \right)$ as the probability 
%of electron transport through the tunneling barrier from the excited state. 
In this way, one arrives at another approximation 
%(below we will refer to it as two-step model) 
often used to explain experimental data where (i) absorption 
of $\ell$-photons results in a non-equilibrium energy distribution of excited 
(hot) electrons, and (ii) an electron tunnels from an initial state within this 
non-equilibrium distribution through the static potential barrier 
\cite{PhysRevLett.103.257603,STM_Light_Schroder_2020,Light_STM_Kumagai23,Luo2024,
MULLER_REVIEW}. Using adjustable parameters allows one to describe the experimental 
data \cite{STM_Light_Schroder_2020,Light_STM_Kumagai23,Luo2024} \textit{a priori} 
without limitation in the pulse duration, as well as to incorporate transport 
and relaxation of excited electrons in solids and thermionic emission
\cite{Light_STM_Muller22}, as broadly discussed nowadays 
in connection with plasmon-induced chemistry \cite{TagliabueGiulia2018,
Dubi2019,Schirato2023,Khurgin2024,TagliabueGiulia2024,Stefancu2024}. 
This said, electron transport appears within this approach as 
an incoherent process comprising sequence of two steps. 
This is while a number of recent experiments report on coherent 
control of electron transport in narrow metal junctions and of electron 
emission from metal surfaces \cite{GargScience2020,maier2025STM2color,
davidovich2025STM2color,PhysRevLett.117.217601,Paschen2017,
DienstbierPaschenHommelhoff,PhysRevLett.126.137403,Dienstbier2023,
Goulielmakis_Nature2023}.

The coherent aspects of electron tunneling and electron emission are 
naturally accounted for within a theoretical framework where 
the excitation and emission (or tunneling) are intimately linked and 
cannot be separated in time. 
%(below we will refer to this description as one-step model). 
Indeed, consider the narrow gap between the metals which allows for tunneling. 
%The one-step model directly incorporates that 
In this situation, the photon 
absorption is associated to the electron excitation into the electronic 
state which is delocalized in the two leads (or in metal and vacuum in the 
case of photoemission). The electron escape from the ``parent" metal 
surface starts therefore as soon as the optical field couples the 
initial and excited electronic states \cite{Yalunin2011,Kruger2012,
Wachter2012,IvanovGap2021,Kim_2021,PhysRevApplied.17.044008,Ritzkowsky2024,
PhysRevBbias,HotElectrons_in_Gaps,ma2025robust,maier2025STM2color,
davidovich2025STM2color}.
The theoretical methods applied in this context are often based on the 
strong-field approximation (SFA) \cite{Reiss2008,Milosevic_2006,RevModPhys.92.025003}, 
which allows to address the situation of short and intense optical 
pulses \cite{Bormann2010,Yalunin2011,Ma_Kruger2024,ma2025robust,GargScience2020} 
also of interest here. Along with the SFA, methods based on semi-analytical 
\cite{Kim_2021,PhysRevApplied.17.044008,Ma_Kruger2024} or 
numerical \cite{Turchetti_21,Thon2004A,Ma_Kruger2024,ma2025robust}
solution of a model one-electron Schr\"{o}dinger equation for a  
single electron active in the transition dynamics have also been reported.

Applying the theoretical techniques described above is very enlightening as it 
allows one to discuss various aspects of optically induced electron transport.  
On the other hand, an extended parametrization of the system, 
\textit{ad hoc} assumptions concerning the field screening in metal, and the 
often imposed reduction of the many-body problem to the one-electron problem 
does not allow for robust quantitative analysis of the main mechanisms 
involved. This points at the interest in studying optically induced electron 
currents in tunneling junctions using quantum time-dependent many-body methods, 
particularly under conditions approaching experimental ones. 
The explicitly time-dependent treatment allows one to naturally describe the 
situation of short and intense optical driving pulses where linear approximations 
fail. Reducing parameters of the theory and avoiding adjustment to the 
experiment allows one to quantify various transport channels with the ultimate 
goal to predict the optimal settings (the waveform of the optical pulse, the size 
of the junction, and an applied bias) for a desired task of USTM.

In this work, we build on our earlier time-dependent density functional 
theory (TDDFT) studies of the dynamics of the electron transport in  
junctions between free-electron metal surfaces. While previous implementations 
focused on few-nanometer and wider gaps, where electron transport is primarily 
governed by over-the-barrier transitions \cite{Ludwig2019,HotElectrons_in_Gaps},
here we explore the regime of PAT by further reducing the gap size, or 
lowering the barrier with an applied bias. In this regime, we can
clearly identify PAT channels corresponding to one-, two-, and higher-order
photon absorption processes, in full agreement with recent experiments.
We address the evolution of the tunneling mechanisms with optical field 
intensity of single-cycle and longer optical pulses, covering situations 
extending from photon-driven to field-driven transport. 
Importantly, once the free-electron description of the metal junction 
is set in the ground state by the choice of the Wigner-Seitz radius and 
the work function, no additional parameters are used in the theory.

To provide  an analytical framework for interpretation of the TDDFT 
results, we extend the strong-field theory (SFT) developed in previous 
work \cite{Ma_Kruger2024, HotElectrons_in_Gaps, ma2025robust} by 
incorporating a static bias. In addition, we perform model calculations 
based on a one-active-electron description of optically induced electron 
transport. By comparing parameter-free, self-consistent, 
many-body TDDFT results with experimental data, and by using additional 
insights from analytical and numerical model approaches, we are able to 
discuss the underlying physical mechanisms. We believe that this 
study contributes to a deeper understanding of ultrafast optically induced 
tunneling, the process of interest for USTM and optoelectronic applications.

The paper is organized as follows. 
Section 1 provides an introduction. Section 2 describes the theoretical 
methods used in this work. Section 3 presents and discusses the results 
for both single-cycle and longer optical pulses, with a focus on the 
underlying intrinsic electron dynamics. Section 4 compares the theoretical 
findings with recent experimental data and discusses several observed 
phenomena in detail. Finally, Section 5 summarizes the main conclusions 
of the work.

Atomic units (a.u.) are used below in this paper unless otherwise 
stated. However, to underline that the change in electron energy 
results from the (multi-)photon absorption, in the figures we 
nonetheless keep the $\hbar \omega$ notation for electron excitation 
energies ($\hbar$=1 in atomic units). The electron energies are 
measured with respect to the vacuum level of the left lead.

%%%%%%%%%%%%%%%%%%%%%%%%%%%%%%%%%%%%%%%%% 

%%%%%%%%%%%%%%%%%%%%%%%%%%%%%%%%%%%%%%%%% 

\section{Model system and methods} \label{sec:Model}

Below, we describe the theoretical methods used in this work to address 
optically induced electron tunneling and electron transport in narrow 
junctions between metal surfaces. 
Readers primarily interested in the results and their comparison with experiment, 
presented in Section~\ref{sec:SingleCyclePulse} and Section~\ref{sec:experiment}, 
may wish to consult the beginning of Subsection~\ref{subsec:TDDFT}, which 
introduces the system under study (see Fig.~\ref{fig:Geometry}), and then 
proceed directly to Subsection~\ref{subsec:TheorySummary}. The latter 
provides a concise comparative overview of these methods 
and clarifies their respective contributions to the present study.

\subsection{TDDFT} \label{subsec:TDDFT}

While many-body approaches (including TDDFT) with a fully atomistic description 
of the leads have been reported in the context of lightwave-driven 
tunneling \cite{STM_CEP_Mukamel,PhysRevB.105.085416,Chen2018,maier2025STM2color}, 
their high computational cost explains the small size of the systems treated 
(far from experimental geometries) and the limited number of the phenomena 
considered so far. These works mainly focused on the optical field emission 
regime and carrier envelope phase (CEP) dependence of the electron transport. 
An identification of the PAT channels associated with $\ell$-photon absorption 
and a study of their evolution with optical field strength, gap width, and  
applied bias have not been performed. Being at the focus of current experimental 
interest \cite{Diesing2004B,GargScience2020,STM_Light_Schroder_2020,
Light_STM_Muller22,Light_STM_Kumagai23,Luo2024,NatureRepp2024,MULLER_REVIEW} 
these aspects can be addressed using TDDFT for the electron density dynamics 
and a free-electron (jellium) model of the metal. The simplification of the 
metal description allows one to address relatively large nanoobjects relevant 
for actual experiments. Thus, finite size effects as they appear, e.g., in the 
earlier 1D study based on the configuration interaction singles method 
\cite{Thon2004A} can be avoided. While lacking the atomistic structure of the 
metal/vacuum interface and d-band electron excitations in noble metals, the 
jellium TDDFT correctly captures the valence electron dynamics triggered by 
the optical pulses. This simple yet representative approach has been shown 
to predict and to semi-quantitatively describe experimental findings on the 
optical response of plasmonic structures with narrow gaps or on electron 
photoemission in strong-field nanooptics \cite{Lemell_CEP_2003,PDombi2004,
Apolonski04,Wachter2012,Ludwig2019,PhysRevLett.117.217601}.

\begin{figure}[t!]
\centering
\includegraphics[width=0.9\linewidth]{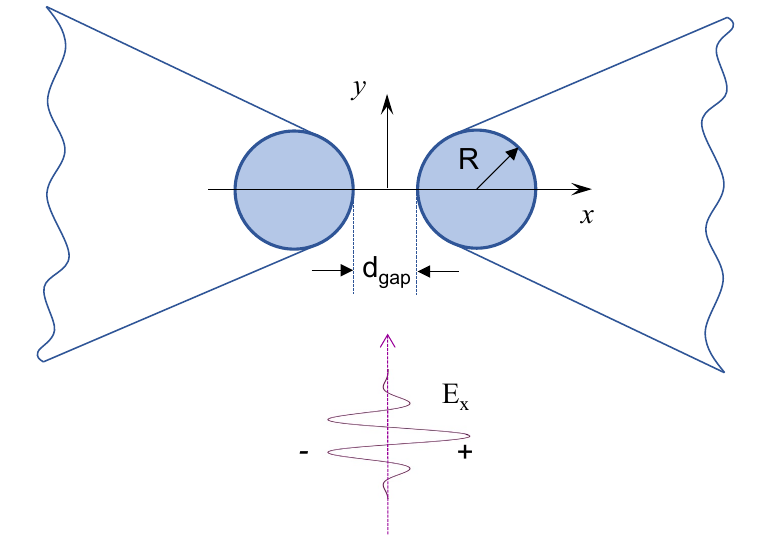}
\caption{Sketch of the studied system. 
Cross section $(x,y)$ of a dimer of two identical parallel cylindrical 
nanowires (nanowire radius $R=5$~nm) infinite along the $z$-axis. The 
nanowires are separated by a narrow gap of width $d_{\rm{gap}}$ in the 
(sub-)nm range. The middle of the gap is located at $(x=0,y=0)$. 
An $x$-polarized optical pulse is incident on the nanowires along the 
$y$-axis leading to optically induced  electron transport across the 
gap. The inset shows the $x$-component of the 
time-dependent electric field of the single-cycle pulse with CEP$=0$.  
}
\label{fig:Geometry}
\end{figure}

The method employed here has been detailed in a number of 
publications \cite{TeperikCylDimer,Ludwig2019,HotElectrons_in_Gaps}. 
We consider a system schematically shown 
in Figure~\ref{fig:Geometry}. Two identical parallel cylindrical 
nanowires of radius $R$ are infinite along the $z$-axis and 
separated by a narrow junction of width $d_{\rm{gap}}$. We use $R=5$~nm,  
which allows us to retrieve the geometry of the junction formed 
by bow-tie antennas as studied in a number of experimental works devoted 
to photoemission and electron transport across nanoantenna gaps 
\cite{rybka2016,Racz2017,Ludwig2019,IvanovGap2021,Luo2024,Ritzkowsky2024}. 
The electronic structure of the nanowires is modelled using the free-electron 
description within the stabilized jellium model \cite{perdew1990stabilized}. 
The ions at the lattice sites are not treated exactly, but represented with 
a uniform positive background density. The density parameter is given by the 
Wigner-Seitz radius $r_s=3.02$~a$_0$ typical for silver and gold 
(a$_0$=1~a.u. $\approx 0.529$~\AA~is the Bohr radius).
%to model e.g. the lightwave induced transport studied in this 
%work (the edge effects because of the finite height of the experimental 
%device are neglected). 

%
% FIGURE 2
%

The valence electron dynamics is driven by an incident 
$x$-polarized optical pulse with electric field
\begin{equation}\label{eq:Gaussianfield}
E(t)=E_0~ e^{-t^2/\tau^2} \cos(\omega t + \varphi).
\end{equation}
Here, $\varphi$ is the carrier envelope phase (CEP) of the pulse, and $\tau$ 
is the pulse duration. In Figure~\ref{fig:Geometry}, we show $E(t)$ for a 
single-cycle optical pulse with $\varphi=0$ and $\tau= 0.85 T$, where 
$T=\frac{2 \pi}{\omega}$ is the optical period. To study the system evolution 
in time we employ the Kohn-Sham (KS) scheme of TDDFT and the adiabatic local 
density approximation (ALDA) \cite{MarquesGross} with the exchange--correlation 
kernel of Gunnarsson and Lundqvist \cite{gunnarsson1976exchange}. Retardation 
effects are neglected because of the small relevant size of the system. 
The time-dependent electron density $n(\mathbf{r},t)$ and the time-dependent 
electron current density $\mathbf{j}(\mathbf{r},t)$ are sought as  
\begin{align}\label{eq:n_and_J}
n(\mathbf{r},t) &= \sum_{k \subset occ} 
\chi_k \left|\Psi_k(\mathbf{r},t)\right|^2, \nonumber \\
\mathbf{j}(\mathbf{r},t) &=\sum_{k \subset occ} 
\chi_k \text{Im}\left[\Psi^*_k(\mathbf{r},t)\mathbf{\nabla}\Psi_k(\mathbf{r},t)\right],  
\end{align}
where $\mathbf{r}=(x,y)$, and the system is invariant with respect to 
translation along the $z$-axis.
The sum runs over the initially occupied KS orbitals $\Psi_k(\mathbf{r},t)$, 
the statistical factor $\chi_k$ accounts for the $\pm \frac{1}{2}$ electron spin 
degeneracy and the degeneracy associated with motion along $z$-axis, $Z^*$ stands  
for the complex conjugate of the complex number $Z$, and $\text{Im}\left[Z\right]$ 
stands for the imaginary part of $Z$.

The time evolution of the KS orbitals is given by the time-dependent KS 
equations 
\begin{align}\label{eq:TDDFT_KS}
 & i \partial_t \Psi_k(\mathbf{r},t) = \nonumber \\
 & \left( -\frac{1}{2} \nabla^2 + V(\mathbf{r},t) +V_{\rm{opt}}(x,t) \right) 
 \Psi_k(\mathbf{r},t).
\end{align}
The potential of the optical pulse is $V_{\rm{opt}}(x,t) = x~E(t)$. 
The self-consistent TDDFT potential $V(\mathbf{r},t)$ is given by the sum 
of several contributions 
\begin{equation}\label{eq:TDDFT_KS_Potential}
  V(\mathbf{r},t) = V_{\rm{H}}(\mathbf{r},t) + V_{\rm{XC}}(\mathbf{r},t) 
  +v_{\rm{stab}}(\mathbf{r}).
\end{equation}
Here, the Hartree potential, $V_{\rm{H}}$, and the exchange correlation potential, 
$V_{\rm{XC}}$, depend on time through the time-dependent electron density.
The time-independent stabilization potential, $v_{\rm{stab}}(\mathbf{r})$, 
allows to fix the desired value of the work function $\Phi$ of the nanowires.

The KS orbitals $\Psi_k(\mathbf{r},t)$ are represented on an equidistant 
mesh in $(x,y)$ coordinates with a typical mesh step of $h_x=h_y=0.8$~a$_0$.  
Eq.~\ref{eq:TDDFT_KS} is then solved using the Fourier-grid 
technique \cite{kosloff1983fourier} and short-time split-operator 
propagation \cite{Leforestier1991} (typical time step $\Delta t=0.125$~a.u.) as detailed 
elsewhere \cite{TeperikCylDimer,Ludwig2019,HotElectrons_in_Gaps}. The initial conditions 
for the time propagation $\Psi_k(\mathbf{r},t \rightarrow -\infty)$ are given by the 
occupied KS orbitals of the ground-state (gs) system, $\psi_{k,\rm{gs}}(\mathbf{r})$, 
calculated from the Density Functional Theory (DFT) \cite{Dreizler1990} as 
\begin{align}\label{eq:DFT_KS}
\left( -\frac{1}{2} \nabla^2 + V_{\rm{gs}}(\mathbf{r}) \right) \psi_{k,\rm{gs}}(\mathbf{r}) 
= \mathcal{E}_k \psi_{k,\rm{gs}}(\mathbf{r}).
\end{align}
The same exchange--correlation kernel, and stabilization potentials as in TDDFT 
are used so that $V_{\rm{gs}}(\mathbf{r})=V(\mathbf{r},t \rightarrow -\infty)$. 
The energies of the occupied orbitals satisfy $\mathcal{E}_k \leq \mathcal{E}_F$.

From the calculated electron current density $\mathbf{j}(\mathbf{r},t)$ we obtain
the net electron transfer $\mathcal{N}$ per optical pulse and per nm length of the 
nanowire dimer. It is given by the time integral of the total 
current through the $(x=0,y,z)$ plane located at the middle of the junction
\begin{equation}\label{eq:NetTransfer}
  \mathcal{N}=\int \,dt \, \int \,dy ~ \hat{e}_x \cdot \mathbf{j}(x=0,y,t) 
   \times \frac{\rm{nm}}{\rm{a}_0}, 
\end{equation}
where $\hat{e}_x$ is the unit length vector in the positive direction of the $x$-axis. 
The last term accounts for the unit conversion. With this definition, $\mathcal{N}$ is 
positive (negative) when electrons are transferred in the 
positive (negative) direction of the dimer axis ($x$-axis).

In addition to the case where the net electron transfer is only possible 
due to the dimer symmetry break by the optical field of a single-cycle laser 
pulse, we also investigate the role of an applied dc bias. The dc bias 
breaks the symmetry of the dimer and enables the net optically induced 
electron transfer even for longer optical pulses. Accordingly, in addition 
to the optical pulse given by Eq.~\ref{eq:Gaussianfield}, the dimer is 
subjected to a slowly varying external THz field directed along the 
negative $x$-axis.
In Eq.~\eqref{eq:TDDFT_KS} this is implemented by replacing 
$V_{\rm{opt}}(x,t)$ with $V_{\rm{opt}}(x,t)+V_{\rm{THz}}(x,t)$, where the 
potential owing to the THz field, $V_{\rm{THz}}(x,t)$, is slowly 
switched on from zero to a constant value reached at time $t_-$,  
prior to the arrival of the optical pulse. The resulting effective dc bias 
$U$ is defined as the difference between the self-consistent potentials 
inside the left and right cylinders at $t=t_-$. 
Further details are given in Appendix~\ref{Apd.THZ}.

When only the THz pulse is present and no optical field is applied, for $t>t_-$ 
the system reaches the steady-state where the effective dc bias $U$ leads to a 
small (in our conditions) tunneling electron current density 
$\mathbf{j}_U(\mathbf{r},t)$. When optical and dc fields are acting on the system 
simultaneously, the electron current density $\mathbf{j}(\mathbf{r},t)$ given by 
Eq.~\eqref{eq:n_and_J} includes the effect of both. Since in this work are 
interested in the optically induced electron transport, we therefore introduce 
the optically induced electron current density as 
$\mathbf{j}_{\rm{opt}}(\mathbf{r},t)=\mathbf{j}(\mathbf{r},t)
-\mathbf{j}_U(\mathbf{r},t)$, and we use 
$\mathbf{j}_{\rm{opt}}(\mathbf{r},t)$ instead of $\mathbf{j}(\mathbf{r},t)$ 
in Eq.~\ref{eq:NetTransfer} to obtain the optically induced net electron transfer 
$\mathcal{N}$. Note that this correction is only essential to disentangle the 
$\ell=1$ electron tunneling channel, assisted by one-photon absorption at low 
optical field strengths. In the multi-photon regime, $\mathbf{j}_U(\mathbf{r},t)$ 
is negligible in our conditions.

To simplify control of the time-dependent electric field in the gap via the 
incident pulse waveform, the present system is designed so that the 
dipolar plasmon of the nanowire dimer is off-resonant with the incident 
optical pulse \cite{HotElectrons_in_Gaps}. As a result, the time-dependent 
electric field along the dimer axis at the center of the narrow gap is 
free from the plasmon ringing effects, and it can be approximated as 
\begin{equation}\label{eq:Egap}
  E_{\rm{gap}}(t) = \mathcal{R} E(t) 
  = E_{\rm{g}}~ e^{-t^2/\tau^2} \cos(\omega t + \varphi), 
\end{equation}
where $\mathcal{R}$ is the field enhancement. To set the order of magnitude, 
$\mathcal{R}\approx 4$ for $d_{\rm{gap}}=1$~nm and $\mathcal{R}\approx 3$ for 
$d_{\rm{gap}}=2$~nm, within the range of other parameters (frequency, bias, 
metal work function) encompassed in this work.

When relating our model system to experimentally accessible
devices, such as bowtie antennas or STM junctions, it is important to note that 
the strongly nonlinear photoemission processes in those systems are primarily 
governed by the gap region, where the electric field enhancement is the 
strongest. In addition, PAT is highly sensitive to the height 
and width of the potential barrier, both of which are minimized along the 
shortest path between metal leads (the $x$-axis in our geometry). One thus 
can conclude that the electron transfer between nanowires owing both to the 
tunneling and photoemission is determined by the current flowing through the 
junction \cite{Ludwig2019,HotElectrons_in_Gaps}.
This observation has two important implications. First, as discussed in our 
earlier work \cite{HotElectrons_in_Gaps}, the field amplitude 
$E_{\rm{g}} = \mathcal{R} E_0$ is the relevant parameter to characterize 
the lightwave-induced electron transfer, which we will refer to throughout this 
paper. Second, the parallel nanowire configuration 
addressed here is representative for understanding ultrafast electron dynamics 
in more complex nanostructures. In particular, the TDDFT results obtained here 
are relevant to narrow-gap bowtie antennas \cite{rybka2016,Ludwig2019,Luo2024} and 
to optically-driven STM junctions \cite{GargScience2020,STM_Light_Schroder_2020,
Light_STM_Muller22,Light_STM_Kumagai23,Luo2024}, which constitute one of 
the main motivations for this work.

%Second consequence concerns generality of the present work since 
%the TDDFT results obtained using the parallel nanowire geometry appear 
%representative not only for the narrow gap bowtie antennas 
%\cite{rybka2016,Ludwig2019,Luo2024} but also for the 
%lightwave induced electron transfer in the STM junctions 
%\cite{GargScience2020,STM_Light_Schroder_2020,Light_STM_Muller22,
%Light_STM_Kumagai23,Luo2024}, one of the objectives of our work. 

%%%%%%%%%%%%%%%%%%%%%%%%%%%%%%%%%%%%%%%%% 
 
\subsection{One-active-electron approach} \label{subsec:WPP}

In addition to the many-body TDDFT calculations, we also use a simplified 
one-dimensional model based on the single-active-electron approximation as 
often employed to analyze multiphoton and optical field emission regimes 
in the context of electron emission from nanotips or transfer across metallic 
gaps \cite{Thon2004A,Yalunin2011,PhysRevLett.97.247402,PhysRevApplied.17.044008,
GargScience2020,IvanovGap2021,Dienstbier2023,Ma_Kruger2024,ma2025robust,
RevModPhys.92.025003,Ciappina2017AttoPhys}. Despite its simplicity, where often 
(but not always \cite{Turchetti_21,Eldar_2022,Yeung2024}) the entire valence 
band is represented by a single orbital, and \textit{ad hoc} 
screening of the optical field is imposed, the method allows one to grasp the 
qualitative physical effects involved in the transport process with minimal 
computational cost. It also provides direct access to intuitive information 
such as the single-particle dynamics, which is masked in fully many-body TDDFT 
simulations.

The method, referred to here as wave-packet propagation (WPP) approach, is 
based on the solution of the one-electron time-dependent Schr\"{o}dinger 
equation (TDSE) for an electron active in the lightwave-induced transport through 
the junction between the two one-dimensional potential wells representing the
metallic leads. A detailed description of this technique can be found 
elsewhere \cite{HotElectrons_in_Gaps}. In brief, the wave function 
$\psi(x,t)$ of an electron ``active" in optically induced 
transitions and initially occupying the Fermi level of one of the leads 
evolves in time according to 
\begin{align}\label{eq:WPP}
 & i \partial_t \psi(x,t) = \nonumber \\
 & \left( -\frac{1}{2} \frac{\partial^2}{\partial x^2} + V_{\rm{m}}(x) 
 +\tilde{V}_{\rm{opt}}(x,t) \right) \psi(x,t),  
\end{align}
where $V_{\rm{m}}(x)$ is the model potential describing the 
electron--metal interaction, and $\tilde{V}_{\rm{opt}}(x,t)$ is the model 
potential of the optical pulse. We use a tilde notation to distinguish 
between the model potential in 1D WPP calculations and that in TDDFT. 
Eq.~\eqref{eq:WPP} is solved in real time using the Fourier-grid 
technique \cite{kosloff1983fourier} 
and short-time split-operator propagation \cite{Leforestier1991}. 
%The initial 
%condition $\psi_{\mathcal{E}_F}(x,t=0)$ corresponds to the situation where an 
%electron is located in one of the metallic leads at the orbital with the Fermi 
%energy $\mathcal{E}_F$.

The model potential describing electron--metal interaction $V_{\rm{m}}(x)$ 
is set using the classical electrostatic theory for an electron moving 
in the vacuum gap between two metal surfaces 
\cite{HotElectrons_in_Gaps,PitarkeMultiIM1990}. This potential accounts for 
the effect of the long-range polarization interaction by incorporating classical 
image potentials created by the metal surfaces
\begin{align}\label{eq:MetalPotWPP}
&V_{\rm{m}}(x)  = - v_{\rm{m}}(x_1) - v_{\rm{m}}(x_2) 
+ \theta\left(d_{\rm{gap}} - 2|x| \right) \nonumber \\
& \times \frac{1}{4} \sum_{f=1}^\infty \frac{1}{f d_{\rm{gap}}}
   \left[ \frac{x_1}{f d_{\rm{gap}}+x_1} + \frac{x_2}{f d_{\rm{gap}}+x_2} \right],
\end{align}
where $x_{1}=d_{\rm{gap}}/2+x$ and $x_2=d_{\rm{gap}}/2-x$ are the electron 
distances from classical image planes at $x_{\rm{im}}=\mp d_{\rm{gap}}/2$, 
and $\theta(x)$ is the Heaviside step function. The free-electron model 
potential $v_{\rm{m}}(x)$ proposed in Ref.~\onlinecite{JJJPotential} to 
describe the photoemission from metals smoothly connects the image 
potential tail $-1/4(x-x_{\rm{im}})$ for an electron in vacuum above the 
metal ($x \gg x_{\rm{im}}$) with a constant potential inside the metal 
($x \ll x_{\rm{im}}$). Obviously, this model is only applicable for sizes 
of the junction $d_{\rm{gap}}$ such that the overlap between the electron 
densities of the nanowires is small, as it is the case in the present study.

The model potential of the optical pulse $\tilde{V}_{\rm{opt}}(x,t)$ 
accounts for the field screening inside the metal. 
\begin{equation} \label{eq:OpticalPotWPP}
\tilde{V}_{\rm{opt}}(x,t) = 
\begin{cases}
  x~E_{\rm{gap}}(t), &|x|\leq \frac{d_{\rm{gap}}}{2}, \\
  \frac{d_{\rm{gap}}}{2}~\text{sgn}(x)~E_{\rm{gap}}(t),
  & |x| > d_{\rm{gap}}/2.
  \end{cases}
\end{equation}
The optical field inside the gap, $E_{\rm{gap}}(t)$, 
is given by $E_{\rm{gap}}(t) = E_{\rm{g}}~ e^{-t^2/\tau^2} \cos(\omega t + \varphi)$. 
Although similar in form to Eq.~\eqref{eq:Egap}, there is an important difference: 
in Eq.~\eqref{eq:Egap}, $E_{\rm{g}}$ is calculated self-consistently 
via TDDFT. It accounts for the incident and induced field. Here, by contrast, 
$E_{\rm{g}}$ is introduced directly into the WPP calculation as a free parameter.

If the bias $U$ is applied across the junction, the static field of the bias 
$E_{\rm{bias}}=-U/d_{\rm{gap}}$ is added to $E_{\rm{gap}}(t)$ 
in Eq.~\ref{eq:OpticalPotWPP}. In this 
situation, similar to the strategy used in TDDFT, we calculate both the total 
probability current density, $j(x,t)$, and the component induced solely by the 
optical field, $j_{\rm{opt}}(x,t)$.  From $j_{\rm{opt}}(x,t)$, we obtain the 
electron transfer probability per optical pulse, 
$\mathcal{P}=\int j_{\rm{opt}}(x=0,t) dt$, which characterizes the lightwave-driven 
transport in the system.

%Below in this work we show 
%the self-consistent $V(x,y=0,t=0)$ potential calculated with DFT/TDDFT, 
%$V(x,y=0,t=0)$, and the model $V_{\rm{m}}^{\rm{1D}}(x)$ potentials along the $x$ axis inside the gap. 

%
\begin{table}[t!]
\begin{tabular}{c|c|c|c|c|c|c|c|}
  $d_{\rm{gap}}$, nm ~ & ~0.8~ & ~1~ & ~1.6~ & ~2~ & ~4~ & ~6~ & ~ $\infty$~ \\
  \hline
  $\Phi_{\rm{M}} - \Phi$, eV~~ & ~0.84~ & ~0.82~ & ~0.61~ & ~0.50~ & ~0.25~ & ~0.17~ & ~0~ 
\end{tabular}
\caption{Differences between the metal work function $\Phi_{\rm{M}}$ and 
work function $\Phi$ to be used in the ALDA TDDFT in order to account for the 
effect of the image charge potential in the gap.}
\label{table1}
\end{table}

\subsection{Consistency between TDDFT and WPP via choice of the metal work 
function}\label{subsec:WorkFun}

As we discussed in Section \ref{subsec:TDDFT}, the present implementation 
of TDDFT employs a local exchange--correlation potential. This choice is 
dictated by numerical efficiency and it makes it possible to address 
metallic nanowires of large radii. The price to pay is that the ALDA fails 
to  reproduce the long-range polarization potential (image potential) felt 
by an electron moving in vacuum in front of metal surface 
\cite{PhysRevLett.68.1359,PhysRevLett.80.4265}.  The electron--surface 
interaction is too short-ranged, resulting in an overestimation of the 
height of the tunneling barrier separating metal surfaces.

In our earlier work 
\cite{HotElectrons_in_Gaps}, we demonstrated that this issue can be 
alleviated to a large extent by a proper choice of the stabilization 
potential \cite{perdew1990stabilized}, and thus of the work function  
$\Phi$ of the nanowires obtained from ground-state DFT calculations. 
In practice, one requires that the height of the tunneling barrier 
$\mathcal{W}_{\rm{tb}}$ within DFT is the same as that between actual 
metals obtained classically with account of the image potential 
(with no bias is applied).
\begin{align}\label{eq:effectiveWF1}
  \mathcal{W}_{\rm{tb}} &= \Phi_{\rm{M}} + V_{\rm{m}}(x=0)  \nonumber \\ 
  &= \Phi + V_{\rm{gs}}(x=0,y=0), 
\end{align}
Here, $\Phi_{\rm{M}}$ is the actual work function of the considered metal 
(in this work, silver or gold), $V_{\rm{gs}}$ is the ground-state potential 
calculated using DFT, and $V_{\rm{m}}$ is the model potential given by 
Eq.~\eqref{eq:MetalPotWPP}. In deriving Eq.~\eqref{eq:effectiveWF1}, 
we consider that the maximum of the tunneling barrier along the shortest 
path between metals is attained at the middle $(x=0,y=0)$ of the 
symmetric junction.
From Eq.~\eqref{eq:effectiveWF1} it follows that
\begin{equation}\label{eq:effectiveWF}
  \Phi=\Phi_{\rm{M}}- \big[ V_{\rm{gs}}(x=0,y=0)-V_{\rm{m}}(x=0) \big].
\end{equation}

The relation between $\Phi$ and $\Phi_{\rm{M}}$ is given in Table~\ref{table1} 
for the $d_{\rm{gap}}$ values relevant for the present work. 
%Below we thus provide either the values of $\Phi$ or $\Phi_{\rm{M}}$ depending on the context. 
It is worth noting that the work function correction does not affect the 
qualitative aspects of the studied processes. It is 
important when comparing TDDFT and WPP results (in WPP, the actual work 
function value $\Phi_{\rm{M}}$ is used), as well as when comparing theoretical 
results with experimental data obtained for a specific metal. 

We verified, that once conditions for $\Phi$ given by Eq.~\ref{eq:effectiveWF1} 
and Eq.~\ref{eq:effectiveWF} are satisfied, the heights  
of the DFT and model tunneling barriers are also approximately equal 
when the dc bias is applied. In this situation $\mathcal{W}_{\rm{tb}}$ obtained 
from the TDDFT is defined as 
\begin{align}\label{eq:BarrierBias}
  \mathcal{W}_{\rm{tb}} &=  -\Phi - V_{max}^{\rm{TDDFT}}, 
\end{align}
where $V_{max}^{\rm{TDDFT}}$ is the maximum self-consistent potential in the 
junction obtained at $t=t_-$, i.e., at constant THz field prior to the 
arrival of the optical pulse (see Eq.~\eqref{eq:THz} and Eq.~\eqref{eq:THz1}).
As a final remark, the self-consistent potential $V_{\rm{gs}}(x,y=0)$ 
and the model potential $V_{\rm{m}}(x)$ within the gap are quite similar 
along the dimer axis up to the energy offset given by 
$V_{\rm{gs}}(x=0,y=0)-V_{\rm{m}}(x=0)$ \cite{HotElectrons_in_Gaps} 
(see Eq.~\ref{eq:effectiveWF}).

%%%%%%%%%%%%%%%%%%%%%%%%%%%%%%%%%%%%%%%%%%%%%%%%%%%%%%%%%%%%%%%%%%%%%%%

% journals \cite{Jennings1988}. 

\subsection{Semi-classical strong-field theory (SFT)}\label{subsec:SFT}

Beyond the many-body self-consistent TDDFT and the one-electron WPP approach 
introduced above, we also employ a strong-field theoretical model developed 
in earlier works~\cite{Ma_Kruger2024,ma2025robust}. One of the prominent 
advantages of SFT is its ability to provide a semi-classical description which 
offers analytical insights into the photoemission and electron transport 
processes. Like the WPP approach (subsection \ref{subsec:WPP}), 
the SFT model is based on the single-active-electron 
approximation and focuses on the evolution of the wavefunction within the junction 
between two metallic leads. By solving the TDSE in Eq.~\eqref{eq:WPP}, the tunneling 
amplitude $M_{\mathcal{E}}$ for the electron transition from an initial state 
in the left lead to the final state at energy $\mathcal{E}$ in the 
right lead is expressed as
\begin{align}
M_{\cal{E}}= \frac{i}{2}\int_{-\infty}^{\infty} 
\bigg[\Psi_{\rm{gap}}(x,t)\frac{\partial}{\partial x}\psi^{*}(x,t)\nonumber\\
-\,\psi^{*}(x,t)\frac{\partial}{\partial x}\Psi_{\rm{gap}}(x,t)\bigg] 
\bigg\vert^{x={d_{\rm{gap}}}/{2}}_{x=-{d_{\rm{gap}}}/{2}}\;dt,
\label{eqSFA}
\end{align}
where $\psi^{*}(x,t)$ and $\Psi_{\rm{gap}}(x,t)$ are wavefunctions that evolve 
inside the contact metal (right) and the junction, respectively. The notation 
$[\cdots]\big\vert^{x={d_{\rm{gap}}}/{2}}_{x=-{d_{\rm{gap}}}/{2}}$ at the 
end of Eq.~\eqref{eqSFA} stands for the subtraction of the term inside the brackets 
evaluated at $x=-{d_{\rm{gap}}}/{2}$ from the term evaluated at 
$x={d_{\rm{gap}}}/{2}$. This formulation captures the instantaneous projection 
of the junction wavefunction onto the metallic lead region throughout the time 
evolution. The transport from the right to the left metal lead is evaluated 
using the same formula, but the signs of both the optical field and the spatial 
coordinate must be reversed.

Here, we only focus on the transfer of the electron initially localized 
at the Fermi level of the left lead to the right lead. 
Eq.~\eqref{eqSFA} provides only a formal solution to the TDSE. To analytically 
derive the dynamical behavior of the system, the wavefunctions 
$\psi^{*}(x,t)$ and $\Psi_\mathrm{gap}(x,t)$ are approximated using the eigenfunctions 
of the metal and the Van Vleck propagator, respectively~\cite{ma2025robust}. By 
applying the saddle-point method~\cite{Yalunin2011,ma2025robust}, the electron 
transfer probability $\mathcal{P}_{\rm{L\rightarrow R}}$ is obtained as
\begin{equation}\label{SFAcurrent}
\mathcal{P}_{\rm{L\rightarrow R}}=\sum_{\mathcal{E}}\vert M_{\mathcal{E}}\vert^2 
\propto \sum_{\mathcal{E}}e^{-2\text{Im}[S(t_{\rm{2s}},t_{\rm{1s}})]},
\end{equation}
where $\text{Im}[S(t_{\rm{2s}},t_{\rm{1s}})]$ is the imaginary part of an 
action function $S(t_{\rm{2s}},t_{\rm{1s}})$ taken at the stationary points 
$t_{\rm{1s}}$ ($t_{\rm{2s}}$) of the emission (transmission) time. The action 
function is given by
\begin{align}\label{actionfunc}
S(t_{\rm{2s}},t_{\rm{1s}})=&\mathcal{E} t_{\rm{2s}} 
+ \frac{{\tilde p}^2}{2}(t_{\rm{2s}}-t_{\rm{1s}}) 
-\int_{t_{\rm{1s}}}^{t_{\rm{2s}}}\frac{A_{\rm{opt}}^2(t')}{2} \;dt'\, \nonumber\\
&-\int_{t_{\rm{1s}}}^{t_{\rm{2s}}} V_{\rm{m}}[x(t')]\;dt'\, 
- \mathcal{E}_{F}  t_{\rm{1s}},
\end{align}
where, $\mathcal{E}$ is the final energy of the transported electron at the transmission 
time $t_{2s}$, $\mathcal{E}_{F}$ is the initial energy (Fermi energy) at the emission 
time $t_{\rm{1s}}$, $A_{\rm{opt}}(t)=-\int^{t} [E_{\rm{gap}}(t')+E_{\rm{bias}}]dt'$ is 
the vector potential of the combined enhanced optical field and static bias, 
and $\tilde p=\frac{-\int_{t_{\rm{1s}}}^{t_{\rm{2s}}}{A_{\rm{opt}}(t')}\;dt'\,+d_{\rm{gap}}}
{t_{\rm{2s}}-t_{\rm{1s}}}$ 
is the canonical momentum of the transported electron driven by the field. 
As a consequence of the saddle-point technique, the times $t_{\rm{1s}}$ and $t_{\rm{2s}}$ 
are not real-valued physical times but complex values obtained by solving the following 
three saddle-point equations:
\begin{align}
\frac{\left[\tilde p+A_{\rm{opt}}(t_{\rm{1s}})\right]^2}{2}+\overline{V}_{\rm{m}}&
= \mathcal{E}_F,\label{SP1}
\\
\int_{t_{\rm{1s}}}^{t_{\rm{2s}}}\left[\tilde p+A_{\rm{opt}}(t')\right]\;dt'\,&
=d_{\rm{gap}},\label{SP2}
\\
\frac{\left[\tilde p+A_{\rm{opt}}(t_{\rm{2s}})\right]^2}{2}+\overline{V}_{\rm{m}}&
=\mathcal{E}.\label{SP3}
\end{align}
where $\overline{V}_{\rm{m}}=
\left(d_{\rm{gap}}\right)^{-1} \int_{-d_{\rm{gap}}/2}^{d_{\rm{gap}}/2} 
V_{\rm{m}}(x)\,dx$ is the average image potential.

%
% FIGURE 3 zeta
%
\begin{figure*}[t!]
\centering
\includegraphics[width=0.95\linewidth]{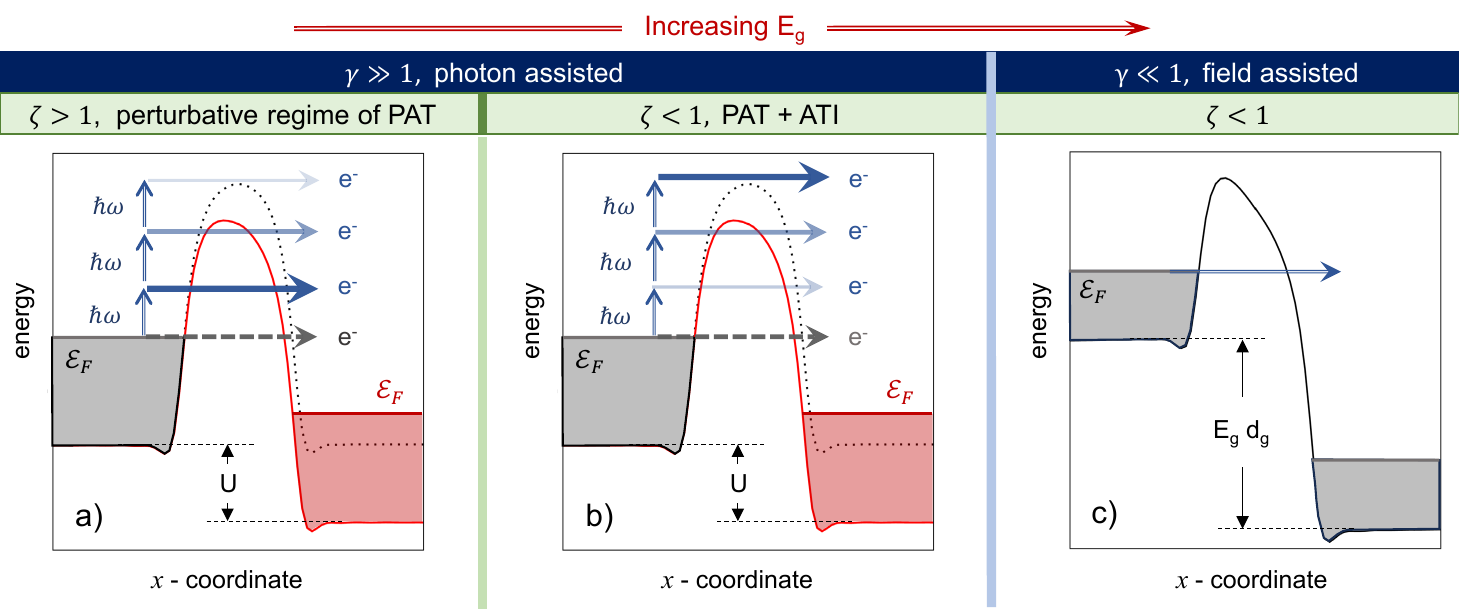}
\caption{Sketch of the electron transport described by the SFT.
\textbf{Panel a}: When the parameter $\zeta > 1$, the electron is excited by 
absorbing a few photons and then tunnels through the junction. In this perturbative 
regime of PAT, the contribution of the $\ell$th channel to the transport decreases 
with $\ell$ as $E_{\rm{g}}^{2\ell}$ (see Eq.~\eqref{eq:Pedersen}).  
The photon-assisted channels do not exceed the barrier threshold so that we 
observe solely the PAT.
\textbf{Panel b}: When $\zeta < 1$ but $\gamma \gg 1$, electron transport occurs 
in the multiphoton regime. The electron can absorb sufficient energy to be excited 
over the barrier and into a Volkov state (above-threshold ionization, ATI). 
The effective photon absorption reaches the barrier threshold. Although 
under-the-barrier PAT still occurs, it is not the dominant transport mechanism.
\textbf{Panel c}: When $\zeta < 1$ and $\gamma \ll 1$, the potential barrier 
is adiabatically suppressed by the field, allowing the electron to efficiently 
tunnel through the junction. 
}
\label{fig:fig_zeta}
\end{figure*}

The three saddle-point equations above provide a three-step description of electron 
transport in the nanojunction: 
\begin{enumerate}
  \item Emission: The bound electron is freed from the parent lead by the 
external field. Equation~\eqref{SP1} describes the energy conservation at this instant. 
The image potential lowers the junction barrier.
%, enhancing electron emission. 
  \item Propagation: The freed electron is accelerated by the driving field, 
  while moving across the nanojunction (Eq.~\eqref{SP2}). The force from the 
  image potential is neglected due to its weak effect in this step.
  \item Transmission: The energy gained in the second step is carried 
  by the electron transferred to the opposite metallic lead (Eq.~\eqref{SP3}).
\end{enumerate}
It should be stressed that the initial energy is negative, leading to complex 
values of $t_{\rm{1s}}$, $\tilde p$, and $t_{\rm{2s}}$. The imaginary part of the 
emission time $t_{\rm{1s}}$ corresponds to purely quantum processes, such as photon 
absorption and tunneling through the barrier, which are not allowed in the classical 
dynamics. In contrast, the imaginary component of the arrival time $t_{\rm{2s}}$ 
is usually related to amplitude attenuation of the transported wavefunction 
component~\cite{Ma_Kruger2024,ma2025robust}.

%%%%%%%%%%%%%%%%%%%%%%%%%%%%%%%%%%%%%%%%%%%%%%%%%%%%%%%%%%%%%%%%

\subsection{Effect of an applied bias within SFT}\label{subsec:PE_bias}

Here we extend the SFT to the scenario where a dc bias field with strength 
$E_\mathrm{bias}=|U|/d_{\rm{gap}}$ is present. As demonstrated in Ref.~\cite{ma2025robust}, 
the saddle-point equations Eq.~\eqref{SP1}-Eq.~\eqref{SP3} allow for two types of 
solutions characterized by different imaginary time durations. These are given by:
\begin{equation}\label{eq:SFtau}
  \text{Im}[t_{\rm{L}}]=
  \frac{E_{\rm{g}}\sqrt{1+\gamma^2}}{E_{\rm{bias}}+E_{\rm{g}}\sqrt{1+\gamma^2}} 
 \frac{ \ln(\gamma+\sqrt{1+\gamma^2})}{\omega},
\end{equation}
and
\begin{equation}\label{eq:Ptau}
  \text{Im}[\tau_{\rm{L}}]=\text{Im}{\bigg[\frac{2d_{\rm{gap}}}
  {\sqrt{2 \left(\mathcal{E}-\overline{V}_{\rm{m}}\right)} 
  +i\sqrt{2 \overline{\mathcal{W}}_{\rm{tb}}}}\bigg]},
\end{equation}
where $\mathcal{E}$ is the final electron energy in the right metallic lead, 
$\mathcal{E}-\overline{V}_{\rm{m}}$ is the effective final energy, and 
\begin{equation}\label{eq:BarrierSFT}
 \overline{\mathcal{W}}_{\rm{tb}}= \overline{V}_{\rm{m}}-\mathcal{E}_F
\end{equation}
is the effective height of the tunneling barrier defined within SFT from the 
average image potential in the junction. Using $\mathcal{E}_F=-\Phi_{\rm{M}}$, 
in the conditions of our study 
 $\overline{\mathcal{W}}_{\rm{tb}} \approx \mathcal{W}_{\rm{tb}}$ where 
 $\mathcal{W}_{\rm{tb}}$ is defined with Eq.~\eqref{eq:effectiveWF1}.
%Due to the weak effect of the image potential, here we have 
%$|{\cal E}_{F}|_{\rm{eff}}\simeq\Phi$. 
The Keldysh parameter \cite{keldysh1965ionization} is 
$\gamma=\sqrt{\overline{\mathcal{W}}_{\rm{tb}}/2U_{\rm{p}} }$ with the 
ponderomotive energy $U_{\rm{p}} = \left[ E_{\rm{g}}/2\omega \right]^2 $. 
Notice that because of the tunneling, this definition of the Keldysh parameter 
is different from the conventional one 
$\gamma=\sqrt{\mathcal{E}_{\rm{B}}/2U_{\rm{p}} }$ used to distinguish between 
photon assisted ($\gamma >> 1$) or optical field assisted ($\gamma <<1$) 
regimes of photoemission 
\cite{Milosevic_2006,RevModPhys.92.025003,Ciappina2017AttoPhys,Park2020}.  
In the conventional definition, 
$\mathcal{E}_{\rm{B}}$ is the electron binding energy, which is given, e.g., 
by the ionization potential in atomic physics or by the work function 
$\Phi_{\rm{M}}=-\mathcal{E}_F$ in the case of photoemission from a metal. 
For a large gap size, the contribution of the electron--metal interaction 
to the tunneling barrier can be neglected, and the two definitions of 
$\gamma$ therefore become equivalent. In the situations relevant 
for the present study, values of $\gamma$ obtained using $\Phi_{\rm{M}}$, 
$\Phi$ (see section \ref{subsec:WorkFun}), or using 
$\overline{\mathcal{W}}_{\rm{tb}}$ differ by less than 15~\%, so that any 
of them is appropriate for the qualitative purpose of distinction between 
electron transport regimes.

The two imaginary time durations given by Eq.~\eqref{eq:SFtau} and 
Eq.~\eqref{eq:Ptau} correspond to distinct types of photoemission processes. 
The time $\text{Im}[t_{\rm{L}}]$ of Eq.~\eqref{eq:SFtau} describes the 
transition of an electron from a bound state to a laser-driven state 
(Volkov state). In contrast, $\text{Im}[\tau_{\rm{L}}]$ corresponds to the 
excitation of a bound electron to an excited eigenstate located under the 
potential barrier~\cite{ma2025robust}. To distinguish these two regimes, we 
compare the time $\text{Im}[t_{\rm{L}}]$ with the extremal value of 
$\text{Im}[\tau_{\rm{L}}]$ ($\text{Im}[\tau_e]=
2d_{\rm{gap}}/\sqrt{2\overline{\mathcal{W}}_{\rm{tb}}}$), 
and define a parameter $\zeta$ as
\begin{equation}\label{eq:zeta}
  \zeta=\frac{\text{Im}[t_{\rm{L}}]}{\text{Im}[\tau_e]}.
\end{equation}
Parameters $\gamma$ and $\zeta$ allow one to characterize the different 
regimes of electron transport in a metallic gap, as we sketch in 
Figure~\ref{fig:fig_zeta} and describe below (for detailed derivations see 
Appendix~\ref{Apd.A}).

\textbf{(1)} For a weak optical field where $\zeta \geq 1$ (Figure~\ref{fig:fig_zeta}a), 
the imaginary time $\text{Im}[\tau_{\rm{L}}]$ 
of Eq.~\eqref{eq:Ptau} becomes dominant. By substituting this solution into the 
saddle-point equations we obtain the action function $S(t_{\rm{2s}},t_{\rm{1s}})$.   
Using $S(t_{\rm{2s}},t_{\rm{1s}})$ in Eq.~\eqref{SFAcurrent}, we obtain the 
following key results (cf. Ref.~\cite{ma2025robust}).
For low bias $U \leq \overline{\mathcal{W}}_{\rm{tb}}$
\begin{align}\label{eq:Transport_En}
\mathcal{P}_{\rm{L\rightarrow R}} \propto 
\exp \left[ -2\sqrt{2 \overline{\mathcal{W}}_{\rm{tb}}-(n\omega+U)}\ d_{\rm{gap}} \right], 
\end{align}
which describes the WKB-like tunneling of an excited electron through the 
potential barrier of a junction in the presence of an applied bias $U$. 
In this perturbative regime PAT is dominated by the low $\ell$ channels  
(see Eq.~\eqref{eq:Pedersen}). Consistently, the effective photon order 
$n=E_{\rm{g}} d_{\rm{gap}}/\omega$ is small. It tends to zero for 
$E_{\rm{g}} \rightarrow 0$ where only an applied bias drives the tunneling. 
Indeed, $\mathcal{P}_{\rm{L\rightarrow R}}$ expressed with Eq.~\eqref{eq:Transport_En} 
includes both electron transport owing to the applied dc bias and optically induced 
transitions.

For high bias $U > \overline{\mathcal{W}}_{\rm{tb}}$, the electron transfer 
probability is given by 
\begin{equation}\label{eq:Transport_FN}
\mathcal{P}_{\rm{L\rightarrow R}} \propto 
\exp \left[ -\frac{2(2\overline{\mathcal{W}}_{\rm{tb}}-n\omega)^{\frac{3}{2}}}{3E_{\rm{bias}}} \right]. 
\end{equation}
Eq.~\eqref{eq:Transport_FN} describes the tunneling of an excited electron through a 
strongly suppressed triangular potential barrier, analogous to photon-assisted 
Fowler--Nordheim tunneling. The effective photon order $n=E_{\rm{g}} d_{\rm{gap}}/\omega$ 
is given by the same expression as for the low bias case.

Eq.~\eqref{eq:Transport_En} and Eq.~\eqref{eq:Transport_FN} are obtained for an  
optical field acting as small perturbation. When $E_{\rm{g}}$ increases such that 
$\zeta < 1$, the imaginary time $\text{Im}[t_{\rm{L}}]$ becomes dominant so that the 
transition of an electron from a bound state to a laser-driven state (Volkov state) 
determines the optically induced transport across the gap (Figure~\ref{fig:fig_zeta}b,c). 
In this situation, the Keldysh parameter $\gamma$ allows for distinguishing between 
the multiphoton absorption and optical field emission regimes.

\textbf{(2)} $\zeta < 1$, $\gamma\gg1$. In this regime, the electron can be directly excited from the 
bound state to a Volkov state via multiphoton absorption (see Figure~\ref{fig:fig_zeta}b). 
The electron transfer probability is given by: 
\begin{equation}\label{eq:MPE}
\mathcal{P}_{\rm{L\rightarrow R}} \propto \left(2\gamma\right)^{-2n}, 
%g(E_{\rm{bias}}) \times 
\end{equation}
where an effective number of absorbed photons 
\begin{equation} \label{eq:MPE1}
n= \left[\frac{\omega\sqrt{2\overline{\mathcal{W}}_{\rm{tb}}}}
{E_{\rm{bias}}+\omega\sqrt{2\overline{\mathcal{W}}_{\rm{tb}}}}\right] 
\frac{\overline{\mathcal{W}}_{\rm{tb}}}{\omega}.
\end{equation}
%
%Unlike the conventional multiphoton photoemission yield with $n=\Phi_{\rm{M}}/ \hbar\omega$, 
For transitions over-the-tunneling barrier one expects 
$n=\overline{\mathcal{W}}_{\rm{tb}}/ \omega$. 
However, the presence of a bias modifies the effective tunneling 
barrier by a factor given in square brackets. For higher-order 
corrections in $E_{\rm{bias}}$, see Eq.~\eqref{eqA20} of Appendix~\ref{Apd.A}.

\textbf{(3)} $\zeta < 1$, $\gamma \ll 1$. In this regime, both the optical 
field and the static bias adiabatically suppress the junction barrier, 
enabling tunneling (Figure~\ref{fig:fig_zeta}c). The electron transfer 
probability follows a Fowler--Nordheim tunneling expression:
\begin{equation}\label{eq:TPE}
 \mathcal{P}_{\rm{L\rightarrow R}} \propto 
 \exp{\left[-\frac{2 (2 \overline{\mathcal{W}}_{\rm{tb}})^
 {\frac{3}{2}}}{3(E_{\rm{bias}}+E_{\rm{g}})} \right] }.
\end{equation}
The key difference between Eq.~\eqref{eq:TPE} and Eq.~\eqref{eq:Transport_FN} 
lies in the role of the fields. 
The former describes electron transfer driven by an adiabatically strong 
optical field and a weak static bias, while the latter describes the regime 
where the static bias is strong and the optical field acts as a perturbation.

We have shown previously \cite{ma2025robust} that in the regime where the 
optical field is strong enough so that $\zeta < 1$, the photoemission exhibits 
a cutoff energy determined by the work function and the parameter $\zeta$:
\begin{equation}\label{eq:cutoff}
\mathcal{E}_{\rm{cutoff}} 
=\frac{\overline{\mathcal{W}}_{\rm{tb}}}{\zeta}+\mathcal{E}_{F}.
\end{equation}
The cutoff energy expression given by Eq.~\eqref{eq:cutoff} allows for a 
physically transparent interpretation of the parameter $\zeta$ as $\zeta=
\overline{\mathcal{W}}_{\rm{tb}}/\left(\mathcal{E}_{\rm{cutoff}}-\mathcal{E}_{F}\right)$, 
and provides a valuable criterion for analyzing photon absorption channels, 
as explored in the following sections. Indeed, the energy difference 
$\mathcal{E}_{\rm{cutoff}}-\mathcal{E}_{F}$ represents the maximum 
energy transfer from the optical field, while the effective 
barrier height $\overline{\mathcal{W}}_{\rm{tb}}$ represents the static binding 
potential. Thus, $\zeta < 1$ admits over-the-barrier, classically allowed 
transitions with final electron energies above the junction barrier. This is 
while $\zeta > 1$ corresponds to the regime where electrons with 
energies below the barrier 
$\mathcal{E}_{\rm{cutoff}}-\mathcal{E}_{F}<\overline{\mathcal{W}}_{\rm{tb}}$ 
tunnel across the junction.
%binding potential dominates and the 
%optical field acts as a weak perturbation, leading to weak photon absorption ($\zeta\geq1$). 
%In contrast, the strong-field perspective treats the optical field as dominant, with 
%the binding potential acting as a perturbation ($\zeta<1$). The parameter $\zeta$ thus 
%serves as an indicator of these two regimes. 

\subsection{Short summary of the theoretical approaches.}
\label{subsec:TheorySummary}

\begin{table*}[t!]
\begin{tabular}{|c|c|c|c|c|l|}
  \hline
  Theoretical &  Formalism  & Electron-metal & Image      & Optical field     & Information \\
  Approach    &  and method & interaction    & potential  & and applied bias  & provided    \\ 
  \hline
  \hline
  ~~          & Numerical       & Self-consistent. & ~   & Self-consistent.  & Quantitative: (1) electron density \\
  Many-body   & solution of the & 2D Free-electron & ~   & Accounts for the  & dynamics (currents, transport); (2) role \\  
   TDDFT      & time-dependent  & metal (FEM).     & No  & screening,        & of various parameters (pulse waveform,\\ 
   ~~         & Kohn-Sham       & Junction of the  & ~   & field enhancement,  & geometry, bias) in electron transport;\\  
   ~~         & equations       & nanowire dimer.  & ~   & (plasmon).          & (3) comparison with experiment. \\ \hline
  ~~          & Numerical       & Model.          & ~      & ~               & Qualitative. Used along with TDDFT: \\
  One active  & solution of the & 1D FEM.          & ~      & Model. Field    & (1) analysis of the symmetry constraints \\
  electron    & time-dependent  & Junction        & Model  & is set to zero  & to observe the one-photon channel of PAT; \\ 
  WPP         & Schr\"{o}dinger & between flat    & ~      & inside metal.   & (2) electron final-state spectroscopy\\
  ~~          & equation        & metal surfaces. & ~      & ~               & revealing $\ell$-photon channels of PAT.\\ \hline
  ~~          & Analytical     & Model.          & ~      & ~               & Analytical. Analysis of three transport\\ 
  One active  & approach       & 1D FEM.         & ~      & Model. Field    & regimes in TDDFT results: \\ 
  electron    & based on the   & Junction        & Model  & is set to zero  & (1) multiphoton perturbative PAT; \\
  SFT         & semiclassical  & between flat    & ~      & inside metal.   &  (2) multiphoton PAT+ATI;\\ 
  ~~          & propagator     & metal surfaces. & ~      & ~               & (3) optical field emission. \\ \hline
\end{tabular}
\caption{\label{table:Methods} Summary of the theoretical approaches used in this work.}
\end{table*}

Given the variety of theoretical tools employed, we 
find it useful to provide in Table~\ref{table:Methods} 
their brief summary. The main difference between the methods consists 
in how the electron dynamics is treated: either as that of a many-body 
system accounting for the entire valence band, or as that of a single 
electron initially at the Fermi energy. Consequently, the potentials 
describing electron-metal and electron-field interactions are either 
obtained self-consistently or introduced as model potentials. In the 
former case, the many-body aspects such as field screening and plasmon 
excitations are naturally included. (Note that under conditions 
considered in our work, the plasmon is off-resonant with the optical 
pulse and remains unexcited, hence the parenthesis in 
Table~\ref{table:Methods}.) In model potentials these effects are 
introduced \textit{ad hoc}, using additional arguments or calculations. 
Model potentials, however, allow one to account for 
long-range polarization interactions (the image potential) acting 
on an electron in front of a metal surface and reducing the 
tunneling barrier of the junction. The present TDDFT implementation 
requires a correction to the metal work function to include this 
effect (Table~\ref{table1}).

The main reason for using different approaches is as 
follows. TDDFT enables self-consistent, quantitative calculations of 
optically induced electron transport in metallic gaps, allowing  
direct comparison with experiment without adjustable parameters. 
Simple models, on the other hand, permit controlled variation of system 
parameters and facilitate the analysis of one-electron dynamics, thereby 
providing valuable information about the studied processes.
Analytical theories are particularly useful for establishing general 
trends and the conditions under which those trends can be observed.
Thus, while the main results presented below are obtained with 
quantitative many-body TDDFT, the simplified numerical (WPP) and 
analytical (SFT) methods provide complementary insights and are 
essential for a general understanding of the underlying physics.

%\vspace{8mm}

%%%%%%%%%%%%%%%%%%%%%%%   START DISCUSSING RESULTS        %%%%%%%%%%%%%%%%%%%%%%
%%%%%%%%%%%%%%%%%%%%%%%%%%%%%%%%%%%%%%%%%%%%%%%%%%%%%%%%%%%%%%%%%%%%%%%%%%%%%%%%%
\section{Electron transport induced by an optical pulse}
\label{sec:SingleCyclePulse}

In this section, we consider electron transport induced by an optical pulse  
across the (sub-)nanometric junction. The electric field profile of the 
pulse is given by Eq.~\eqref{eq:Gaussianfield}. Unless otherwise stated, 
the pulse frequency is $\omega=0.95$~eV, and $\tau= 0.85 \times 2\pi/\omega$ 
corresponds to a single-cycle optical pulse with a duration of $5.2$~fs 
(intensity full width at half maximum). The choice of the frequency and 
envelope of the optical pulse allows for connecting current results with 
earlier theoretical and experimental studies of electron transport in 
symmetric 1~nm and 2~nm gaps \cite{HotElectrons_in_Gaps} as well as wider 
6~nm gaps \cite{Ludwig2019}. The work function of the metal $\Phi_{\rm{M}}$ 
is set within the range $5-5.5$~eV as reported for gold and 
used to describe multiphoton and optical field emission from gold 
surfaces \cite{PDombi2004,Kusa2015,Putnam2017,Luo2024,chulkov1999image}.

%%%%%%%%%%%%%%%%%%%%%%%%%%%%%%%%%%%%%%%%%%%%%%%
\subsection{Symmetric system}
\label{subsec:SymmetricSystem}

We start our discussion with results obtained in the absence of an applied bias. 
Because of the symmetric geometry of the metallic gap, the net electron transfer 
results solely from the symmetry break induced by the optical field.
We then obtain the well-documented sinusoidal dependence of the 
net electron transport on the CEP, as observed for broad vacuum gaps 
\cite{Ciappina2019,Heide2024,RevModPhys.92.025003}, in 
STM junctions \cite{GargScience2020,Brida2025STM}, and in our recent calculations 
for 1~nm and 2~nm gaps \cite{HotElectrons_in_Gaps} (see also \cite{Ma_Kruger2024}). 
We therefore do not show these results here. 
We focus our study on identifying and discussing signatures of PAT. To this end, 
we present results calculated for CEP$=\pi$ and CEP$=0$, which lie approximately 
at the extrema of net electron transport in the positive and negative 
$x$-directions, respectively. Notably, in addition to the discussion of PAT,
these data allow an estimate of the oscillation amplitude in the CEP dependence 
of the net electron transport.

%\onecolumngrid

%\vspace{5mm}

%
% FIGURE 4  Power curves no bias, gap=0.8, 1, 2 nm, thresholds
%
\begin{figure*}[t!]
%\begin{figure}[h!]
\centering
\includegraphics[width=0.85\linewidth]{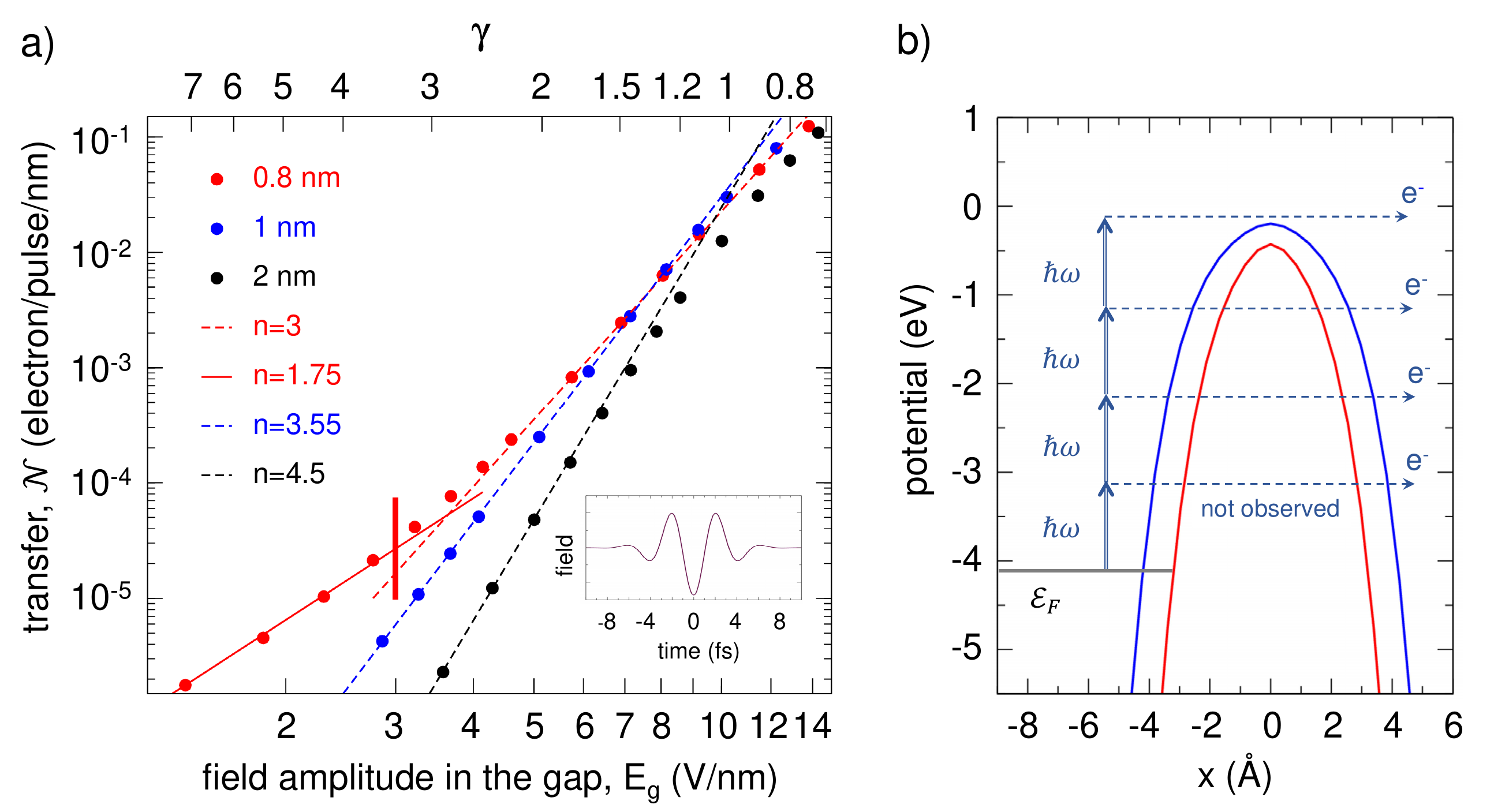}
\caption{Electron transfer induced by a $x$-polarized single-cycle optical pulse 
($\omega=0.95$~eV, CEP$=\pi$) across the gap of a nanowire dimer 
with work function $\Phi_{\rm{M}}=5$~eV. 
The electric field of the pulse is shown in the inset of panel a.
\textbf{Panel a:} Electron transfer $\mathcal{N}$ defined as the net number of 
electrons transferred per pulse and per nm length of the dimer. TDDFT results (dots) 
calculated for a width of the gap of 0.8~nm (red), 1~nm (blue), and 2~nm (black) 
are shown as a function of the electric field in the gap $E_{\rm{g}}$ and of the 
Keldysh parameter $\gamma$. The lines of the corresponding color display the fit by 
the $\mathcal{N} \propto E_{\rm{g}}^{2n}$ dependence characteristic for the 
multiphoton regime. The effective number of photons $n$ is given in the legend.
The vertical red bar marks the regime transition at $\zeta=1$ for the 0.8 nm gap. 
%Single-cycle optical pulse. TDDFT transfer across 
%the 0.8, 1, 2 nm gaps, $\Phi=$ 4.1 eV, 4.1 eV, 4.5 eV.  
%representing $\Phi_{\rm{M}}=$ 4.9 eV, 4.9 eV, 5.0 eV
\textbf{Panel b:} Self-consistent ground-state potential $V_{\rm{gs}}(x,y=0)$ 
in the gap region as a function of the $x$-coordinate along the dimer axis. 
Results are shown for a gap of the width of 0.8~nm (red), and 1~nm (blue).
The energy is measured with respect to the vacuum level. $\mathcal{E}_F=-\Phi$ 
is the Fermi level energy corresponding to the effective work function 
$\Phi=4.2$~eV used in the TDDFT (see Table~\ref{table1}). 
Dashed arrows indicate electron tunneling and over-the-barrier transitions 
induced by $\ell$-photon absorption. The latter is indicated with vertical 
arrows. The $\ell=1$ tunneling channel is not observed in TDDFT results in panel a.
}
\label{fig:Transfer_08_1_2}
\end{figure*}
%\end{figure}

%\vspace{5mm}

%\twocolumngrid

In Figure~\ref{fig:Transfer_08_1_2}a we analyze the net electron transfer $\mathcal{N}$ 
calculated with TDDFT for gap sizes $d_{\rm{gap}}=0.8$~nm, $1$~nm,~and~$2$~nm. 
The CEP$= \pi$ of the single-cycle optical pulse is such that the maximum $\mathcal{N}$ 
is reached in the positive direction of the dimer $x$-axis \cite{HotElectrons_in_Gaps}. 
In the inset of Figure~\ref{fig:Transfer_08_1_2}a we show the time evolution of the 
corresponding electric field of the pulse. Obviously, the sign of $\mathcal{N}$ is 
reversed for CEP$=0$.
The TDDFT results for $\Phi_{\rm{M}}=5$~eV are shown as a function of the amplitude 
of the field in the gap $E_{\rm{g}}$ (lower $x$-axis) and of the Keldysh parameter 
defined as $\gamma=\sqrt{\Phi/2U_{\rm{p}} }$. 
%The multiphoton emission (weak field) corresponds to $1 \ll \gamma $ while 
%adiabatic optical field emission regime (strong field) corresponds to 
%$\gamma \ll 1 $. The transition between the two regimes is at $\gamma \approx 1$. 
%We note, however, that Keldysh theory was originally developed for the photoemission 
%and has to be applied with care when discussing the lightwave-induced tunneling 
%because of the difference in the final states of these two processes. 

For the 2~nm wide junctions the potential barrier is too broad to allow for 
notable electron tunneling. Within the studied range of the optical fields 
the electron transfer between the nanowires is then dominated by classically 
allowed over-the-barrier transitions where the $n$-photon absorption leads to 
photoexcited electron energies above the potential barrier of the junction, 
similar to that discussed earlier \cite{Ludwig2019,HotElectrons_in_Gaps}.
Indeed, fitting the optically induced net electron transfer $\mathcal{N}$ 
with $\mathcal{N} \propto E_{\rm{g}}^{2n}$ dependence, we obtain in 
Figure~\ref{fig:Transfer_08_1_2}a that in the multiphoton regime the effective 
number of photons absorbed by transferred electron (the photon order) is $n=4.5$. 
This roughly corresponds to $\Phi/\omega$. At large field strength 
($\gamma \lesssim 1$), a smaller nonlinearity of $\mathcal{N}(E_{\rm{g}})$ 
(reduced slope) reflects the onset of the optical field emission. 
For the $1$~nm junction we obtain $n=3.55$. In this situation, the potential 
barrier of the junction is lower and narrower, and as a consequence, the 
electron transfer involves both over-the-barrier transitions requiring at least 
4-photon absorption, and under-the-barrier tunneling transitions, dominated 
by 3-photon absorption (see  Figure~\ref{fig:Transfer_08_1_2}b). A smaller 
numbers of absorbed photons (lower electron excitation energy, higher 
tunneling barrier) implies a too small tunneling exponent. The above analysis 
is supported by the TDDFT and model studies in Ref.~\cite{HotElectrons_in_Gaps} 
where the lightwave-induced transport in $d_{\rm{gap}}=1$~nm and 2~nm junctions 
was addressed, albeit for somewhat different workfunction $\Phi$.

To clearly identify the contribution of PAT to electron transport, the TDDFT 
results should reveal a change of the leading electron transport channel(s) 
when varying $E_{\rm{g}}$. Indeed, consider the perturbative regime shown 
in Figure~\ref{fig:fig_zeta}a. As follows from Eq.~\eqref{eq:Pedersen}, the 
probability of absorbing $\ell$ photons scales as $\propto E_{\rm{g}}^{2\ell}$. 
Obviously, for a very small $E_{\rm{g}}$ the lowest $\ell$ transport channel  
allowed by the symmetry (see below) will dominate. 
Increasing field strength in the gap leads to the involvement of higher-$\ell$ 
channels \cite{ma2025robust} for which the tunneling probability is higher. 
One might expect that the transition between dominant $\ell$ channels should then 
appear as a change of the slope in the log-log plot of $\mathcal{N}(E_{\rm{g}})$ 
\cite{STM_Light_Schroder_2020,Light_STM_Kumagai23,Luo2024}. Moreover, as 
follows from the semiclassical theory in subsection~\ref{subsec:PE_bias}, 
above certain $E_{\rm{g}}$ the tunneling regime of electron transport evolves 
into a regime dominated by over-the-barrier transitions. This change of 
the transport regime operates for parameter $\zeta = 1$ (see Eq.~\eqref{eq:zeta}), 
and it should also be evidenced by a change of the photon order $n$.

However, within the studied range of $E_{\rm{g}}$, the TDDFT results 
for $d_{\rm{gap}}=1$~nm and 2~nm feature linear 
log-log plot of $\mathcal{N}(E_{\rm{g}})$ with constant 
slope (except for the onset of the optically assisted tunneling). 
%As we pointed out above, 
%this result is fully consistent with our semiclassical theory since $\zeta<1$ 
%does not allow reaching the perturbative regime of tunneling for these gap 
%sizes and within the studied range of $E_{\rm{g}}$.
This can be explained by our semiclassical theory 
outlined in subsection~\ref{subsec:SFT} and in subsection~\ref{subsec:PE_bias}. 
The $\zeta$-parameter values $\zeta<1$ do not allow reaching the perturbative 
regime of tunneling for these gap sizes within the studied range of 
$E_{\rm{g}}$. The multiphoton regime of electron transport is then operative 
with an onset of the optical field emission at $E_{\rm{g}} \gtrsim 10$~V/nm. 
As follows from Eq.~\eqref{eq:MPE} and Eq.~\eqref{eq:MPE1}, and in accord with 
TDDFT results, the photon order $n$ in this situation is given by the 
threshold value $n=\frac{\mathcal{W}_{\rm{tb}}}{\omega}$, where the tunneling 
barrier height $\mathcal{W}_{\rm{tb}}$ is defined with  Eq.~\eqref{eq:effectiveWF1}.
Reaching the perturbative regime at $\zeta \geq 1$ even with $d_{\rm{gap}}=1$~nm 
would require too low optical fields where $\mathcal{N}$ is then too small to 
be calculated with TDDFT.

To observe the change of the electron tunneling channels bringing the main 
contribution to PAT one thus needs a smaller size of the gap which allows 
for sufficient tunneling probability at high electron binding energies. This 
is confirmed in Figure~\ref{fig:Transfer_08_1_2}a by the results obtained 
for a narrow $d_{\rm{gap}}=0.8$~nm gap where the change of the slope of the 
$\mathcal{N}(E_{\rm{g}})$ dependence clearly appears close to 
$E_{\rm{g}} \approx 3.5$~V/nm. The effective number of photons is $n \approx 2$ 
at lower $E_{\rm{g}}$. At higher $E_{\rm{g}}$ it is $n \approx 3$, close to 
the $\frac{\mathcal{W}_{\rm{tb}}}{\omega}$ threshold. It is tempting to 
associate the photon order $n$ with the leading electron transfer channel 
$\ell_{\rm{Main}}$ where both two- and three-photon absorption are associated 
with the tunneling transition. Indeed, the electron energies of the $\ell=2$ 
and $\ell=3$ channels of PAT are below the potential barrier separating the 
nanowires as sketched in Figure~\ref{fig:Transfer_08_1_2}b. In this respect, 
as we demonstrate below, the assignment of the leading electron 
transfer channel $\ell_{\rm{Main}}$ from effective photon order $n$ should be 
done with care. This is because the $\mathcal{N} \propto E_{\rm{g}}^{2n}$ 
dependence generally stems from the simultaneous contribution of several 
electron transfer channels $\ell$. It is only at lowest $E_{\rm{g}}$ values, 
in the perturbative regime, that the $n \simeq \ell_{\rm{Main}}$ link can be 
made. In this respect $n \approx 3$ involves not only $\ell=3$, but also 
higher $\ell$ channels corresponding to over-the barrier transitions.

The evolution of the electron transport regime with $E_{\rm{g}}$ can 
be explained with the help of the SFT. For a junction width of 0.8~nm,  
the field amplitude $E_{\rm{g}}=3$~V/nm results for the present system 
in $\zeta=1$. As discussed in subsection~\ref{subsec:PE_bias}, the 
over-the-barrier electron excitation is not permitted within the 
semiclassical theory for field strengths below this value. In this 
situation, the electron transport can be described by the perturbative 
multiphoton regime of PAT. For $E_{\rm{g}}$ above this threshold, marked 
with a vertical red bar in Figure~\ref{fig:Transfer_08_1_2}a, the 
over-the-barrier electron transport is possible. An effective photon 
order $n$ quickly converges to 
$\frac{\overline{\mathcal{W}}_{\rm{tb}}}{\omega} \approx
\frac{\mathcal{W}_{\rm{tb}}}{\omega}$, where the effective height 
$\overline{\mathcal{W}}_{\rm{tb}}$ of the tunneling barrier introduced 
in the semiclassical theory is defined with Eq.~\eqref{eq:BarrierSFT}. 
The observed change in the slope of the $\mathcal{N}(E_{\rm{g}})$ 
dependence near $\zeta = 1$ thus signals a shift in the underlying 
transport channels.

%
% FIGURE 5   WPP for the gap 0.8 nm, no bias, thresholds 
%
\begin{figure}[t!]
\centering
\includegraphics[width=0.95\linewidth]{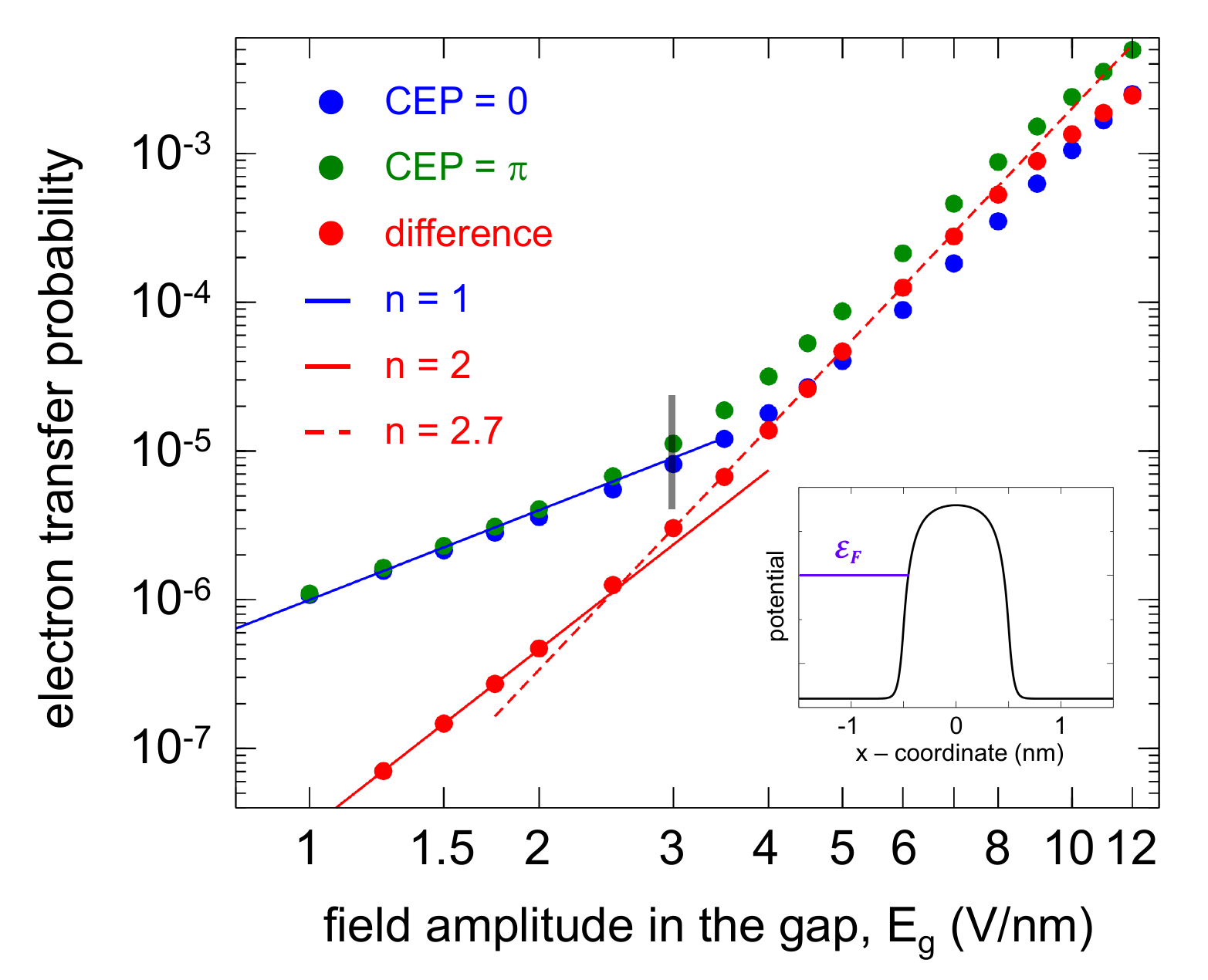}
\caption{Model 1D WPP study of electron transfer induced by the $x$-polarized 
single-cycle optical pulse ($\omega=0.95$~eV, CEP$=0$ and CEP$=\pi$) across metal 
leads separated by a 0.8~nm gap. The metal work function is $\Phi_{\rm{M}}=5.0$~eV. 
Initially, an electron occupies an orbital at the Fermi energy $\mathcal{E}_F$ in 
the left lead. The electric field of the single-cycle pulse is given by 
Eq.~\eqref{eq:Egap}. The probability $\mathcal{P}_{L \rightarrow R}$ of electron
transfer from the left to the right electrode across the junction is shown with 
solid dots as a function of the field amplitude in the gap $E_{\rm{g}}$. 
Blue dots: $\mathcal{P}_{L \rightarrow R}(0)$ obtained with CEP$=0$. 
Green dots: $\mathcal{P}_{L \rightarrow R}(\pi)$ obtained with CEP$=\pi$.  
Red dots show the difference 
$\mathcal{N}=\mathcal{P}_{L \rightarrow R}(\pi)-\mathcal{P}_{L \rightarrow R}(0)$
representing the net electron transfer induced by the CEP$=\pi$ single-cycle 
optical pulse between the two leads. Lines: fit by the $E_{\rm{g}}^{2n}$ dependence. 
The effective number of photons $n$ is given in the legend. Clear changes of $n$ 
occur around $E_{\rm{g}}=3$~V/nm ($\zeta=1$) marked by a vertical grey bar. The 
inset shows the dependence of the model potential in the gap region on the 
$x$-coordinate. Energy zero corresponds to the vacuum level. For further details 
see the text. 
}
\label{fig:WPP_gap_08}
\end{figure}

Noteworthy, the TDDFT results in Figure~\ref{fig:Transfer_08_1_2}a do not show 
electron transfer associated with one-photon absorption as might be expected 
from the perturbative regime of PAT at low $E_{\rm{g}}$ and as reported 
in recent experiments \cite{Light_STM_Kumagai23,Luo2024}. At this point it is 
important to recall that a dc bias was applied to the junction in experimental 
conditions. For a symmetric junction at zero bias as considered here, the 
cw PAT theories predict zero net electron transfer (consider 
Eq.~\eqref{eq:Pedersen} for $U=0$). The rectification becomes possible owing 
to the single-cycle duration of the optical pulse, where the symmetry is 
broken by the optical field itself. The absence of PAT associated with 
one-photon absorption at low $E_{\rm{g}}$ can be then understood using 
perturbation theory, rotating wave approximation, and the linear response 
formalism as developed in Appendix~\ref{Apd.B}. We demonstrate that $\mathcal{N}$ 
is zero when driven by one-photon absorption in a symmetric system. 
The non-zero net electron transfer in the symmetric nanowire dimer requires 
at least a second-order process associated with two-photon absorption as 
we indeed observe in the TDDFT results.
%
%  This is the interference between the first and second order terms

This finding can be further illustrated with a one-dimensional 
single-active-electron model WPP study.  In Figure~\ref{fig:WPP_gap_08} we 
analyze the probability $\mathcal{P}_{L \rightarrow R}(\varphi)$ for an 
electron initially located in the left metal lead to be transferred across 
the $d_{\rm{gap}}=0.8$~nm junction. Results of the WPP calculations for 
the single-cycle optical pulse with CEP $\varphi=0$ (blue dots) and 
$\varphi=\pi$ (green dots) are shown as a function of the amplitude of the 
optical field in the gap $E_{\rm{g}}$. As follows from the fit of 
$\mathcal{P}_{L \rightarrow R}$ by the $E_{\rm{g}}^{2n}$ dependence,  
clear changes of $n$ occur around $E_{\rm{g}}=3$~V/nm ($\zeta=1$), marked 
by a vertical grey bar. The underlying change of the transport regime is 
discussed in subsection~\ref{subsec:PE_bias}. 
At low $E_{\rm{g}} < 3$~V/nm ($\zeta>1$), in the perturbative regime of PAT, 
electron transfer across the junction is dominated by one-photon 
absorption ($n=1$ reflects the $\ell=1$ channel of PAT). In this situation 
the effect of CEP is very small, in full accord with predictions of the 
first-order perturbation theory outlined in Appendix~\ref{Apd.B}. At high 
$E_{\rm{g}}>3$~V/nm ($\zeta<1$), below the onset of optical field emission, 
nonlinearity increases ($n \approx 2.7$) and the results start to notably 
depend on CEP. This corresponds to the multiphoton regime with 
electron transport owing to PAT plus over-the-barrier transitions.

The \textit{net} electron transfer $\mathcal{N}$ (red dots) is given by 
the difference of electron transfer probabilities from one lead to 
the other. Using the symmetry of the system, one can write 
$\mathcal{N}=\mathcal{P}_{L \rightarrow R}(\pi)-\mathcal{P}_{R \rightarrow L}(\pi) 
= \mathcal{P}_{L \rightarrow R}(\pi)-\mathcal{P}_{L \rightarrow R}(0)$. 
Since for $E_{\rm{g}} < 3$~V/nm the one-photon absorption channels cancel each 
other, as explained above, the two-photon absorption ($n=2$) determines the 
net electron transfer with $\mathcal{N} \propto E_{\rm{g}}^{4}$. 
This is followed by the transition to higher nonlinearity with $n=2.7$ for 
$3 \lesssim E_{\rm{g}} \lesssim 8 $~V/nm. Finally, upon further increase of 
$E_{\rm{g}}$, the optical field emission regime sets in, leading to lower 
nonlinearity. It is remarkable that the simple one-state single-electron 
model calculations capture the main trends of the calculated with TDDFT 
dependence of the net electron transfer on the optical field (see 
Figure~\ref{fig:Transfer_08_1_2}a).

%
% FIGURE 6  gap 1 nm, transport for different bias, potential barrier, thresholds
%

%\onecolumngrid

%\vspace{3mm}

\begin{figure*}[t!]
%\begin{figure}[h]
\centering
\includegraphics[width=0.85\linewidth]{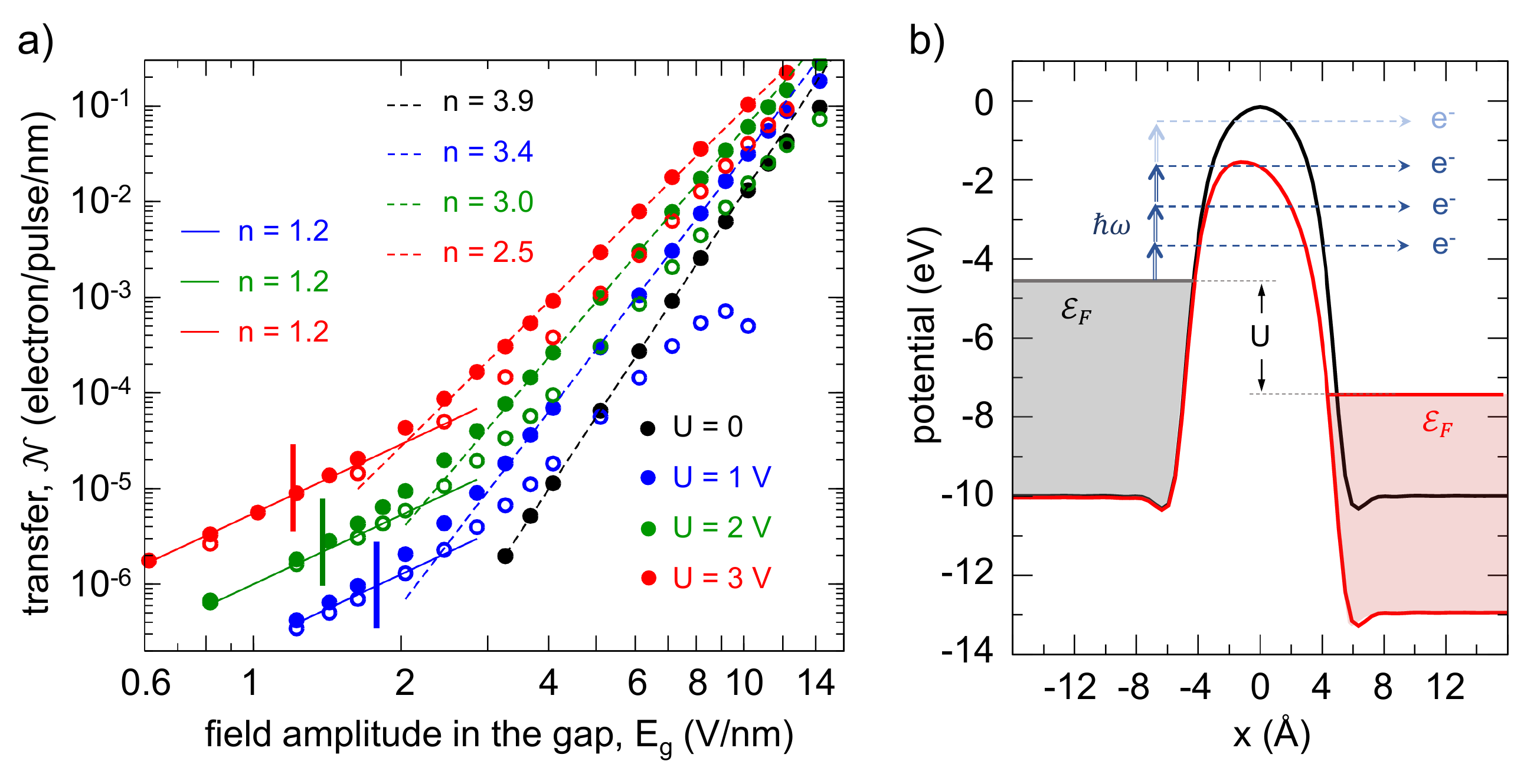}
\caption{
%Single-cycle optical pulse. TDDFT transfer across the 1~nm gap. The effect 
%of an applied bias. In the TDDFT calculations we use $\Phi=4.5$~eV resulting 
%from the polarisation correction and aimed to reproduce the 1 nm junction 
%between metals with work functions $\Phi_{\rm{M}}=5.4$~eV. In panel b $U=3$~eV.
%
%
Effect of an applied bias on electron transfer induced by a single-cycle 
$x$-polarized optical pulse ($\omega=0.95$~eV, CEP$=0$, CEP$=\pi$), across a  
1~nm wide gap of a cylindrical nanowire dimer with work function 
$\Phi_{\rm{M}}=5.4$~eV. 
\textbf{Panel a:} Optically induced net electron transfer $\mathcal{N}$ 
calculated with TDDFT per pulse and per nm length of the dimer is shown 
as a function of the field in the gap $E_{\rm{g}}$. 
Color dots:  results obtained for different values of an applied dc bias $U$. 
Filled dots are used for CEP$=\pi$, and open dots of the same color are used 
for CEP$=0$ except for $U=0$ where only the CEP$=\pi$ result is shown. 
Lines: fit by the $\mathcal{N} \propto E_{\rm{g}}^{2n}$ 
dependence characteristic for the multiphoton regime. The effective number of 
photons $n$ is given in the legend. The vertical bars of the corresponding 
color mark the change of the transport regime at $\zeta=1$.
For further details see the legend.  
\textbf{Panel b:} One-electron potential in the 1~nm gap region 
as a function of the $x$-coordinate along the dimer axis. The energy is 
measured with respect to the vacuum level of the left cylinder. 
Black line: zero applied bias ($U=0$); 
red line: applied bias $U=3$~V as measured between the Fermi level energies 
of the nanowires indicated with $\mathcal{E}_F$. The effective work function 
$\Phi=4.6$~eV is used in the TDDFT calculation (see Table~\ref{table1}). 
The hatched areas indicate occupied electronic states of the valence band. 
Dashed horizontal arrows indicate electron tunneling and over-the-barrier 
transitions induced by $\ell$-photon absorption. The latter is indicated 
with vertical arrows. 
Light color is used for $\ell=4$-photon absorption with electron 
energy close to the top of the potential barrier, dominating 
transport in the $U=0$ case. 
}
\label{fig:TransportBias1nm}
%\end{figure}
\end{figure*}

%\twocolumngrid

\subsection{Symmetry breaking by an applied bias}
\label{subsec:AppliedBias}

The $\ell=1$ electron tunneling channel associated with one-photon 
absorption emerges in the net electron transfer for asymmetric 
systems. Such asymmetry occurs, e.g., in break-junctions or STM 
tip-to-flat-surface junctions \cite{ward2010optical,Lenner11,
Dasgupta2018}, as well as in junctions formed by different 
metals \cite{ning2024}. Symmetry breaking can also be controllably 
introduced by applying a dc bias across the junction
\cite{STM_Light_Schroder_2020,Light_STM_Kumagai23,Luo2024}. For a 
dimer of identical parallel nanowires, we adopt the latter strategy, 
which enables direct comparison with recent experiments in tunneling 
junctions \cite{STM_Light_Schroder_2020,Light_STM_Kumagai23,Luo2024}. 
Along with breaking the symmetry, the applied bias in these 
experiments was used to vary the tunneling barrier and, consequently, 
the interplay between tunneling channels with different $\ell$.

The effect of the applied bias $U$ on the net electron transfer 
$\mathcal{N}$ induced by the single-cycle optical pulse across the 1~nm wide 
gap is shown in Figure~\ref{fig:TransportBias1nm}a. 
TDDFT calculations were performed for different values of the optical field 
amplitude in the gap $E_{\rm{g}}$. 
For zero bias and within the $\mathcal{N}$ range addressed here, we obtain 
a multiphoton character of the net electron transfer with an effective 
number of absorbed photons $n=3.9$, close to $\Phi/\omega$. This indicates 
that overall, the electron transport is dominated by photoexcited electrons 
with energies close to the top of the potential barrier separating the metals 
(see Figure~\ref{fig:TransportBias1nm}b), in full agreement with the 
semiclassical theory (parameter $\zeta < 1$). 
Note that the slightly larger workfunction $\Phi_{\rm{M}}=5.4$~eV (also 
representative for gold) used here explains the somewhat larger $n$ 
obtained for $U=0$ when compared with the results reported for 
the 1~nm gap in Figure~\ref{fig:Transfer_08_1_2}.

The finite bias $U$ applied across the gap reduces the potential barrier 
of the junction so that the net electron transfer increases. 
Moreover, the symmetry breaking introduced by the applied bias enables 
the contribution of electron tunneling assisted by one-photon absorption 
($\ell=1$ PAT channel) into $\mathcal{N}$. 
It follows from the TDDFT results in Figure~\ref{fig:TransportBias1nm}a 
that the one-photon absorption dominates for 
$E_{\rm{g}} \lesssim 2.5$~V/nm with $\mathcal{N} \propto E_{\rm{g}}^{2n}$ 
where $n \approx 1.2$. In agreement with the discussion in subsection 
\ref{subsec:SymmetricSystem}, the CEP dependence tends to be washed out in 
this situation. As predicted by the semiclassical theory and sketched in 
Figure~\ref{fig:fig_zeta}, with increasing $E_{\rm{g}}$ we observe the 
transition from the perturbative regime of PAT ($\zeta > 1$) dominated by 
the $\ell=1$ channel to the multiphoton regime ($\zeta < 1$, PAT+ATI). 
The values of $E_{\rm{g}}$ corresponding to the parameter $\zeta = 1$,  
as introduced by our semiclassical theory, are marked in 
Fig.~\ref{fig:TransportBias1nm}a by vertical color bars. The applied bias 
enhances the above threshold ionization
% from the bound state to the Volkov state 
so that the transition shifts to lower $E_{\rm{g}}$ with increasing $U$.

In the multiphoton regime of electron transport ($E_{\rm{g}} > 3$~V/nm), 
we obtain that the effective number of absorbed photons $n$ is in the 
$2.5-3.4$ range and it is approximately equal to 
$n \approx \mathcal{W}_{\rm{tb}}/ \omega$. Similar to the situation with 
$U=0$, on average, the transferred electron energies approach the threshold 
value determined by the height of the potential barrier of the junction 
$\mathcal{W}_{\rm{tb}}$ given by Eq.~\eqref{eq:BarrierBias}. The smaller 
$n$ calculated for larger $U$ reflects the reduction of the tunneling 
barrier by the applied bias (see Figure~\ref{fig:TransportBias1nm}b). 
This result is fully consistent with the prediction of the SFT 
(Eq.~\eqref{eq:MPE1}), and it was also reported in model calculations with 
simplified form of the potential barrier of the junction 
\cite{PhysRevApplied.17.044008,SciReportsZhang2016}. 
It is worth noting that in TDDFT we define the tunneling barrier height 
from the maximum of the self-consistent potential in the junction including 
the dc field owing to the applied bias (see Eq~\eqref{eq:BarrierBias}). 
In the semiclassical theory, $\overline{\mathcal{W}}_{\rm{tb}}$ is determined   
by the average image potential (see Eq.~\eqref{eq:BarrierSFT}). The 
dc field owing to the applied bias is accounted for via vector potential.

Notably, the TDDFT results reveal that only the transition from the 
one-photon $\ell=1$ to the multiphoton regime of electron transport around 
$E_{\rm{g}} = 2.5$~V/nm is well marked. In the multiphoton regime, the 
contributions of individual transport channels $\ell$ cannot be resolved. 
With increasing $E_{\rm{g}}$, the effective photon order $n$, extracted 
from the power-law dependence of net electron transfer on the optical field, 
$\mathcal{N} \propto E_{\rm{g}}^{2n}$, becomes an averaged characteristic. 
It rapidly converges to the threshold given by the height of the tunneling 
barrier in full accord with Eq.~\eqref{eq:MPE1} of the semiclassical theory.

The degeneracy of the results obtained with CEP$=\pi$ and CEP$=0$ at low 
$E_{\rm{g}}$, where one-photon absorption ($\ell=1$ PAT channel) dominates 
electron transport, is lifted in the multiphoton regime at 
$E_{\rm{g}} > 3$~V/nm. Similar to the unbiased case, the net electron 
transport between parallel nanowires in the presence of a dc bias exhibits a 
sinusoidal dependence on the CEP. However, as reported earlier 
\cite{HotElectrons_in_Gaps,PhysRevBbias}, and therefore not analyzed here, 
an applied bias leads to a vertical offset $\overline{\mathcal{N}}$, such that 
$\mathcal{N} \approx \mathcal{N}_0 \sin(\varphi-\pi/2)+\overline{\mathcal{N}}$. 
For a positive bias, for CEP$=\pi$, the optical and dc fields point in the same 
(negative) $x$-direction, maximizing the net positive electron transport. 
For CEP$=0$, the optical field is along the positive $x$-direction, while the dc 
field points in the opposite direction, so that the net electron transport 
is close to minimum. The difference between the results for CEP$=\pi$ 
and CEP$=0$ provides access to both the oscillation amplitude $\mathcal{N}_0$ 
and the offset $\overline{\mathcal{N}}$ of the CEP-dependent transport.
Notably, at the lowest bias $U=1$~eV, a strong optical field with CEP$=0$ 
can reverse the direction of the net electron transfer. This 
is manifested in the $\mathcal{N}(E_{\rm{g}})$ dependence, which reaches a 
maximum at $E_{\rm{g}} \approx 9$~V/nm and changes sign at higher 
$E_{\rm{g}}$ (not shown in the log-log scale plot used here). For larger 
values of the applied bias, the net electron transfer is always from the 
left to the right cylinder because of the positive offset
$\overline{\mathcal{N}}$ induced by $U$.

%
% FIGURE 7  gap 2 nm, transport for different bias
%
\begin{figure}[t!]
\centering
\includegraphics[width=0.95\linewidth]{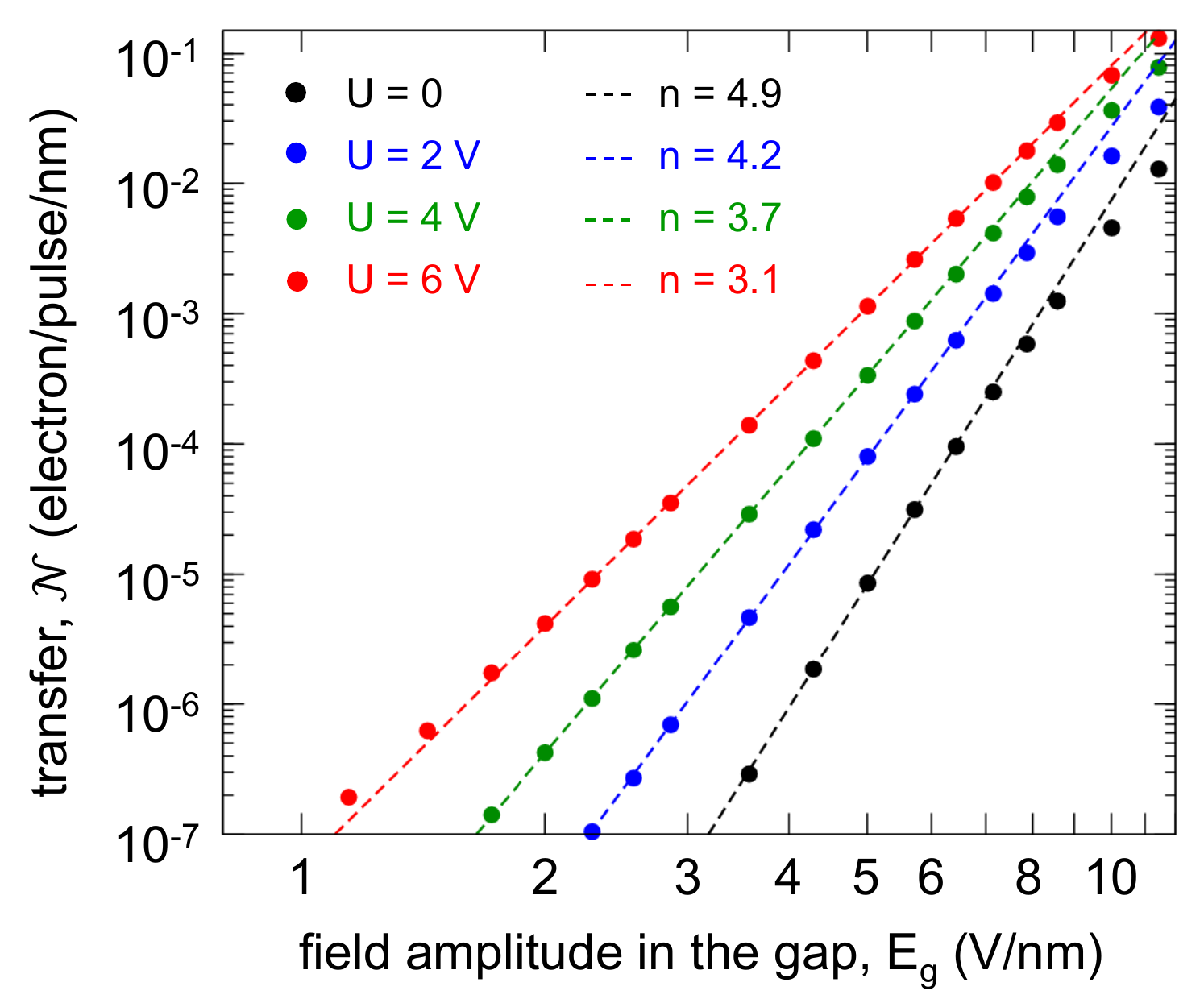}
\caption{
Same as in Figure~\ref{fig:TransportBias1nm}a, but for a  
2~nm wide junction. Only results obtained with CEP$=\pi$ are 
presented.
%In the TDDFT calculations we use $\Phi=5$~eV resulting from the 
% polarisation correction and aimed 
%to reproduce the 2 nm junction between metals with work functions 
%$\Phi_{\rm{M}}=5.4$~eV.
%
%The effect of an applied bias on the electron transfer induced 
%by the single-cycle optical pulse ($\omega=0.95$~eV, CEP$=\pi$) 
%across the 2~nm wide junction of the cylinder nanowire dimer 
%with $\Phi_{\rm{M}}=5.4$~eV. 
%The net electron transfer $\mathcal{N}$ calculated with TDDFT 
%for different values of an applied dc bias $U$ is shown 
%with color dots as a function of the field in the gap $E_{\rm{g}}$. 
%The lines display the fit by the 
%$\mathcal{N} \propto E_{\rm{g}}^{2n}$ dependence characteristic for 
%the multiphoton regime. For further details see the legend.  
}
\label{fig:TransportBias2nm}
\end{figure}

The suppression of PAT associated with the $\ell=1$ single-photon channel with 
increasing gap size is illustrated in Figure~\ref{fig:TransportBias2nm}. 
The TDDFT calculations are performed for the 2~nm gap using an optical pulse 
with CEP$=\pi$. The applied bias $U$ is set in such a way that the dc field 
in the junction $E_{\rm{bias}}$ is the same as for the 1~nm gap discussed in 
Figure~\ref{fig:TransportBias1nm}. Thus, the height of the tunneling 
barrier is similar in both cases. However, since the gap size increases from 1~nm 
to 2~nm the tunneling barrier broadens. 
This favors the multiphoton transitions with electron energies close to the 
threshold value $\mathcal{E}_F + \ell \omega \approx \mathcal{W}_{\rm{tb}}$.  
Small values of $E_{\rm{g}}$ are then required for the 
$\ell=1$ channel to be observed. Consequently, the corresponding values of 
$\mathcal{N}$ would be extremely small, and lie outside the convergence range 
of the present implementation of the TDDFT.
Thus, the results in Figure~\ref{fig:TransportBias2nm} reflect only the 
multiphoton regime of optically induced electron transport ($\zeta<1$ in 
the semiclassical SFT) with an effective photon order $n$ smaller for larger 
applied bias because of the smaller potential barrier. The suppression of 
below-the-barrier transitions also explains the overall higher $n$ obtained 
for the 2~nm gap as compared to the 1~nm gap.

%
% FIGURE 8, WPP gap 1 nm, detailed result, spectra, thresholds, Long Pulse O.95 eV
% the fluxes inside original lead
%

%\onecolumngrid

%\begin{center}
%\begin{figure}[h]
\begin{figure*}[t!]
\includegraphics[width=0.95\linewidth]{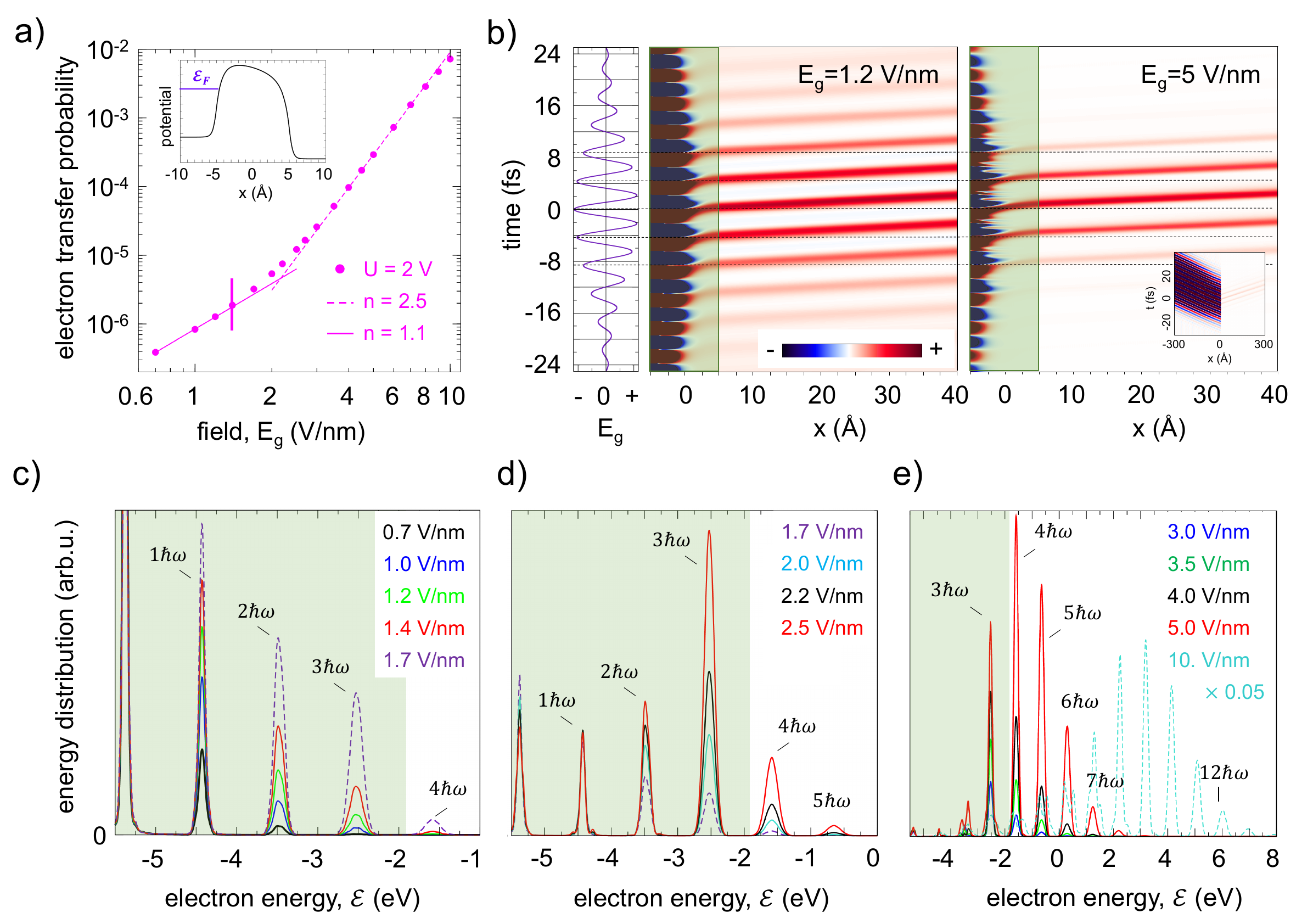}
\caption{
%WPP, $\Phi_{\rm{M}}=5.4$~eV.
Model 1D WPP study of the optically induced electron transfer between metal leads 
separated by a 1~nm vacuum gap. Metal work function $\Phi_{\rm{M}}=5.4$~eV.
A dc bias $U=2$~V is applied across the gap. The electric field of the 
several-cycle optical pulse is given by Eq.~\eqref{eq:Egap} with 
$\tau=3.4 \times 2\pi/\omega$, and CEP$=\pi$. 
\textbf{Panel a:} Probability of optically induced electron transfer 
(for this duration of the optical pulse it is independent 
of CEP) as a function of the optical field 
amplitude in the gap $E_{\rm{g}}$. Initially, the active electron occupies an  
orbital with Fermi energy $\mathcal{E}_F$ in the left lead. WPP results 
are shown with solid dots. The vertical bar marks $\zeta=1$. 
The lines display the fit by the $E_{\rm{g}}^{2n}$ 
dependence. The effective number of photons $n$ is given in the legend.  
The inset shows the $x$-coordinate dependence of the model potential 
in the gap region. Energy zero corresponds to the vacuum level of the left lead.
\textbf{Panel b:} 2D maps of the normalized electron probability current 
density $j_{\rm{opt}}(x,t)$ induced by an optical pulse with 
$E_{\rm{g}}=1.2$~V/nm (left sub-panel),  
and with $E_{\rm{g}}=5$~V/nm (right sub-panel), as indicated in the legends. 
The WPP results are shown as a function of the $x$-coordinate (horizontal 
axis) and time (vertical axis). The color code is given in the inset of 
the left sub-panel. The line plot at the left shows the 
time evolution of the electric field in the gap. The shaded green region 
of $x$-coordinates indicates the junction. 
The dashed lines mark the crests of the field in the gap with negative 
polarity. The inset of the right sub-panel shows the 2D map of $j_{\rm{opt}}(x,t)$ 
using large $x$-scale.
\textbf{Panels c,d,e:} Energy spectra of the electron transferred across the gap 
into the right lead calculated for different $E_{\rm{g}}$ as indicated in the legends. 
Results are shown as a function of electron energy $\mathcal{E}$ measured 
with respect to the vacuum level of the left lead. In panel e results for 
$E_{\rm{g}}=10$~V/nm are scaled by $\times 0.05$. 
The shaded green region indicates electron 
energies which are below the barrier of the junction, and therefore correspond 
to tunneling transitions. For further details see the text.
}
\label{fig:Figure_WPP_w0.95}
\end{figure*}
%\end{figure}
%\end{center}

%\twocolumngrid

Additional insights into the interplay between electron tunneling channels 
associated with the absorption of different number of photons $\ell$ can be 
obtained from the analysis of the energy spectra of the transferred electron 
for a given initial state. To this end, we perform one-dimensional, 
one-electron WPP calculations (see subsection~\ref{subsec:WPP}) of electron 
transfer across a 1~nm gap between semi-infinite jellium metal leads. 
A bias $U=2$~V is applied. The electron active in the optically assisted 
transitions initially occupies the orbital at the Fermi energy 
$\mathcal{E}_F$ in the left lead. To clearly resolve sidebands associated 
with the absorption of different numbers of photons $\ell$ in the energy 
spectra of the transmitted electron, 
we increase the optical pulse duration from 5.2~fs to 21~fs 
($\tau=3.4 \times 2\pi/\omega$, see Eq.~\eqref{eq:Gaussianfield}). 
The WPP results are detailed in Figure~\ref{fig:Figure_WPP_w0.95}. 
It is worth noting that for a several-cycle pulse, such as that used here, 
the electron transport does not depend on CEP.

In Figure~\ref{fig:Figure_WPP_w0.95}a, we show the dependence of the 
optically induced electron transfer on the amplitude of the optical field 
in the gap $E_{\rm{g}}$. Notice that although the WPP calculations address  
only the electron transfer from the left to the right lead, the results in 
Figure~\ref{fig:Figure_WPP_w0.95} are representative for the net electron 
transfer. Indeed, for $U=2$~eV the electron transfer in the opposite
direction  (from the right lead to the left lead) would require absorption 
of at least two additional photons and is therefore essentially smaller 
under the present conditions. Similar to the findings in 
subsection~\ref{subsec:SymmetricSystem}, comparison of the TDDFT 
results in Figure~\ref{fig:TransportBias1nm} with the WPP results in 
Figure~\ref{fig:Figure_WPP_w0.95}a shows that the one-dimensional  
single-active-electron model reproduces the trends obtained in TDDFT 
calculations with an applied bias. The WPP results are also consistent 
with an analysis based on the semi-analytical theory 
(see Ref.~\cite{ma2025robust} and subsection~\ref{subsec:SFT}).  
Specifically, at low  $E_{\rm{g}}$ ($\zeta>1$), photon-assisted electron 
transport is dominated by one-photon absorption ($n=1.1$), whereas for 
$E_{\rm{g}} \gtrsim 3$~V/nm ($\zeta<1$) a higher nonlinearity 
characteristic for the multiphoton regime sets in. In this regime, 
an effective number of absorbed photons quickly converges to $n=2.5$ 
close to the tunneling barrier threshold 
$\mathcal{W}_{\rm{tb}} / \omega = 2.7$, and remains essentially 
constant with increasing $E_{\rm{g}}$ up to the onset of 
the optical field emission.

In Figure~\ref{fig:Figure_WPP_w0.95}b we show the electron probability 
current density $j_{\rm{opt}}(x,t)$ induced by the optical pulse.  
The WPP results reveal electron bursts emitted from the surface 
of the left lead at the crests of the optical field with negative polarity. 
These bursts traverse the junction (hatched green area) in less than 
1~fs, and are subsequently injected into the right lead. Since the optical 
and dc fields are screened inside the metal, the injected electron follows 
a straight-line trajectory after crossing the metal/vacuum interface of the 
right lead. Under the present conditions, the applied bias and narrow 
junction impede the quiver motion that would reverse the direction of 
electron propagation in the gap 
(see also Ref.~\cite{HotElectrons_in_Gaps}).

The stronger nonlinearity of electron transport with electric-field 
strength in the gap, or equivalently, the larger effective number of 
absorbed photons in the multiphoton regime, results in a shorter time 
interval during which the optical pulse drives the transport between 
the leads. This is evidenced in Figure~\ref{fig:Figure_WPP_w0.95}b by 
the reduced number of electron bursts emitted at the crests of the 
optical field and traversing the junction for 
$E_{\rm{g}} \gtrsim 5$~V/nm (multiphoton regime) compared to the data 
obtained for $E_{\rm{g}} \gtrsim 1.2$~V/nm 
(tunneling dominated by single-photon absorption). Importantly, in all  
cases considered, we find that the optically induced electron transfer 
dynamics instantaneously follows the temporal variation of the optical 
field. The probability current density $j_{\rm{opt}}(x,t)$ in the gap 
region is nonzero only while the optical pulse acts on the system. 
This result is in full agreement with previous TDDFT and one-electron 
studies of electron transport in metallic gaps and photoemission from 
metal surfaces driven by single cycle and longer laser pulses 
\cite{Yalunin2011,Kruger2012,Wachter2012,IvanovGap2021,Kim_2021,
PhysRevApplied.17.044008,Ritzkowsky2024,PhysRevBbias,
HotElectrons_in_Gaps,ma2025robust}. It highlights the contribution 
of a fast, one-step process in which electron excitation by the laser 
pulse and transport across the gap cannot be separated in time.

It is also worth noting that, because of optical-field screening inside 
the metal and electron momentum change at the jellium surface, electron 
excitation occurs within the surface layer (analogous to the Landau 
damping mechanism of surface plasmons \cite{Khurgin2024}).  A fraction 
of the excited electrons crosses the gap, leading to electron transport 
between the leads. However, the majority remains within the ``parent" 
left lead, propagating from the surface into the bulk, as evidenced by 
the inset of the right sub-panel of Figure~\ref{fig:Figure_WPP_w0.95}b). 
In the TDDFT calculations we obtain qualitatively similar trends. 
However, because of contributions from 
both surfaces across the junction, and because of the overlap between 
current densities associated with the electron transport and those 
arising from polarization of the cylinders, the data are considerably 
more difficult to analyze \cite{HotElectrons_in_Gaps}. The relaxation 
of the nonequilibrium carriers generated at the surface and propagating 
into the bulk involves electron-electron and electron-phonon scattering 
events \cite{Voisin2001,Grua2003,BAUER2015,Besteiro2017,Dubi2019,
Schirato2023,Khurgin2024}, which are not included in the one-electron 
WPP calculations and lie beyond the ALDA-TDDFT framework employed 
here \cite{ullrich2019time,PhysRevLett.95.086401,PhysRevLett.77.2037}.

Finally, in Figure~\ref{fig:Figure_WPP_w0.95}c,d,e we show the energy 
spectra of the transferred electron given by the energy-resolved outgoing 
electron probability current density in the asymptotic region inside the 
right lead $j(x\rightarrow \infty, \mathcal{E})$. Results are obtained 
for different values of the optical field in the gap. The electron energy 
$\mathcal{E}$ is measured with respect to the vacuum level of the left 
lead. The spectra display peaks at energies $\mathcal{E}_F+\ell \omega$ 
corresponding to electron transfer assisted by absorption of $\ell$ 
photons. For optical fields $E_{\rm{g}} \lesssim 1.4$~V/nm 
(Figure~\ref{fig:Figure_WPP_w0.95}c), the $\ell=0,1,2,3$ contributions 
are visible. Electron transfer occurs primarily via tunneling with 
electron energies below the potential barrier of the junction (the 
corresponding energy range is indicated by the hatched green area 
in Figure~\ref{fig:Figure_WPP_w0.95}c,d,e). The intensity of the sidebands 
scales as $\propto E_{\rm{g}}^{2\ell}$ with the $\ell=0$ peak (dc tunneling) 
and the $\ell=1$ peak dominating the spectrum (tunneling assisted by one-photon 
absorption, i.e., the $\ell=1$ channel of PAT). These results fully  
agree with WPP results reported in Figure~\ref{fig:Figure_WPP_w0.95}a 
for this range of $E_{\rm{g}}$. Indeed, the effective photon order $n=1.1$ 
obtained using fit by $\mathcal{N} \propto E_{\rm{g}}^{2n}$ dependence 
corresponds to the $\ell=1$ channel leading the PAT.

The electron spectra in Figure~\ref{fig:Figure_WPP_w0.95}c shed light on 
why, in Figure~\ref{fig:Figure_WPP_w0.95}a, the theoretical parameter $\zeta=1$,  
indicating a change of the transport regime, appears at smaller
$E_{\rm{g}}$ than the transition from lower ($n=1$) to higher 
($n=2.5$) nonlinearity in the simulated $\mathcal{N}$. Similar trends are 
observed in Figure~\ref{fig:Transfer_08_1_2}a, in Figure~\ref{fig:WPP_gap_08}, 
and in Figure~\ref{fig:TransportBias1nm}a. According to the definition of the 
cutoff energy in Eq.~\eqref{eq:cutoff}, the condition $\zeta=1$ (here at 
$E_{\rm{g}} \sim 1.4$~V/nm) corresponds to the point where ${\cal E}_{\rm{cutoff}}$ 
reaches the barrier threshold, so that at higher $E_{\rm{g}}$ electronic states 
above the potential barrier can be populated. As shown in 
Figure~\ref{fig:Figure_WPP_w0.95}c, clear evidence of photon-assisted excitation 
above the barrier emerges only when $E_{\rm{g}}>1.4$~V/nm. Because the 
contribution of high-energy sidebands near the cutoff is statistically small, 
a larger field strength is required to increase $n$ and, in particular, to 
reverse the perturbative evolution of the sideband weights with $\ell$.

This change in the $\ell$ progression of the sidebands is clearly observed 
in Figure~\ref{fig:Figure_WPP_w0.95}d for 
$1.7$~V/nm $\leq E_{\rm{g}} \leq 2.5$~V/nm. The sequence evolves from the 
decreasing perturbative $E_{\rm{g}}^{2\ell}$ trend described by 
Eq.~\eqref{eq:Pedersen} to a reversed progression with dominance of the 
three-photon absorption process ($\ell=3$ channel of PAT). The electron 
energies corresponding to the $\ell=1,2,3$ PAT channels are below the 
top of the potential barrier, so that the tunneling transitions continue 
to dominate the transport. Note that within this range of $E_{\rm{g}}$, 
the nonlinearity of the $\mathcal{N}(E_{\rm{g}})$ dependence progressively 
increases. For higher fields (Figure~\ref{fig:Figure_WPP_w0.95}e), the 
leading contributions to the energy spectrum shift to the $\ell=4,5$ 
sidebands. For $E_{\rm{g}} \geq 5$~V/nm electron transport becomes 
dominated by classically allowed over-the-barrier transitions. At the 
onset of the optical field emission regime ($E_{\rm{g}}=10$~V/nm), 
photon orders up to $\ell=12$ can be clearly observed, with electron 
energies reaching up to 7~eV above the barrier 
(theoretical cutoff is 7.63 eV according to Eq.~\eqref{eq:cutoff}).

To summarize the results obtained in this Section, TDDFT calculations show, 
in agreement with semiclassical theory, that establishing a one-to-one 
correspondence between the dominant channel $\ell_{\rm{Main}}$ of optically
induced transport and the effective photon order $n$ extracted from the 
$\mathcal{N} \propto E_{\rm{g}}^{2n}$ dependence must be done with care. 
Strictly speaking, assignment is only possible in the limit 
$E_{\rm{g}} \rightarrow 0$, where $n$ match the lowest possible $\ell$. 
In this regime, one can resolve 
one-photon assisted tunneling ($\ell = 1$) for non-symmetric junctions, 
and the two-photon assisted tunneling ($\ell= 2$) for symmetric junctions. 
In the multiphoton regime of electron transport, the contributions of 
higher photon orders overlap. The transition between the leading transport 
channels $\ell_{\rm{Main}}$ with changing $E_{\rm{g}}$ cannot be resolved. 
The effective photon order $n$ is then represents an average characteristic. 
With increasing $E_{\rm{g}}$, it quickly converges to the threshold value 
set by the barrier height, $n \approx \mathcal{W}_{\rm{tb}}/\omega$. 
This behavior contrasts with photoemission into vacuum, as in the case of 
an individual surface or wide junction between metals, where PAT with 
electron energies below the barrier is suppressed. In this situation, 
the $E_{\rm{g}}^{2n}$ dependence of electron emission or transport over 
the entire $E_{\rm{g}}$ range prior to the onset of the optical field 
emission, is determined by $n \approx \Phi_{\rm{M}}/ \omega$ 
corresponding to the multiphoton photoemission threshold.

%%%%%%%%%%%%%%%%%%%%%%%%%%%%%%%%%%%%%%%%%%%
%%%%%%%%%%%%%%%%%%%%%%%%%%%%%%%%%%%%%%%%%%%

\section{Comparison with available experimental data}
\label{sec:experiment}

A number of recent experiments reported on the effect of an applied bias and 
of the strength of the optical field on lightwave-driven electron tunneling 
in metallic junctions \cite{Diesing2004B,GargScience2020,STM_Light_Schroder_2020,
Dai2021,Luo2024,Luo2023,MULLER_REVIEW,Light_STM_Kumagai23,Light_STM_Muller22}. 
In particular, the relative contributions of tunneling channels 
involving single-photon and multiphoton absorption have been examined 
\cite{STM_Light_Schroder_2020,Light_STM_Kumagai23,Luo2024}. 
The interpretation of these experiments typically involves parametrization 
of the electron excitation and tunneling processes. To gain clarity in 
the underlying physical mechanisms it is thus of interest to examine the TDDFT 
results for experimentally studied systems.

%%%%%%%%%%%%%%%%%%%%%%%%%%%%%%%%%%%%%%%%%%%

\subsection{Gold bowtie antenna}
\label{subsec:bowtite}

The contributions of one-photon and multiphoton assisted tunneling  
in the gap of a gold bowtie antenna were investigated by Yang Luo and 
collaborators \cite{Luo2024}. In their experiment, electron transport across 
the gap was induced by 7~fs laser pulses, resonant 
with the nanoantenna plasmon at $\omega=1.6$~eV. The radii of the 
nanoantenna tips forming the gap were estimated to be 5~nm. Since this matches 
the radii of the metallic nanowires used in our calculations, a comparison 
of the experimental data with TDDFT calculations becomes particularly relevant.

\begin{figure}[t!]
\centering
\includegraphics[width=0.95\linewidth]{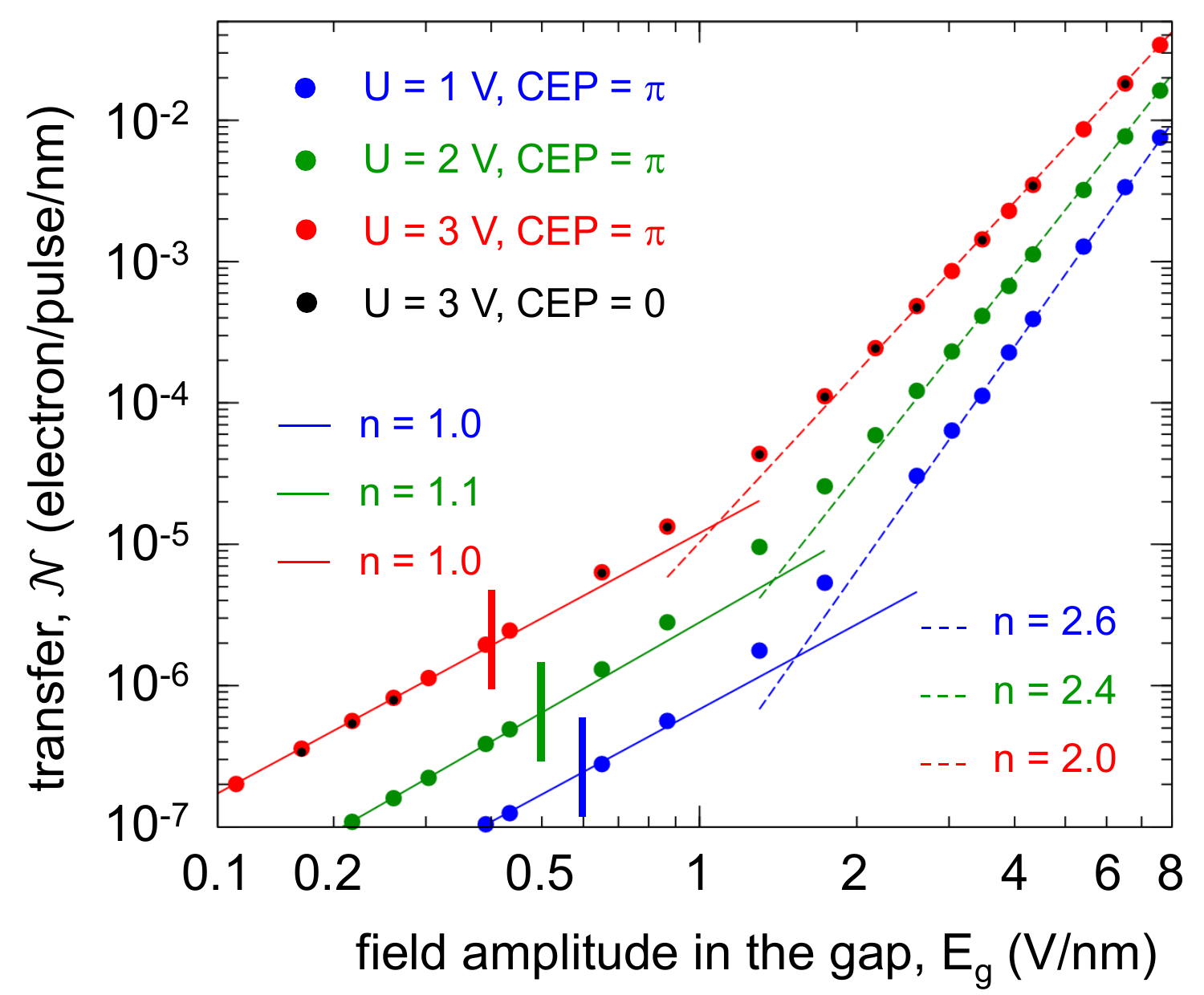}
\caption{
%7~fs optical pulse, $\omega=1.6$~eV. TDDFT transfer across 
%the 1 nm gap. The effect of an applied bias. In the TDDFT calculations 
%we use $\Phi=4.5$~eV resulting from the polarisation correction and aimed 
%to reproduce the 1 nm junction between metals with work functions 
%$\Phi_{\rm{M}}=5.3$~eV.
%
Effect of an applied dc bias on electron transfer induced by a 7~fs 
long $x$-polarized optical pulse ($\tau=4.2$~fs, $\omega=1.6$~eV, CEP$=0$ 
and CEP$=\pi$ in Eq.~\eqref{eq:Gaussianfield}) across a 1~nm wide gap of 
a nanowire dimer. The work function is $\Phi_{\rm{M}}=5.3$~eV. 
Dots: optically induced net electron transfer $\mathcal{N}$ calculated with TDDFT 
per pulse and per nm length of the dimer as a function of the field in the 
gap $E_{\rm{g}}$. The different colors indicate different values of the 
applied dc bias $U$ and CEP, as explained in the legend.  
Lines: fit by the $\mathcal{N} \propto E_{\rm{g}}^{2n}$ dependence. 
The effective number of photons $n$ is given in the legend. The vertical 
bars of the corresponding color mark the transport regime change at $\zeta=1$.
}
\label{fig:Garg_PowerCurveBias}
\end{figure}

In Figure~\ref{fig:Garg_PowerCurveBias} we show the net electron transfer 
$\mathcal{N}$ across the gap of a nanowire dimer calculated with TDDFT as 
a function of the optical field strength in the gap, $E_{\rm{g}}$, for 
different values of the applied bias $U$. Following the discussion of 
experimental results in Ref.~\onlinecite{Luo2024}, we consider a gap of 
1~nm and a metal work function $\Phi_{\rm{M}}=5.3$~eV representative for 
gold. The incident optical pulse is described by Eq.~\eqref{eq:Gaussianfield} 
with $\omega=1.6$~eV and $\tau=4.2$~fs  (corresponding to an intensity 
FWHM of 7~fs). In contrast to the single-cycle optical pulse addressed in 
the previous section, the several-cycle pulse used here results in a 
negligible dependence of $\mathcal{N}$ on CEP. This is illustrated in 
Figure~\ref{fig:Garg_PowerCurveBias} were calculations performed for an 
applied dc bias $U=3$~V are shown for CEP$=\pi$ 
(red dots) and CEP$=0$ (black dots).

Qualitatively, the results reported in Figure~\ref{fig:Garg_PowerCurveBias} 
are similar to those discussed in Figure~\ref{fig:TransportBias1nm}a. 
For small $E_{\rm{g}}$, optically induced electron transport is dominated 
by single-photon absorption. The multiphoton absorption prevails for 
$E_{\rm{g}} \gtrsim 1.5$~V/nm. A higher applied bias $U$ lowers the potential 
barrier between the nanowires so that $\mathcal{N}$ becomes larger over 
the entire $E_{\rm{g}}$ range. For the same reason, in the multiphoton regime, 
the effective number of photons $n$ driving the transport decreases with 
increasing $U$. Note that the fundamental frequency $\omega=1.6$~eV 
corresponding to the experiment of Luo \textit{et al.} \cite{Luo2024} is used 
here, whereas the results in Figure~\ref{fig:TransportBias1nm}a are obtained 
with $\omega=0.95$~eV. The difference in photon energy accounts for the 
different effective photon orders $n$ obtained in the two sets of results.

%
% FIGURE 9  The gold bowtie, Garg Pulse, Different bias, thresholds
%
%

%
% FIGURE 10   Garg pulse, effect of d gap and work function
%
\begin{figure*}[t!]
\centering
\includegraphics[width=0.9\linewidth]{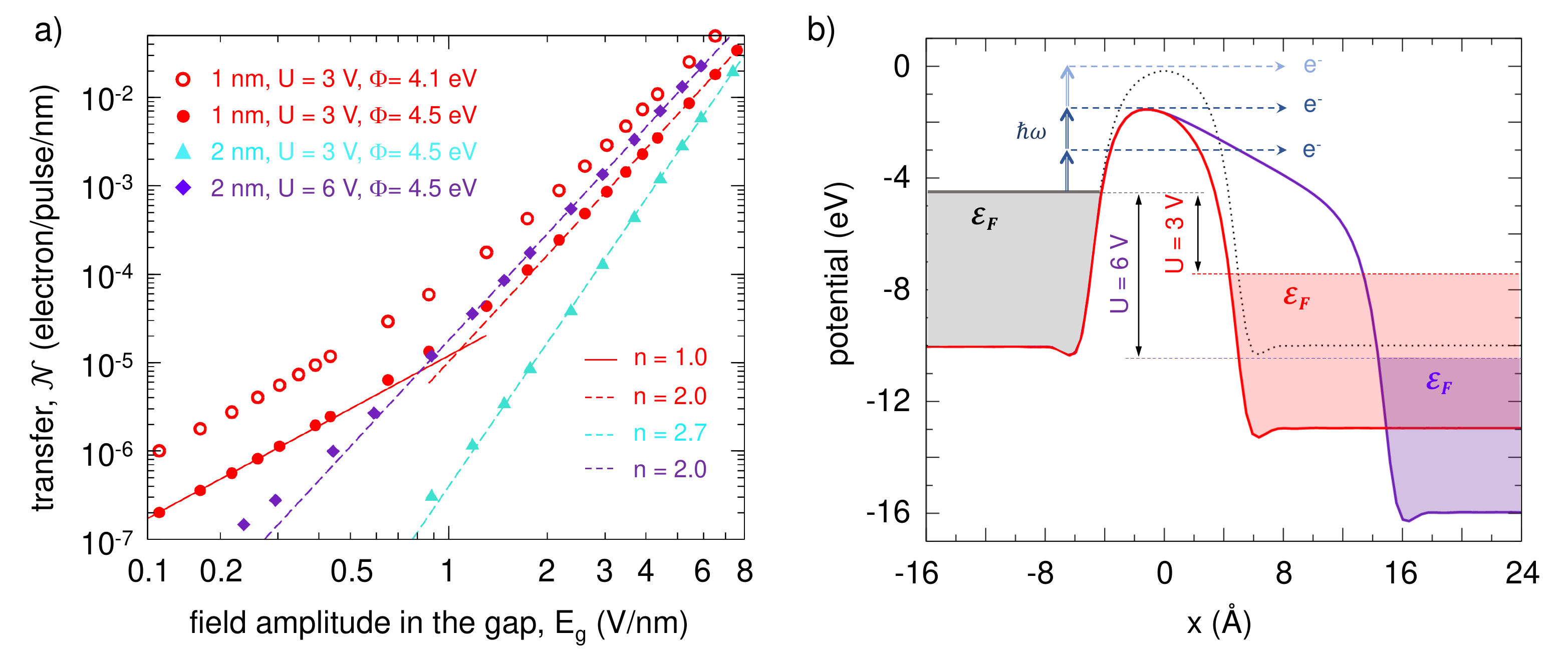}
\caption{
%In the TDDFT calculations 
%we use $\Phi=4.5$~eV, $\Phi=4.1$~eV resulting from the polarisation correction and aimed 
%to reproduce the 2 nm junction between metals with work functions 
%$\Phi_{\rm{M}}=5.3$~eV, $\Phi_{\rm{M}}=4.9$~eV.
%
Effect of the work function $\Phi$, gap size $d_{\rm{gap}}$, and applied bias 
$U$ on the electron transfer induced by a 7~fs long $x$-polarized optical pulse 
($\tau=4.2$~fs, $\omega=1.6$~eV in Eq.~\eqref{eq:Gaussianfield}) across the gap 
of a nanowire dimer.
\textbf{Panel a:} Symbols show the net optically induced electron transfer 
$\mathcal{N}$ calculated with TDDFT per pulse and per nm length of the   
dimer as a function of the field in the gap $E_{\rm{g}}$. Calculations are 
performed for different values of $\Phi$, $d_{\rm{gap}}$, and $U$ as explained 
in the legend. 
The lines display the fit by the $\mathcal{N} \propto E_{\rm{g}}^{2n}$ 
dependence. The effective number of photons $n$ is given in the legend.  
\textbf{Panel b:} One-electron potential in the gap region as a function of 
$x$-coordinate along the dimer axis. The energy is measured with respect to 
the vacuum level of the left cylinder. 
Black dotted line:  $d_{\rm{gap}}=1$~nm, $U=0$~V; 
red line: $d_{\rm{gap}}=1$~nm, $U=3$~V; 
violet line: $d_{\rm{gap}}=2$~nm, $U=6$~V. 
The applied bias is measured between the Fermi levels $\mathcal{E}_F$ of the 
metal nanowires. The hatched areas indicate occupied electronic states of the 
valence band.
The dashed horizontal arrows indicate electron tunneling and over-the-barrier 
transitions induced by $\ell$-photon absorption (vertical solid arrows). 
Light color is used for the $\ell=3$-photon absorption with electron 
energy close to the top of the potential barrier dominating 
transport in the $U=0$ case.
}
\label{fig:Garg_PowerCurve_d_WorkFunction_Bias}
\end{figure*}

In line with experimental findings \cite{Luo2024}, for an applied
bias of $U=3$~V we find that: 
\begin{enumerate}
  \item The optically assisted electron tunneling with $\ell=1$ photon absorption 
at low $E_{\rm{g}}$ evolves into a multiphoton regime with effective photon 
order $n=2$ as $E_{\rm{g}}$ increases; 
  \item No higher-order nonlinearity $n>2$ is observed over a wide range of optical 
  field strengths in the gap, up to the onset of the optical field emission regime,  
  which occurs at $E_{\rm{g}} \approx 12$~V/nm (not shown).
\end{enumerate}
%
%The optical field emission regime corresponds to well-documented reduction of $n$ 
%\cite{Bormann2010,dombi2010observation,piglosiewicz2014,Zimmermann2019}. 
According to the SFT, TDDFT and model WPP 
calculations performed in section~\ref{sec:SingleCyclePulse} 
(see Figure~\ref{fig:Figure_WPP_w0.95}), the second finding implies that while 
several $\ell$-channels (photon absorption orders) contribute to electron 
transport in the multiphoton regime, the overall net electron transfer follows 
a scaling $\mathcal{N} \propto E_{\rm{g}}^{2n}$ with $n=2$. This effective 
number of photons $n$ is given by the threshold for over-the-barrier transitions 
$n \approx \mathcal{W}_{\rm{tb}}/ \omega$. It remains constant with 
increasing $E_{\rm{g}}$ up to the onset of the optical field emission regime, 
in agreement with Eqs.~\eqref{eq:MPE},\eqref{eq:MPE1}.

For an applied bias $U=3$~V, a clear change of the effective photon 
order $n$ from the one-photon to the multiphoton regime occurs in the TDDFT 
calculations for $E_{\rm{g}} \approx 1$~V/nm. 
In the experiment \cite{Luo2024}, under the same bias, this transition 
is observed for a free-space electric field of $\approx 0.09$~V/nm. 
The theory is consistent with experiment when considering that the field 
enhancement due to the plasmon excitation in the gap of the actual 
device is of the order of $\mathcal{R}=10$. 
This enhancement is smaller than that predicted by classical electrodynamics 
simulations \cite{Luo2024}. However, for 1~nm gaps, quantum effects such as 
nonlocality become significant \cite{zhu16,Monticone25}, so that classical 
results should be taken with caution.

In the experiment, the transition between the two regimes corresponds to a 
current of $20-30$~pA, i.e., to approximately two electrons transferred per 
pulse. Considering that the experimental device contains a parallel circuit 
of seven bowtie antennas, each 30~nm in height, this translates to about 
$10^{-2}$ electrons transferred per pulse and per nanometer of antenna height, 
approximately 3 orders of magnitude higher than the TDDFT prediction. One of 
the reasons behind this discrepancy is that, because of plasmon ringing, the 
effective duration of the optical pulse in the gap of the experimental 
antenna is about twice as long as that in free space \cite{Luo2024}. This 
effect is not captured in the TDDFT simulations because of the use of a 
free-electron metal model and a simplified geometry with the plasmon mode off 
resonance with the optical frequency, as discussed in Section~\ref{sec:Model}. 
An approximate account for plasmon ringing through scaling the calculated net 
electron transfer $\mathcal{N}$ by a factor of 2 does not change the 
substantial difference between the experimental and our theoretical results.

Along with the pulse duration, the discrepancy between theory and 
experiment may also stem from differences in the optical transitions 
of the free-electron metal used in the present TDDFT calculations 
and of the actual gold studied experimentally. In addition to possible 
band-structure effects, there is also uncertainty in the geometry of 
the actual experimental device, particularly in the gap size,  
as acknowledged by Luo \textit{et al.} \cite{Luo2024}, as well as 
in the work function of the nanoantennas. These last two factors 
strongly affects the electron tunneling and, consequently, the 
net electron transfer.

In Figure~\ref{fig:Garg_PowerCurve_d_WorkFunction_Bias}, we explicitly 
address the effect of gap size and metal work function on 
electron transport in the system. One can use table~\ref{table1} to relate 
the work function $\Phi$ used in the ALDA TDDFT and the actual work function 
$\Phi_{\rm{M}}$ of the metals across the gap that the theory is designed to 
represent. When the gap size changes from 1~nm to 2~nm at fixed $U$ and 
fixed $\Phi$, the tunneling barrier becomes broader and higher. 
Consequently, the effective photon order $n$ increases, and the net electron 
transfer $\mathcal{N}$ drops off by several orders of magnitude, 
in particular at low $E_{\rm{g}}$ (compare filled red circles and light blue 
triangles in Figure~\ref{fig:Garg_PowerCurve_d_WorkFunction_Bias}a). 
The drop-off of the tunneling probability is most pronounced for electrons 
with high binding energies, rendering the $\ell=1$ PAT channel undetectable 
within the computed $\mathcal{N}$ range.

As an interesting observation, if a bias $U=6$~V is applied across the 2~nm gap, 
the dc field in the gap is identical to that in the $1$~nm gap under an applied 
bias of $U=3$~V. The tunneling barrier then has the same height for both gap 
sizes. However, for electrons with $\mathcal{E}_{\rm{F}}+ \omega$ energies, the 
width of the tunneling barrier is larger for $d_{\rm{gap}}=2$~nm 
(see Figure~\ref{fig:Garg_PowerCurve_d_WorkFunction_Bias}b). As a result, while 
in the multiphoton regime the effective $n=2$ is the same for both cases, the 
one-photon PAT channel at low $E_{\rm{g}}$ is observed clearly only for 
$d_{\rm{gap}}=1$~nm, while for the 2~nm gap it becomes marginally 
distinguishable at very low $\mathcal{N}$ (compare violet diamonds and filled 
red circles in Figure~\ref{fig:Garg_PowerCurve_d_WorkFunction_Bias}a).

%
% FIGURE 11 Garg dependence on applied bias for fixed Eg Also Semiclassical
%

\begin{figure}[t!]
\centering
\includegraphics[width=0.95\linewidth]{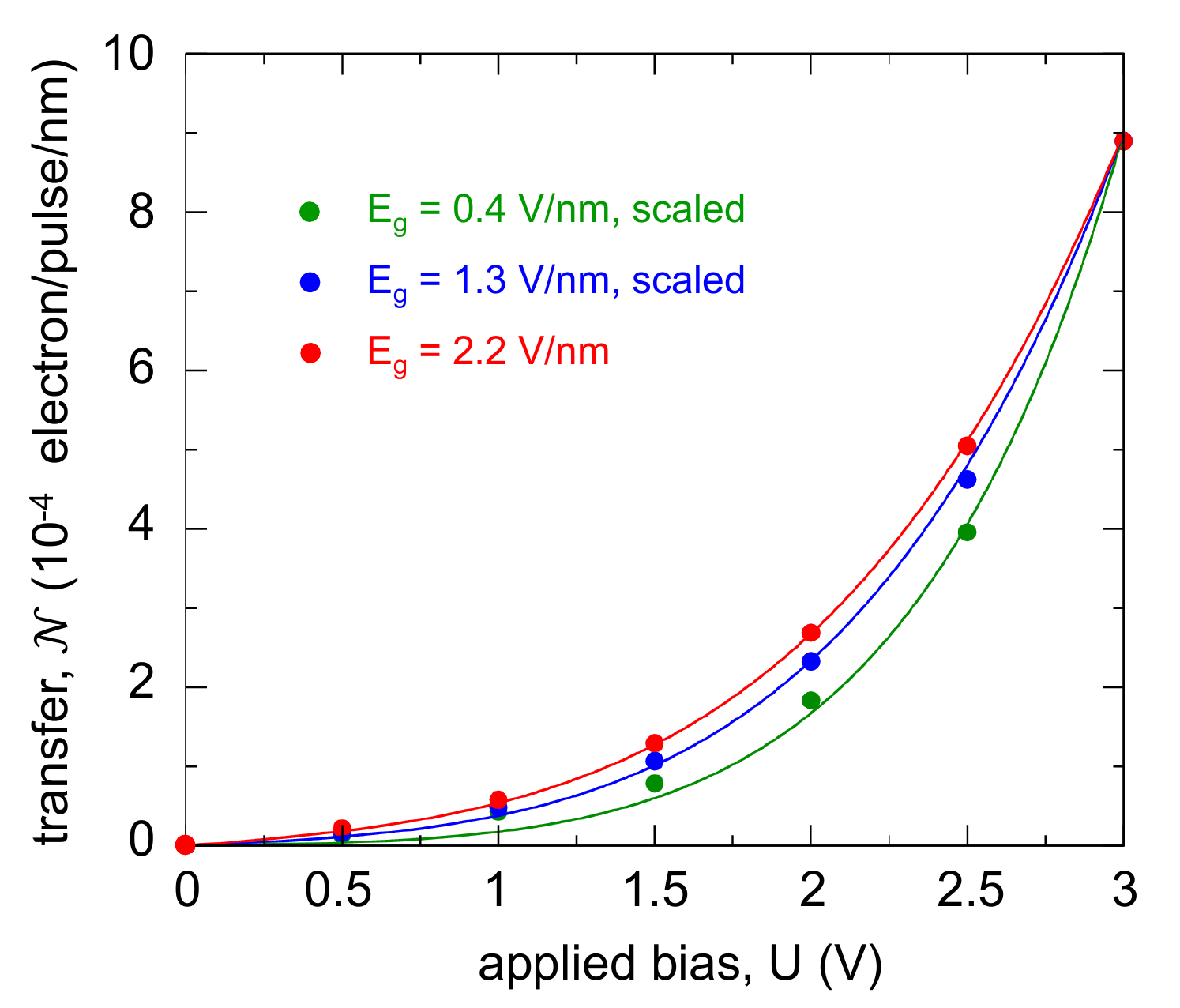}
\caption{
%Single-cycle optical pulse. TDDFT transfer across 
%the 1 nm gap. The effect of an applied bias. 
%In the TDDFT calculations 
%we use $\Phi=4.1$~eV resulting from the polarisation correction and aimed 
%to reproduce the 1 nm junction between metals with work functions 
%$\Phi_{\rm{M}}=4.9$~eV. 
Effect of an applied bias on optically induced electron transfer 
across a 1~nm gap between metal cylinders with work function 
$\Phi_{\rm{M}}=4.9$~eV (see also open red circles 
in Figure~\ref{fig:Garg_PowerCurve_d_WorkFunction_Bias}). 
Results are shown for three values of the optical field in the 
gap $0.4,~1.3, ~\text{and}~2.2$~V/nm as displayed in the legend. 
Dots: net optically induced electron transfer $\mathcal{N}$ calculated 
with TDDFT per pulse and per nm length of the dimer as a function 
of an applied bias $U$. The $x$-polarized optical pulse has 7~fs 
duration and $\omega=1.6$~eV. The results for $E_{\rm{g}}=0.4$~V/nm 
and $E_{\rm{g}}=1.3$~V/nm are scaled to match $\mathcal{N}$ calculated 
for $E_{\rm{g}}=2.2$~V/nm, and $U=3$~V.
Solid lines of the corresponding color: results obtained within the 
semi-classical SFT framework and scaled to match the $\mathcal{N}$ 
calculated with TDDFT for $E_{\rm{g}}=2.2$~V/nm, and $U=3$~V.  
See the text for further details. 
}
\label{fig:Garg_FixEg_BiasDependence}
\end{figure}

Similarly to the pronounced effect of changing the gap size, a reduction 
of the work function from $\Phi=4.5$~eV to $\Phi=4.1$~eV leads to an order 
of magnitude increase in the net electron transfer (compare filled and open 
red circles in Figure~\ref{fig:Garg_PowerCurve_d_WorkFunction_Bias}a). This 
brings our theoretical results closer to the experiment \cite{Luo2024}. Given 
experimental uncertainties, in particular concerning essential parameters such 
as gap geometry and work function, we consider that the qualitative agreement 
between theory and experiment is sound. 
It is worth noting that since in the multiphoton regime the nonlinearity 
of the $\mathcal{N}(E_{\rm{g}})$ dependence is the same in theory and 
experiment ($n=2$), the height of the tunneling barrier should be 
captured reasonably well in our calculations.

We close our comparison with the experiment by 
Luo \textit{et al.} \cite{Luo2024} by showing in 
Figure~\ref{fig:Garg_FixEg_BiasDependence} the calculated dependence 
of the net electron transfer, $\mathcal{N}$, on the applied bias, $U$, 
for three representative values of the amplitude of the optical field 
in the gap: $E_{\rm{g}}=0.4$~V/nm corresponding to the one-photon 
absorption regime of electron transfer, $E_{\rm{g}}=2.2$~V/nm 
corresponding to the multiphoton regime, and $E_{\rm{g}}=1.3$~V/nm 
near the transition between these two regimes. 
For $E_{\rm{g}}=0.4$~V/nm and $E_{\rm{g}}=1.3$~V/nm the results are 
normalized  by a scaling factor 
$X=\mathcal{N}_{E_{\rm{g}}}(U)/\mathcal{N}_{E_{\rm{ref}}}(U)$, 
where $E_{\rm{ref}}=2.2$~V/nm and $U=3$~V. We find that the overall 
shape of the $\mathcal{N}(U)$ dependence is only weakly sensitive to 
$E_{\rm{g}}$, and it closely resembles the experimental data 
(see Figure~4a in Ref.~\cite{Luo2024}).

This behavior can be explained using the SFT framework. Since the 
field amplitudes $E_{\rm{g}}$ considered here roughly satisfy $\zeta>1$ and 
$\gamma\gg1$, the multiphoton photoemission described by Eq.~\eqref{eq:MPE}, 
Eq.~\eqref{eq:MPE1}, and in a more precise form by Eq.~\eqref{eqA20}, becomes 
applicable. Taking into account the symmetry of the system the net transfer 
can be analytically expressed as:
\begin{align}\label{eq:netcurrent}
 \mathcal{N}(U)&= 
 \mathcal{P}_{\rm{L\rightarrow R}}(U)-\mathcal{P}_{\rm{R\rightarrow L}}(U) 
 \nonumber \\
&= \mathcal{P}_{\rm{L\rightarrow R}}(U)-\mathcal{P}_{\rm{L\rightarrow R}}(-U),
\end{align}
where $\mathcal{P}_{\rm{L\rightarrow R}}(U)$ can be obtained from 
Eq.~\eqref{eqA20} with $E_{\rm{bias}}=U/d_{\rm{gap}}$, and using 
$\mathcal{W}_{\rm{tb}} \approx \Phi$. Importantly, considering the leading 
power-law dependence of the transition probability on the optical field 
(Eq.~\eqref{eq:MPE1}), the photon number $n$ is independent of $E_{\rm{g}}$, 
which explains the weak sensitivity of the shape of $\mathcal{N}(U)$ to 
variations in $E_{\rm{g}}$. For comparison, we normalize in 
Figure~\ref{fig:Garg_FixEg_BiasDependence} the SFT results such that for 
$U=3$~V all the results coincide with the value of $\mathcal{N}$ calculated 
with TDDFT for  $E_{\rm{g}}=2.2$~V/nm.

%
% FIGURE 12
%

%\onecolumngrid

%\vspace{5mm}

%\begin{center}
\begin{figure*}[t!]
%\begin{figure}[h]
\includegraphics[width=1.0\linewidth]{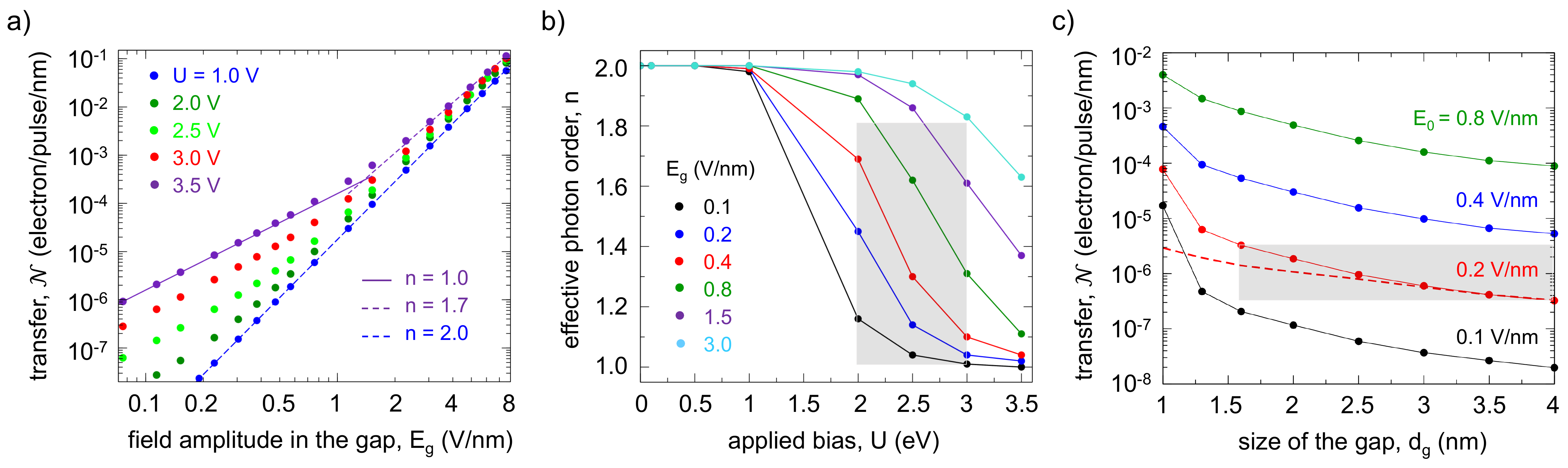}
\caption{
%TDDFT, $\Phi=3.9$~eV, $\Phi_{\rm{M}}=4.5$~eV, 
%gap = 1.6 nm, Field enhancement for $d_{\rm{gap}}=1.6$~nm $\mathcal{R}=3.0$,  
%$\omega=2.33$~eV, 6~fs long laser pulse. Panel b data obtained for the 
%0.5 V bias. Panel c, dashed line, fit by the $\mathcal{R}^{4}$ dependence.
%fit by the $e^{-0.7 d_{\rm{gap}}}$ dependence.
Effect of an applied bias $U$ and gap size $d_{\rm{gap}}$ on optically 
induced electron transfer across the gap of the nanowire dimer. The several-cycle 
$x$-polarized optical pulse is 6~fs long ($\tau=3.6$~fs, and $\omega=2.33$~eV 
in Eq.~\eqref{eq:Gaussianfield}). The TDDFT calculations are performed with 
CEP$=\pi$. Notice however, that for this several-cycle pulse, results are 
independent of CEP. In all the panels, results are shown per pulse and per nm 
length of the dimer.
\textbf{Panel a:} Gap size $d_{\rm{gap}}=1.6$~nm. 
Dots: optically induced net electron transfer $\mathcal{N}$ calculated with 
TDDFT as a function of the field in the gap, $E_{\rm{g}}$, for different 
values of the applied dc bias $U$. 
Lines: fit by the $\mathcal{N} \propto E_{\rm{g}}^{2n}$ dependence, 
where $n$ is the effective number of absorbed photons. For further details 
see the legend. 
\textbf{Panel b:} Gap size $d_{\rm{gap}}=1.6$~nm. 
Lines with dots: effective local photon order (effective number of photons 
driving the transport for a given $E_{\rm{g}}$) defined from the 
$\mathcal{N}(E_{\rm{g}})$ calculated with TDDFT as  
$n= \frac{E_{\rm{g}}}{2\mathcal{N}(E_{\rm{g}})} 
~\frac{\partial \mathcal{N}(E_{\rm{g}})}{\partial E_{\rm{g}}}$. Results are 
shown as a function of the applied dc bias, $U$, for 
different values of the optical field amplitude in the gap, $E_{\rm{g}}$,   
as indicated in the legend. 
The grey shaded region indicates the typical variation range of $\mathcal{N}$ 
with $U$ obtained experimentally \cite{Light_STM_Kumagai23}. 
\textbf{Panel c:} Applied dc bias $U=0.5$~V. 
Lines with dots: optically induced net electron transfer $\mathcal{N}$ 
calculated with TDDFT as a function of gap size, $d_{\rm{gap}}$, 
for different values of the amplitude of the incident optical field, $E_0$,  
as indicated in the legend. The amplitude of the optical field in the gap  
is $E_{\rm{g}}=\mathcal{R} E_0$, where the field enhancement $\mathcal{R}$ 
depends on $d_{\rm{gap}}$ (e.g. $\mathcal{R}=3.0$ for $d_{\rm{gap}}=1.6$~nm).  
Dashed red line: fit by the $\mathcal{R}^{4}$ dependence expected for 
electron transport driven by two-photon absorption. The grey shaded region 
indicates the typical variation range of $\mathcal{N}$ with 
$d_{\rm{gap}}$ obtained experimentally \cite{Light_STM_Kumagai23}.
}
\label{fig:Ag_Wolf_Muller}
\end{figure*}
%\end{figure}
%\end{center}

%\twocolumngrid

%%%%%%%%%%%%%%%%%%%%%%%%%%%%%%%%%%%%%%%%%%%%%%%%%%%%%%%
\subsection{Tunneling junction between silver surfaces}
\label{subsec:STMSilver}

Finally, we apply our approach to study electron transport induced 
by a several-cycle optical pulse with carrier frequency  $\omega=2.33$~eV 
and a duration of 6~fs. The sufficiently long pulse duration ensures that 
the net electron transfer $\mathcal{N}$ is independent of the CEP, 
similarly to the behavior discussed in subsection~\ref{subsec:bowtite}. 
We use a typical work function of silver $\Phi_{\rm{M}}=4.5$~eV 
\cite{chulkov1999image} with the aim of comparing the present results with 
recent cw experiments performed by Lin and 
collaborators \cite{Light_STM_Kumagai23}, who studied multiphoton 
photocurrents in a STM junction between a silver tip and an Ag(111) 
surface. Similar experiments involving gold tip and Cu(100) surface have 
been conducted by Schr\"{o}der \textit{et al.} \cite{STM_Light_Schroder_2020}. 
The authors of Ref.~\cite{Light_STM_Kumagai23} argued that the band structure 
of Ag(111) has no effect on the lightwave-induced electron transport, which 
supports the applicability of the free-electron model in this case. We further 
comment on the validity and limitations of this approximation below.
It is worth noting that the STM junction between the tip and the surface 
is non-symmetric. However, an applied bias used in experiments  
overrides the effect of the symmetry break because of the geometry 
\cite{Light_STM_Kumagai23}. This validates the symmetric junction geometry 
used in our calculations, provided a dc bias is applied as the main source 
of asymmetry.

The different panels in Figure~\ref{fig:Ag_Wolf_Muller} show the net electron 
transfer $\mathcal{N}$ calculated as a function of optical field amplitude 
in the gap, $E_{\rm{g}}$, applied bias, $U$, and gap size, $d_{\rm{gap}}$. 
To facilitate comparison with experiment, we adopt the form 
of the data analysis used in the experimental work \cite{Light_STM_Kumagai23}. 
The dependence of $\mathcal{N}$ on $E_{\rm{g}}$ for several values of $U$ 
(Figure~\ref{fig:Ag_Wolf_Muller}a), and the dependence of the effective photon 
order $n$ on $U$ for several values of $E_{\rm{g}}$ 
(Figure~\ref{fig:Ag_Wolf_Muller}b) are calculated for a fixed size of the gap 
$d_{\rm{gap}}=1.6$~nm, consistent with the experimental 
configuration \cite{Light_STM_Kumagai23}.

The theoretical results reported in Figure~\ref{fig:Ag_Wolf_Muller}a are in 
qualitative agreement with experiment and can be interpreted using the 
general trends of optically assisted electron transfer discussed above and 
in Ref.~\cite{ma2025robust}. Thus, because of the relatively 
large gap width, tunneling is suppressed at low bias, and electron 
transport is dominated by electrons excited above the barrier. The effective 
number of absorbed photons, $n=2$, is therefore set by the energy 
threshold for the multiphoton process, $n=\Phi/\omega$. Increasing $U$ 
lowers the potential barrier, leading to the progressive emergence of the 
$\ell=1$ one-photon assisted tunneling channel for small 
$E_{\rm{g}}\lesssim 1.5$~V/nm. In this range of $E_{\rm{g}}$, the interplay 
between the $\ell = 1$ and higher-order transport channels manifests itself 
in a gradual decrease of the observed nonlinearity $n$ with increasing 
bias. Ultimately, $n = 1$ is reached for $U = 3$~V and for $U = 3.5$~V. 
Above $E_{\rm{g}} \approx 1.5$~V/nm, the multiphoton regime persists 
due to the significant contribution of $\ell > 1$ channels, albeit the 
effective photon order $n$ is somewhat reduced for large $U$.

The effective local (with respect to $E_{\rm{g}}$) photon order $n$ is 
shown in Figure~\ref{fig:Ag_Wolf_Muller}b as a function of the applied 
bias for several $E_{\rm{g}}$. The field amplitudes $E_{\rm{g}}$ used here 
for the TDDFT calculations match the field amplitudes in the junction 
within the $0.2-0.9$~V/nm range as we obtained from the near-field intensity 
estimations performed by Lin and collaborators \cite{Light_STM_Kumagai23}, 
who accounted for the plasmon enhancement in the system. The effective 
local photon order $n$ is obtained via numerical differentiation of the 
$\mathcal{N}(E_{\rm{g}})$ curve at fixed bias $U$ according to 
$n= \frac{E_{\rm{g}}}{2\mathcal{N}(E_{\rm{g}})} 
~\frac{\partial \mathcal{N}(E_{\rm{g}})}{\partial E_{\rm{g}}}$. 
The results in Figure~\ref{fig:Ag_Wolf_Muller}b reflect the transition from 
the multiphoton-dominated transport to single-photon assisted tunneling 
($\ell=1$) due to the reduction of the tunneling barrier by 
the applied bias. This transition is more abrupt for low $E_{\rm{g}}$ since 
weak optical fields favor lower photon-order processes (recall the 
$E_{\rm{g}}^{2\ell}$ scaling of different $\ell$-channels at low optical 
fields). Overall, we find that the present calculations reproduce 
the typical variation range of $\mathcal{N}$ with $U$ reported experimentally 
\cite{Light_STM_Kumagai23} and indicated with the grey shaded region.

In Figure~\ref{fig:Ag_Wolf_Muller}c, we study the evolution of the 
lightwave-induced electron transport with gap size $d_{\rm{gap}}$. For the 
sake of comparison with the experimental data \cite{Light_STM_Kumagai23}, 
we show the results as a function of the free-space field amplitude $E_0$. 
For the same reason, we performed the TDDFT calculations using relatively 
low applied bias ($U=0.5$~eV) such that the tunneling barrier is only 
weakly affected by the dc field. Under these conditions, the evolution of 
the dc tunneling current with junction size can be estimated as 
$\propto \exp\left(-2\sqrt{2\Phi_{\rm{M}}-U}\ d_{\rm{gap}} \right)$.
According to this estimate, the dc tunneling current decreases by an order 
of magnitude when $d_{\rm{gap}}$ increases by 1~\AA~ (0.1~nm). Therefore, 
within the $d_{\rm{gap}}$ range shown in Figure~\ref{fig:Ag_Wolf_Muller}c, 
the lightwave-induced current dominates electron transport in the 
system \cite{STM_Light_Schroder_2020,Light_STM_Kumagai23}.
Because of the reduction of the tunneling barrier for optically excited 
electrons, the contribution of the $\ell=1$ PAT channel associated with 
one-photon absorption decays more slowly with increasing junction size. 
It follows the $\propto \exp\left(-2\sqrt{2 \left( \Phi_{\rm{M}}-
\omega \right)-U}\ d_{\rm{gap}} \right)$ dependence, which is indeed 
observed in Figure~\ref{fig:Ag_Wolf_Muller}c, for the lowest $E_0$ and 
for $1$~nm~$\leq d_{\rm{gap}} \leq 1.25$~nm. This trend is consistent 
with the dominance of one-photon-assisted electron tunneling as also 
evidenced by the log-log plot of $\mathcal{N}(E_0)$ calculated with 
TDDFT (not shown).

For large gap sizes $d_{\rm{gap}} \gtrsim 2$~nm, tunneling is 
negligible. As follows from the results shown in 
Figure~\ref{fig:Ag_Wolf_Muller}a, optically induced electron transport 
is dominated by over-the-barrier transitions via multiphoton absorption. 
The effective photon order $n=2$ is determined by the photoemission 
threshold. The calculated net electron transport is thus described by 
the $\mathcal{N}(d_{\rm{gap}}) \propto E_{\rm{gap}}^4$ dependence. 
Since $E_{\rm{gap}} = \mathcal{R} E_0$, for a fixed $E_0$ we obtain that 
$\mathcal{N}(d_{\rm{gap}})$ follows the variation of the local field 
enhancement in the gap 
$\mathcal{N}(d_{\rm{gap}}) \propto \left[\mathcal{R}(d_{\rm{gap}})\right]^4$ 
(see red dashed line in Figure~\ref{fig:Ag_Wolf_Muller}c). Interestingly, 
the TDDFT results reproduce the variation of $\mathcal{N}$ by an 
order of magnitude measured experimentally by Lin and collaborators 
\cite{Light_STM_Kumagai23} as the gap size increases from about $1.5$~nm 
to $4$~nm (highlighted by the gray rectangle).

All in all, we have found that the TDDFT calculations performed in our 
model system are capable of reproducing the major trends observed 
experimentally for optically induced electron transport. These include 
the data obtained with $7$~fs laser pulses used to drive electron transfer 
in the gap of a gold bowtie antenna \cite{Luo2024}, and the data obtained 
with cw illumination of a silver-based STM 
junction \cite{Light_STM_Kumagai23}. Importantly, our theoretical TDDFT 
approach is parameter-free. The size of the junction (the key geometric 
parameter) has been set equal to that reported in the experiments. Our 
TDDFT study describes the ``fast" \cite{Light_STM_Muller22} coherent 
component of the optically induced transport. It corresponds to the 
one-step process driven by the field of an optical pulse and it cannot 
be separated in time into excitation and tunneling (or over-the-barrier 
transfer) events. This said, electron relaxation leading, e.g., to thermionic 
emission \cite{Light_STM_Muller22}, cannot be captured within the ALDA-TDDFT 
scheme \cite{ullrich2019time,PhysRevLett.95.086401,PhysRevLett.77.2037}. 
Further theoretical work that explicitly incorporates excitation and 
relaxation events \cite{TagliabueGiulia2018,Dubi2019,Schirato2023,Khurgin2024,
TagliabueGiulia2024,Stefancu2024,Voisin2001,Grua2003,Besteiro2017}, as well 
as experimental investigations are then needed in order to provide a more 
accurate picture of the process.

One remark is in order regarding possible effects of the band structure 
of the substrate. While d-electrons in noble metals have binding energies  
several eV below the Fermi level, so that electron transport is dominated 
by the valence electrons, the projected band structure of the surface may 
affect electron transport by changing the surface reflectivity. 
Indeed, the Ag(111) surface used in Ref.~\cite{Light_STM_Kumagai23}, and the 
Cu(100) surface used in Ref.~\cite{STM_Light_Schroder_2020,Light_STM_Kumagai23}
are known to possess a projected band gap in the direction perpendicular 
to the surface. This gap extends to low binding energies in the case 
of Ag(111), or even above the vacuum level for Cu(100)  
\cite{chulkov1999image}. The high reflectivity of the surface for  
electrons with energies inside the gap is behind the observation of  
field-emission resonances in STM \cite{PhysRevLett.55.991,PhysRevB.75.165326,
PhysRevB.76.195404,PhysRevLett.121.226802}. While for Ag(111), at high applied 
bias, the transferred electron energies are above the band gap, this is not the 
case for $U<1.6$~eV. A treatment of band structure effects is beyond the 
scope of the present work and is noted for future investigation.

%\textcolor{red}{
%$1$~pA $= 6.28 \times 10^6$~electrons/s \\
%$80$~MHz $= 80 \times 10^6$~pulse/s \\
%$1$~pA $= 6.28/80=0.08$~electrons/pulse \\
%Since in experiment we have 7 nanontennas of the height of 30~nm this gives\\
%$1$~pA $\rightarrow  0.0004$~electrons/pulse/nm/antenna \\
%Plasmon ringing effectively doubles the duration \\
%We end up with $\boxed{\rightarrow  0.0002~\text{electrons/pulse/nm/antenna}}$
%}

\section{Summary and conclusions and outlook}

In conclusion, we have theoretically investigated optically induced electron 
transport in metallic gaps ranging from sub-nanometer to several nanometers, 
with a particular focus on photon-assisted electron tunneling. 
Our work is relevant to STM junctions, optical nanoantennas, and other tunneling 
devices, as demonstrated by the comparison between our theoretical results 
and recent experimental observations.

As a representative system we considered a junction formed by a dimer of 
parallel free-electron metal nanowires exposed to few-cycle and several-cycle 
laser pulses. The electron dynamics following optical excitation has been 
described using many-body time-dependent density functional theory (TDDFT). 
Along with TDDFT we also applied model approaches based on direct numerical 
and semiclassical strong-field theory solutions of the one-dimensional, 
one-electron time-dependent Schr\"{o}dinger equation. 
These model approaches, and in particular the analytical semiclassical theory, 
are found to be extremely useful in providing guidelines for understanding 
the main trends calculated with TDDFT.

The conditions for photon-assisted electron tunneling are achieved by 
considering a sub-nm gap (here 0.8~nm), or considering a situation where 
an external dc bias is applied across wider (1~nm and 2~nm) gaps. 
The dc bias lowers the potential barrier between metals and enables 
optically assisted electron tunneling, as exploited in recent experiments. 

%Both cases
%are in line with the strong-field theory model which predicts that
%photon-assisted electron tunneling occurs for parameters
%corresponding to $\zeta > 1$.}

In line with earlier findings, we confirm that an optical wave drives 
electron transport even in a more direct manner than merely producing 
electron excitation. The very electron ejection from the metal is locked to 
the waveform of the optical field. Electrons are launched at the 
metal/vacuum interface during the half-cycles of the optical field 
with the corresponding polarity.

While for single-cycle optical pulses this opens a possibility of CEP 
control of the net electron transfer, $\mathcal{N}$, this control is only 
effective in the multiphoton regime, where the photon absorption orders  
$\ell$ driving electron transport satisfy $\ell \geq 2$. 
As we have demonstrated, in the single-photon regime of photon-assisted 
tunneling ($\ell=1$), the CEP sensitivity of $\mathcal{N}$ vanishes. 
As a consequence, in a symmetric gap, the single-photon channel does not 
contribute to the net electron transfer across the gap, and $\ell=2$ is 
the lowest photon order observed in the $\mathcal{N}(E_{\rm{g}})$ 
dependence at low fields in the gap $E_{\rm{g}}$. 
In this situation, applying a dc bias breaks the symmetry 
of the system and allows the $\ell=1$ channel of the photon-assisted 
tunneling to be observed provided the tunneling barrier is sufficiently 
low. In turn, this barrier can be controlled by an applied bias $U$ and 
the gap size $d_{\rm{gap}}$.

We have demonstrated that at low optical field strengths in the gap, in the 
perturbative regime, the optically induced net electron transfer scales as  
$\mathcal{N} \propto E_{\rm{g}}^{2\ell_{\rm{min}}}$. This dependence reflects the 
lowest symmetry-allowed channel of photon-assisted tunneling (i.e., the 
lowest photon order). This lowest photon order is $\ell_{\rm{min}}=1$ for an 
asymmetric gap and $\ell_{\rm{min}}=2$ for a symmetric gap.

With increasing $E_{\rm{g}}$, the perturbative regime of optically induced 
electron transport evolves into the multiphoton regime with 
a $\mathcal{N} \propto E_{\rm{g}}^{2n}$ dependence of the net electron 
transfer on the optical field strength. The \emph{effective} photon order 
$n$ reflects an average value over all active transport channels (tunneling 
and over-the-barrier) with $\ell \geq \ell_{\rm{min}}$ photon absorption. 
It follows from the TDDFT calculations, that the transition between the 
perturbative and multiphoton regimes is relatively prompt, with 
$n$ converging to the threshold of photo-emission above the tunneling 
barrier $n \approx \mathcal{W}_{\rm{tb}} / \omega$. This finding is in 
accord with semiclassical theory.

Using several-cycle optical pulses, representative of the experimental 
conditions, we qualitatively (and in some cases semi-quantitatively) 
reproduce recent experimental data that dissect different photon channels 
in optically assisted tunneling. Because of the relatively long pulse 
duration, the CEP dependence is suppressed in this case. 
The contribution of different $\ell$-photon channels of electron transport 
is controlled by the applied bias $U$ and the gap size $d_{\rm{gap}}$, 
which we set according to experiment. No fitting parameters are used, 
which strengthens the validity of approach and the mechanisms it 
uncovers. Furthermore, our results confirm that because of the 
localization of electron transfer within the junction region, the nanowire 
dimer model system is suitable for addressing photo-assisted tunneling 
in narrow gaps. This said, TDDFT captures the ``fast" component 
of the photoemission, while quantifying the role of the relaxation of 
photoexcited electrons, as well as the ``slow" thermionic part of 
electron emission \cite{Light_STM_Muller22}, requires further 
theoretical work.

As possible future developments, we also foresee studying the role of the 
band structure of the substrate and addressing electron 
tunneling induced by short optical pulses with controlled frequency chirp. 
%where the photon energy associated with different temporal portions of 
%the pulse varies with time.}

We believe that the results presented in this work contribute to a deeper 
understanding and to the discussion of the physical mechanisms at play 
in electron transport induced in metallic junctions by optical 
pulses of single-cycle and longer duration. The processes and their control 
strategies, as studied here, are relevant for petahertz optoelectronic 
devices, as well as for scanning probe techniques targeting ultimate spatial 
and temporal resolution.

\hspace{5 mm}

\begin{acknowledgments}
A.G.B gratefully acknowledges the warm hospitality of DIPC. 
B.M.~and M.K.~acknowledge funding from the European Union's Horizon 2020 research 
and innovation program under grant agreement No 853393-ERC-ATTIDA and from the 
Israel Science Foundation (ISF) under grant 1504/20, as well as partial financial 
support from the Helen Diller Quantum Center at the Technion. 
J. A. and A. B acknowledge financial support from grant PID2022-139579NB-I00 funded 
by MICIU/AEI/10.13039/501100011033 and by ERDF/EU, and from grant IT 1526-22 funded 
by the Department of Science, Universities and Innovation of the Basque Government.
\end{acknowledgments}

\vspace{10mm}

\begin{center} 
\textbf{DATA AVAILABILITY}
\end{center}
The data that support the findings of this article are not publicly 
available upon publication because it is not technically feasible 
and/or the cost of preparing, depositing, and hosting the data would 
be prohibitive within the terms of this research project. The data 
are available from the authors upon reasonable request.

\appendix

\section{TDDFT calculations with an applied dc bias }\label{Apd.THZ}

As we discussed in subsection~\ref{subsec:TDDFT}, in order to study the 
role of an applied dc bias the dimer is subjected to slowly varying 
THz field. The corresponding potential, introduced into the time-dependent 
KS equations of the TDDFT is given by 
\begin{equation} \label{eq:THz}
V_{\rm{THz}}(x,t)=-x~\frac{\tilde{U}}{d_{\rm{gap}}}~ \mathcal{U}(t-t_-).
\end{equation}
The switching function $\mathcal{U}$ is defined as
\begin{equation} \label{eq:THz1}
\mathcal{U}(\xi) = 
\begin{cases}
  0, &\xi < -\tau_{\rm{THz}}, \\
  \sin^2\left( \frac{\pi}{2}~\frac{\xi+\tau_{\rm{THz}}}{\tau_{\rm{THz}}} \right),
  & -\tau_{\rm{THz}} \leq \xi \leq 0,\\
    1, &0 < \xi. 
  \end{cases}
\end{equation}
Typically, we choose a switching time $\tau_{\rm{THz}}=50$~fs and $t_-=-3.5 \tau$ 
so that at the moment of the arrival of the optical pulse the system is polarized 
by a nearly constant electric field $\tilde{U}/d_{\rm{gap}}$. The effective dc bias 
$U$ is given by the difference between the self-consistent potentials inside the 
left and the right cylinders at $t=t_-$
\begin{equation}\label{eq:effectiveU}
  U=V(-R-d_{\rm{gap}}/2,0,t)-V(R+d_{\rm{gap}}/2,0,t), 
\end{equation}
and it equals to the difference between the Fermi energies of the cylinders.  
Note that $U \ne \tilde{U}$  because of the cylindrical geometry and small 
charge transfer during the rise of the THz field.

\section{Photoemission formulas}\label{Apd.A}
With Eq.~\eqref{eqSFA} and Eq.~\eqref{actionfunc}, we give expressions 
for electron transfer probability $\mathcal{P}_{\rm{L}\rightarrow \rm{R}}$ 
and the action function $S(t_{\rm{2s}},t_{\rm{1s}})$, respectively. 
For convenience we repeat the definition of the latter here.
\begin{align}\label{eqA1}
S(t_{\rm{2s}},t_{\rm{1s}})=&\mathcal{E} t_{\rm{2s}} 
+ \frac{{\tilde p}^2}{2}(t_{\rm{2s}}-t_{\rm{1s}}) 
-\int_{t_{\rm{1s}}}^{t_{\rm{2s}}}\frac{A_{\mathrm{opt}}^2(t')}{2} \;dt'\,\nonumber\\
&-\int_{t_{\rm{1s}}}^{t_{\rm{2s}}} V_{\rm{m}}[x(t')]\;dt'\, -\mathcal{E}_{F} t_{\rm{1s}},
\end{align}
and the corresponding electron transfer probability from the left to the right is 
expressed as:
\begin{align}\label{eqA2}
\mathcal{P}_{\rm{L}\rightarrow \rm{R}}\propto \sum_\mathcal{E} 
e^{-2\text{Im}[S(t_{\rm{2s}},t_{\rm{1s}})]}.
\end{align}

For a strong optical field where $\zeta<1$, we have 
\begin{align}
&\text{Im}[t_{\rm{1s}}] \equiv\text{Im}[t_{\rm{L}}] \nonumber \\
&=\frac{E_{\rm{g}}\sqrt{1+\gamma^2}}{E_{\rm{bias}}+E_{\rm{g}}\sqrt{1+\gamma^2}} 
\frac{\ln(\gamma+\sqrt{1+\gamma^2})}{\omega}, 
\end{align}
and
\begin{align}
\text{Im}[t_{\rm{2s}}]=0,\ ~~~~ \text{Im}[\tilde p]\approx 0.
\end{align}
Substituting these solutions into Eq.~\eqref{eqA1}, we obtain:
\begin{align}\label{eqA17}
\text{Im}\big[\mathcal{E} t_{\mathrm{2s}}\big] = 
\text{Im}\bigg[\frac{{\tilde p}^2}{2}(t_{\rm{2s}}-t_{\rm{1s}})\bigg]=0,
\end{align}
as well as
\begin{align}\label{eqA18}
&\text{Im}\bigg[\int_{t_{\rm{1s}}}^{t_{\rm{2s}}}\frac{A^2(t')}{2} \;dt'\,\bigg]=
\frac{\overline{\mathcal{W}}_{\rm{tb}}}{2\omega\gamma^2} \times \\
&\bigg\{ \sinh[\omega \text{Im}(t_{\rm{1s}})] \cosh[\omega \text{Im}(t_{\rm{1s}})]
-\omega \text{Im}(t_{\rm{1s}}) \bigg\} \nonumber\\
&-\frac{E_{\rm{g}}E_{\rm{bias}}}{\omega^3} \bigg\{ \sinh[\omega \text{Im}(t_{\rm{1s}})]-
\omega \text{Im}(t_{\rm{1s}})\cosh[\omega \text{Im}(t_{\rm{1s}})] \bigg\}\nonumber,
\end{align}
and
\begin{align}\label{eqA19}
\text{Im}\bigg[\int_{t_{\rm{1s}}}^{t_{\rm{2s}}} V_{\mathrm{m}}[x(t')]\;dt'\,\bigg]= 
- \overline{V}_{\mathrm{m}}\ \text{Im}[t_{\mathrm{1s}}].
\end{align}

Depending on the value of the Keldysh parameter $\gamma$ we distinguish two situations:  

\vspace{2em}
\noindent
\textbf{Case:} $\gamma \gg 1$
\vspace{2em}

In the multiphoton regime where $\gamma\gg1$, substituting the above solutions into the 
Eq.~\eqref{eqA2} and replacing the emission time with 
\begin{align}
\text{Im}[t_{\rm{1s}}]=
\frac{\omega\sqrt{\overline{\mathcal{W}}_{\rm{tb}}}}{E_{\rm{bias}} 
+\omega\sqrt{\overline{\mathcal{W}}_{\rm{tb}}}}\frac{\ln(2\gamma)}{\omega}, 
\end{align}
we obtain the formula describing the multiphoton regime:
\begin{align}\label{eqA20}
\mathcal{P}_{\rm{L\rightarrow R}} &\propto~ (2\gamma)^
 {-2\ \left(\Upsilon \frac{\overline{\mathcal{W}}_{\rm{tb}}}{\omega}\right)} \nonumber \\
 & \times(2\gamma)^{\frac{2 \Upsilon E_{\rm{bias}}E_{\rm{g}}}{\omega^3} 
 \cosh \big[\Upsilon \ln(2\gamma)\big]}\cdots,
\end{align}
where
\begin{align}
\Upsilon \equiv  \frac{\omega\sqrt{2\overline{\mathcal{W}}_{\rm{tb}}}}{E_{\rm{bias}} 
+\omega\sqrt{2\overline{\mathcal{W}}_{\rm{tb}}}}.
\end{align}

\vspace{2em}
\noindent
\textbf{Case:} $\gamma \ll 1$
\vspace{2em}

In the tunneling regime, $\gamma \ll 1$. As a result, we obtain 
\begin{align}
\text{Im}[t_{\rm{1s}}]= 
\frac{\sqrt{2 \overline{\mathcal{W}}_{\rm{tb}}}}{E_{\rm{bias}}+E_{\rm{g}}}.
\end{align}
Substituting this approximation into Eq.~\eqref{eqA2}, we obtain the formula 
for the probability of the field-driven tunneling:
\begin{equation}\label{eqA21}
 \mathcal{P}_{\rm{L\rightarrow R}} \propto 
 \exp{\left[-\frac{2(2 \overline{\mathcal{W}}_{\rm{tb}})^{\frac{3}{2}}}
 {3(E_{\rm{bias}}+E_{\rm{g}})}\right]}.
\end{equation}

\vspace{2em}

\section{Transport assisted by one photon absorption, 
perturbative treatment for the small optical field}\label{Apd.B}

Let us assume for simplicity that the electronic coupling between the nanowires 
and thus tunneling can be neglected in the ground state system. This implies that 
$\langle \psi_k^p | \psi_{k'}^{p'} \rangle = \delta_{k,k'} \delta_{p,p'}$ 
where $k$, and $k'$ label the ground-state bound K-S orbitals 
localized in the left (L) and right (R) nanowires as indicated with $(p,p')=(L,R)$ 
(see Figure~\ref{fig:Geometry} for geometry). 
Here, $\delta$ is the Kronecker delta. For identical cylindrical nanowires 
the energies $\mathcal{E}_k$ of the K-S orbitals are given by 
$\hat{H}_0 \psi_k^{p} = \mathcal{E}_k \psi_k^{p}$ where $\hat{H}_0$ is the ground 
state Hamiltonian. For an initial state given by the K-S orbital localized in the 
left nanowire $\psi_k^L(\mathbf{r})$, the 
time-dependent wave function $\Psi_k^L(\mathbf{r},t)$ is obtained within the TDDFT 
from the solution of the time-dependent K-S equation
\begin{equation}\label{eq:1}
  i \partial_t \Psi_k^L(\mathbf{r},t) = \left\{\hat{H}_0 +  
  \delta V(\mathbf{r},t)\right\} \Psi_k^L(\mathbf{r},t),
\end{equation}
where $\delta V(\mathbf{r},t)=V_{\rm{opt}}(\mathbf{r},t)+V_{\rm{ind}}(\mathbf{r},t)$ 
is the (small) time-dependent potential. It is given by the sum of the potential 
$V_{\rm{opt}}(x,t)=x~ a(t) \cos(\omega t+\varphi)$ owing to the incident field 
($a(t)$ is the slow envelope function) and the self-consistent response of the system 
$V_{\rm{ind}}(\mathbf{r},t)= V(\mathbf{r},t)- V_{\rm{gs}}(\mathbf{r})$. The latter 
accounts for the difference between the self-consistent time-dependent 
and ground state potentials. For the sake of brevity, below we drop off the 
coordinate variables. We are interested in the weak-field regime. 
Therefore, we use perturbation theory and linear response to describe the 
one-photon absorption process. With the wave function sought in the form 
 $\Psi_k^L(t)=e^{-i \mathcal{E}_k t} \psi_k^L + \delta \Psi_k^L(t)$, where  
$ \delta \Psi_k^L(t)$ is the perturbation, and keeping only the first 
order terms in $\delta V$ we obtain
\begin{equation}\label{eq:2}
  \left\{ i \partial_t - \hat{H}_0 \right\} \delta \Psi_k^L(t) 
  =  \delta V(t) e^{-i \mathcal{E}_k t} \psi_k^L.
\end{equation}
The potential $ \delta V$ can be found from
%
%\begin{equation}\label{eq:3}
%V_{\rm{ind}}(t)=\int_{-\infty}^{t} dt' \hat{\alpha}(t-t') V_{\rm{opt}}(t'),
%\end{equation}
%
%
\begin{equation}\label{eq:31}
 \delta V(t)= \int_{0}^{\infty} d \omega' ~e^{-i \omega' t}~ 
\left[1+\hat{\alpha}(\omega')\right] ~V_{\rm{opt}}(\omega') + CC,
\end{equation}
where $\hat{\alpha}(\omega,\mathbf{r},\mathbf{r}')$ is the response tensor, 
$CC$ stands for the complex conjugate, and $V_{\rm{opt}}(\omega')$ is obtained from 
\begin{align}\label{eq:32}
V_{\rm{opt}}(\omega') 
&= \frac{1}{2\pi} \int_{-\infty}^{\infty} dt~e^{i \omega' t} V_{\rm{opt}}(t) \nonumber \\
&= \frac{x}{2}~ \left[e^{i \varphi} a(\omega'+\omega) + e^{-i \varphi} a(\omega'-\omega) \right].
\end{align}
Since $a(t)$ is a slowly varying function only the second term of Eq.~\eqref{eq:32} 
will contribute to $\delta V(t)$ given by Eq.~\eqref{eq:31}. 
% so that approximately, 
%
%\begin{align}\label{eq:34}
%\delta V(t) &=  \frac{e^{-i \varphi}}{2} \int_{0}^{\infty} d \omega' 
%~e^{-i \omega' t}~ \left[1+\hat{\alpha}(\omega')\right]  x  A(\omega'-\omega) \nonumber \\
%&+ CC,
%\end{align}
%
Applying the rotating-wave approximation we arrive at
\begin{align}\label{eq:35}
  \left\{ i \partial t - \hat{H}_0 \right\} \delta \Psi_k^L(t) 
  =  e^{-i \varphi} e^{-i \mathcal{E}_k t} \delta v(t) ~\psi_k^L, 
  \end{align}
where  
\begin{align}\label{eq:36}  
 \delta v(t) =  \frac{1}{2} \int_{0}^{\infty} d \omega' 
~e^{-i \omega' t}~ \left[1+\hat{\alpha}(\omega') \right]\ x\ a(\omega'-\omega).
\end{align}
Finally, from Eq.~\eqref{eq:35} $\delta \Psi_k^L(t)$ can be found from
\begin{align}\label{eq:6}
\delta \Psi_k^L(t) = \frac{e^{-i \varphi}}{i} \int_{-\infty}^{t} dt'
~ e^{-i \hat{H}_0 (t-t')} e^{-i \mathcal{E}_k t'} \delta v(t') ~\psi_k^L.
\end{align}

The electron transfer probability to the right nanowire is given by
\begin{equation}\label{eq:7}
\mathcal{P}_{(L,k) \rightarrow R}(\varphi) 
= \sum_\nu \left| \langle \psi_\nu^R | \delta \Psi_k^L(t \rightarrow \infty) \rangle \right|^2.
\end{equation}
From Eq.~\eqref{eq:6}, and Eq.~\eqref{eq:7}, it follows that 
$\mathcal{P}_{(L,k) \rightarrow R}$ is independent of CEP given by $\varphi$. 
Consider now an initial state given by the ground-state K-S orbital localized 
in the right nanowire. From the symmetry of the system 
$\mathcal{P}_{(R,k) \rightarrow L}(\varphi) = 
\mathcal{P}_{(L,k) \rightarrow R}(\varphi+\pi)$, and from the demonstration 
above it follows then that $ \mathcal{P}_{(R,k) \rightarrow L}(\varphi) = 
 \mathcal{P}_{(L,k) \rightarrow R}(\varphi)$. The net electron transfer 
$\mathcal{N}=\sum_k \chi_k \left\{\mathcal{P}_{(L,k) \rightarrow R}(\varphi)-
\mathcal{P}_{(R,k) \rightarrow L}(\varphi)\right\}=0$. Here, the summation runs 
over occupied KS orbitals and $\chi_k$ accounts for the spin statistical 
factors and degeneracy associated with $z$-motion. We therefore conclude that 
if only one-photon absorption is considered, then $\mathcal{N}$ is zero for 
a symmetric system.

%\bibliography{../GAPSbiblio_Short}     % Produces the bibliography via BibTeX.


\end{document}